# Modelling income, wealth, and expenditure data by use of Econophysics

by

**Elvis Oltean**

Doctoral Thesis

Submitted in partial fulfilment of the requirements

for the award of

Doctor of Philosophy of Loughborough University

July 2015

i

*To my mother, Victoria Oltean*

# Acknowledgements

I wish to thank my parents and my PhD advisor, Professor Fedor Vasilievichi Kusmartsev.

# Modelling income, wealth, and expenditure data by use of Econophysics

Elvis Oltean

## Abstract

In the present paper, we identify several distributions from Physics and study their applicability to phenomena such as distribution of income, wealth, and expenditure. Firstly, we apply logistic distribution to these data and we find that it fits very well the annual data for the entire income interval including for upper income segment of population. Secondly, we apply Fermi-Dirac distribution to these data. We seek to explain possible correlations and analogies between economic systems and statistical thermodynamics systems. We try to explain their behaviour and properties when we correlate physical variables with macroeconomic aggregates and indicators. Then we draw some analogies between parameters of the Fermi-Dirac distribution and macroeconomic variables. Thirdly, as complex systems are modelled using polynomial distributions, we apply polynomials to the annual sets of data and we find that it fits very well also the entire income interval. Fourthly, we develop a new methodology to approach dynamically the income, wealth, and expenditure distribution similarly with dynamical complex systems. This methodology was applied to different time intervals consisting of consecutive years up to 35 years. Finally, we develop a mathematical model based on a Hamiltonian that maximises utility function applied to Ramsey model using Fermi-Dirac and polynomial utility functions. We find some theoretical connections with time preference theory. We apply these distributions to a large pool of data from countries with different levels of development, using different methods for calculation of income, wealth, and expenditure.

# Contents

# CHAPTER 1: INTRODUCTION

In the present thesis, we aim first to identify several distributions from statistical physics and study their applicability to phenomena such as distribution of income, wealth, and expenditure. After their identification and study of their applicability, we seek to explain possible correlations and analogies between economic systems and thermodynamic systems. We will try to explain their behaviour and properties when applied to macroeconomic aggregates and indicators. We seek to find links with economic theory.

Income, wealth, and expenditure are important variables that in the last twenty years became object of study using statistical mechanics distribution. This approach became an important part of Econophysics, involving more Economics than business and finance, which were initially the fields were Physics started to be applied in the social realm.

## 1.1 Short History

The interest of physicists for social and economic phenomena is longer than normally believed. Famous physicists and mathematicians were interested in economic or financial phenomena. The most notorious examples are:

- unsuccessful predictions of stock prices made by sir Isaac Newton and, consequently, his terrible loss in 1720 of 20000 pounds in South Sea speculation bubble.
- in 1738, Daniel Bernoulli introduced the idea of utility in order to describe preferences of people and consumer satisfaction.
- successful management of the fund for the widows of Goettingen professors, performed by Carl Friedrich Gauss.
- Giovanni Ceva published an essay "A mathematical approach of money" in 1711.
- Laplace in his work "Essai philosophique sur les probabilites" (1812) showed that what apparently might seem random and unpredictable (such as number of letters in the Paris dead-letter office) is predictable and obeys a simple law.
- Adolphe Quetelet (a former student of Fourier) studied the existence of patterns in data sets ranging from the frequency of different methods for committing murder to the chest size of Scottish men. It was him who coined the term "social physics" in 1835.

- explanation of the Brownian random walk and the formulation of the Chapman-Kolmogorov condition for Markovian processes by Louis Bachelier in his PhD thesis on the theory of speculation. This was done 5 years before the work of Smoluchowski and Einstein on diffusion, based on the observations of price changes at Paris stock-market.
- Italian physicist Ettore Majorana wrote in 1936 a paper based on analogies between statistical physics laws and the ones from social sciences [1-5].

## 1.2 Relation between Economics and Physics

The systematic relationship between Economics and Physics is as long as 130 years, when the marginalists began to use massively Mathematics, borrowing their tools from Physics [4]. Economics as a science was constantly criticised. There are plenty of reasons for people who criticise Economics, but the most important reason is about homo oeconomicus. This abstraction is accused of removing from human behaviour any cultural inclination, which implies that human beings act mechanically [6]. The old view in science was based on the attitude whereby mechanics is the only way to divine knowledge as Laplace claimed [7].

The other sciences which borrowed tools and methods from Physics had to give up this view not because of their repeated failures outside of Physics, but because in Physics it had to be abandoned. This view was continued by Walras and Jevons, the fathers of this view in Economics. Walras claimed that, in order to turn Economics into a science similar to Mathematics and Physics, one needs to be able to measure utility in a new different way such that utility can be measured mathematically and to measure exactly the influence of utility on prices [8].

Jevons stated that he tried to rebuild Economics as a mechanics of utility and of self-interest. However, Jevons did not explain how variables from mechanical equations can be replaced with common statistical data. He hoped that statistics would become more complete and exact such that formulae can get a precise sense [9].

Thus, Alfred Marshall and William Jevons, the main theoreticians of the Marginalist school, aimed of making Economics as a second-Physics and the fundamental notion of utility to be treated as mechanics of human interest. Implementation of this interdisciplinary field was necessary to overpass the limits of Economics. Its introduction was due to the mechanicist epistemology, which influenced to a large extent the modern economic thinking. This limitation was determined primarily by the decisive influences of human behaviour and human interrelations on economic activity, which change throughout the time and make impossible to explain them with mechanical-type methods and laws [6].

According to the theoreticians who followed the line of reasoning of Walras and Jevons, Economics uses the method whereby given means are used in order to accomplish given purposes. Thus, at any point in time there are given the means and the purposes of an individual; also, there are given technical and social ways by which the available means can be used accordingly in order to achieve the goals (partially or totally). The economy is reduced to a "mechanics of utility and self-interest". Consequently, any system which includes a conservation principle (given means) and a maximisation rule (optimal satisfaction) is a mechanic analogue [10].

Economics was built so far as a mechanic analogue. Continuing the same line of reasoning, we could represent the economic process through a new system of equations created similarly with thermodynamics. In principle, one could write the equations of two systems (production and consumption). Then, we can assemble these equations in a giant system or to transform them into a smaller one. But in order to write a set of initial equations, we have to know the exact nature of the process considered. And the difficulty related to this process is that the economic process on the long term (just as the biologic one) is inevitably under the influence of qualitative changes, which cannot be predicted. This is true since life has to rely on new mutations to continue its existence in an environment which is changed irrevocably by this. Therefore, no system of equations can describe the development of an evolutionary process [6].

There are some differences between economic process and physical process in nature. Thus, the product of the economic process does not consist of a physical flux of waste but of pleasure of living. One can see that a good description of economic process is not viable as long as it contains only physical attributes. Pleasure of living does not correspond to any characteristics of elementary material properties. As Frank Fetter and Irving Fisher stated, this psychological flux is the right notion for income in economic analysis [11-12].

## 1. 3 Econophysics

The word Econophysics was coined by Henry Stanley in order to categorise a large number of research papers written by physicists following the year 1990. These papers aimed at the problems arising when dealing with capital markets, financial derivatives, macroeconomic development, and other topics related to socio-economic area [13-15]. Consequently, most researchers categorised it at the border between Physics and Finance.

Econophysics was introduced by analogy with other branches of physics such as Astrophysics, Geophysics, and Biophysics, which deal with applications of Physics to different areas. However, Econophysics does not apply the laws of Physics literally (for instance Newton's laws or quantum mechanics) to humans, but rather uses mathematical methods developed in statistical physics to study statistical properties of complex economic systems composed of a large number of individuals. Also, it can be categorised as a branch of applied theory of probabilities. However, it is important to keep in mind that statistical physics is different from mathematical statistics in its focus, methods, and results. Originating from Physics as a quantitative science, Econophysics deals with quantitative analysis of large amounts of economic and financial data, which is possible due to the massive introduction of computers and of the Internet. Econophysics distances itself from the style of political economy and is closer to Econometrics in its focus. Dealing with mathematical models of a large number of interacting economic agents, Econophysics is related to agent-based modelling and simulation. Consequently, it distances itself from the mainstream Economics, which ignores statistical and heterogeneous aspects of the economy [16].

Paradoxically, the ones who introduced for the first time methods from Physics in order to study economic systems were economists. A Monte Carlo simulation of a capital market was published in 1964 by G.J. Stiegler, a representative of economics school from Chicago [17]. In 1989, Harry Markowith published along with G.W. Kima a paper with regard to the market crash on the Wall-Street in 1987, which includes a model that contains two types of investors, similar with many other models which were published later on by econophysicists [18]. Keynes has already stated in his book written in 1934 "The general theory of employment, interest, and money" [19] that changes in stock shares prices originate in collective behaviour of numerous interacting agents rather than in fundamental values that can be deducted from the analysis of the current situation and from future prospects of companies. Thomas Lux started to model explicitly the idea and to propose a new theoretical model which relates the stock market price crashes to phase transition studied in statistical mechanics. Thus, he explains the emergence of speculative bubbles and crashes in terms of a self-organising process consisting of influences among heterogeneous traders. Following the same line of reasoning, Brian Arthur's research [20] leads to a "historic memory" (hysteresis) of a system, which can lead to multiple equilibria instead of a unique solution of uncooperative equilibrium [5].

Throughout the entire 20$^{th}$ century, three related fields: mathematical statistics, economic statistics, and statistical physics evolved on parallel trajectories creating their own identity, after having defined very clearly their borders. Removal of these artificial borders is a phenomenon that occurred in last two decades, when the paradigm of science of complexity brought to attention processes whose investigation required the tools of certain theoretical frameworks from several sciences. One can notice that the field of Econophysics developed exponentially in the last ten years given tens of books and thousands of articles published at prestigious publishing houses and journals. Numerous famous universities included on their research topics related to the Physics of socioeconomic systems and an increasing number of physicists worked in financial system. Actually, the branch of Physics which is applicable to social sciences is statistical physics. This is based on the principle that properties of a system are not just the sum of the properties of their elements. The key to this difference is owed to the interaction among the consisting elements and stochastic nature of their behaviour, which requires proper averaging statistical methods. Most of the methods developed in statistical physics do not impose restrictions regarding the components, which are an important aspect of the applicability of these methods in describing the evolution of human collectiveness. What Econophysics can bring in a scientific area more than financial mathematics is not very clear, as the border between theoretical Physics and some branches of Mathematics is also not clear. The differences are about the formulation of the problems, generality of the utilised equations and, mainly, about the interpretation regarding the obtained results. Dealing with observables, the gap between theoretical Physics and economic modelling is narrowed, having defined a meaning for the variables considered. Similarly with Physics, Econophysics can be divided in experimental and theoretical. First one has as an objective the analysis of series of data originated from real stock market and decoding the information that these contain. The second one builds microscopic models which can lead to some values of observables, which are similar with empirically determined values. What econophysicists achieved was to make a systematic categorisation of a process, interpret, and model empirical data, which allowed the access of more researchers to these databases and introduction of new specialised methods less known in Economics. From this point of view, statistical physics can be very helpful as it operates with dynamic collective systems, which consist of many interacting components. The main object of study is the probability distribution function of price variations at a given temporal scale [5].

Historically, the evolution in this area began with collection and study of "social numbers", such as the rates of death, birth, and marriage [21]. This has been growing progressively starting from the seventeenth century [22]. The term "statistics" came up to designate these studies dealing with the civil "states" in the eighteenth century and its practitioners were called "statists". A turning point can be considered the work of the Belgian astronomer Adolphe Quetelet. Initially, statistics was a part of political economy, but later on turned into a general method of quantitative analysis applicable to all disciplines. Thus, physicists created statistical mechanics in the second half of the nineteenth century. This phenomenon was described by Philip Ball [22]: "Today, physicists regard the application of statistical mechanics to social phenomena as a new and risky venture. Few, it seems, recall how the process originated the other way around, in the days when physical science and social science were the twin siblings of a mechanistic philosophy and when it was not in the least disreputable to invoke the habits of people to explain the habits of inanimate particles." Historical studies show [22] the important role of the widespread popularity of social statistics in developing statistical mechanics. Boltzmann was very explicit [22]: "The molecules are like individuals, ... and the properties of gases only remain unaltered, because the number of these molecules, which on average has a given state, is constant. In "Populäre Schriften" [23-24], Boltzmann said "This opens a broad perspective, if we do not only think of mechanical objects. Let us consider the application of this method to the statistics of living beings, society, sociology and so forth." Also, Mandelbrot observed [25]: "There is a great temptation to consider the exchanges of money which occur in economic interaction as analogous to the exchanges of energy which occur in physical shocks between gas molecules." Mandelbrot [26] had for the first time the idea of comparing properties of price distribution on different temporal scales. The phenomenon highlighted by his research is the scale invariance: the price variation for a temporal scale can be obtained from the one obtained for a smaller temporal scale. In September 1987, the Santa Fe Institute hosted the first international conference for economists and physicists. The research papers and opinions presented were published during the next year under the title "The economy as an adaptive evolutionary complex system" [5].

Econophysics emerged as a consequence of the application of methods from statistical physics to financial markets, but subsequent scientific works in Econophysics focussed on three areas. The first area is about prices on capital market, currency exchange rates, and the prices of goods. The second one is about size of firms, macroeconomic aggregates, individual income and wealth. The third one is about analysis of economic phenomena using network type models [1-2].

As an example for further analogies, in Table 1 we find analogies between Mechanics and Economics [4], [27]

| Table I | |
|---|---|
| Mechanics | Economics |
| There are elementary entities called "material points", whose behaviour is suitable for experimental observation | There are elementary entities called "agents", whose behaviour is suitable for experimental observation |
| The behaviour of the material points is completely described by a set of generalised coordinates, $x_1, x_2, …, x_N$. | The behaviour of the agents is completely characterised by a set of parameters, $x_1, x_2, …, x_N$., so that the system is completely described by setting the numerical values of the parameters. |
| There is a quantity called potential energy $U$, which is measurable in the sense that, given the coordinates of the systems (the quantities $x_1, x_2, …, x_N$.), one can determine the potential energy of the system place in such position. | There is a quantity called utility $U$, that is measurable in the sense that, given the parameters of the systems (the quantities $x_1, x_2, …, x_N$.), one can determine the utility of the system found in such conditions. |
| The evolution of the system is given by | The evolution of the system is given by |

| the derivative of potential energy $U$, which exactly means that the second derivative of each coordinate is proportional to the derivative of potential energy $U$ with respect to the coordinate: $$\frac{\partial x_i^2}{\partial t^2} = -\frac{1}{m}\frac{\partial U}{\partial x_i}, \; i = 1,....,n$$ | the derivative of potential energy $U$, in the somewhat vague meaning that the system tends to the maximum of the energy U, or that there is a generic force, proportional to the marginal utility $U$, or the derivative of $U$, that drives the system. Only in some cases does this meaning become precise and the second derivative of each coordinate is proportional to the derivative of the potential energy $U$ with respect to the coordinate: $$\frac{\partial x_i^2}{\partial t^2} = -\frac{1}{m}\frac{\partial U}{\partial x_i}, \; i = 1,....,n$$ |
|---|---|

As for theoretical side of Econophysics, in the last 10 years many models were created and published for modelling the microstructure of stock market, each model outlining explanation of the results of the empirical data. There is no unique model unanimously accepted of stock market.

Diffusion-reaction model (Bak-Paczuski-Shubik) considers that agents form a market where big fluctuations of prices exist due to the nature and behaviour of two types of agents: first one is represented by "noise" traders, whose prices are volatile, dependent upon recent variations of market and which tend to influence mutually (by imitation) in establishing prices; the second type is represented by "rational" traders that set the transaction prices on the basis of maximisation of their own utility function. In spite of the significant simplifications, these models lead to a statistical structure of prices variations similar to the empirical one.

Hierarchy model (Johansen-Sornette) tries to describe mainly transitory behaviour of stock market and financial crash. One considers that individual agents (trader of level 0) are organised in groups of individuals, each group being a trader of level 1. These are organised likewise, in groups of m individuals forming traders of level 2 and so on. An important consequence of this type of organisation is that decision of a trader can influence only a limited number neighbouring traders, situated at the same level or to lower levels of hierarchy. Because of a waterfall effect, decisions to a lower level can exert powerful effects on upper levels. The model highlights a collapse at a well-defined critical moment, similar to the one noticed in the study of phase transition in statistical physics.

Percolation model (Cont-Bouchard) is characterised by simplicity of its principles and by profound physical sense. Each node of a network is randomly occupied by a trader (with probability p) or vacant (with probability 1-p). Neighbouring traders in the network form a cluster that acts as a company: at every moment each cluster selects to sell (probability a) or not to take part at transactions (probability 1-a). The volume of transactions part of a cluster is proportional with number of nodes that are part of that cluster. Difference between overall demand and supply (both calculated for all clusters) determines the upward or downward evolution of price.

Generalized Lotka-Volterra model (Levy-Solomon) is a development of auto-catalytic model from Physics and Chemistry. Considering the accumulation of the earning as a stochastic and multiplicative process, this model explains the emergence of power-laws in asymptotic parts of probability distribution function associated with variations of prices. The model is applicable to financial markets and to social and macroeconomic aggregates.

**1.4 Methodological aspects regarding income, wealth, and expenditure distribution**

Income, wealth, and expenditure distribution are among the most important issues in a society considering that an optimal level ensures social stability, while a high degree of inequality causes multiple problems.

A recent book by Wilkinson [28] shows that for the developed countries there is a direct relation between economic inequality and all problems that impact negatively on society such as criminality, social trust, obesity, infant mortality, violence, child poverty, mental illness, imprisonment, and many others that characterise the quality of life. Thus, the countries with the lowest inequality, such as Scandinavian countries and Japan, have the best indicators regarding these social phenomena that affect the quality of life. The opposite is represented by the USA, which has the highest inequality among developed countries. Thus, the USA is characterised by the highest impact of negative phenomena affecting the society among the developed countries.

The study of income distribution defined broadly is important for three reasons. The first reason is to know income distribution and how this is related to the way societies are organised. The second one is about the need to know the impact of public policies on different socio-economic groups. Examples where data regarding income distribution can be considered important are welfare, taxation and other fiscal policies, housing, education, labour market, and health. The third interest is about how different patterns of income distribution are related to well-being and the ability of individuals to acquire goods and services necessary for their needs. Consequently, these are used to study poverty, social exclusion, and the consumer behaviour.

The questions asked in the study of income distribution address:

- the inequality in a country, its evolution over time, and comparison with other countries
- the characteristics of groups with low income or at risk of poverty and, also, which groups are in most financial need and how is this related to their previous evolution and with the evolution of similar groups from other countries
- the evolution of real income and how is this related to fiscal and monetary policies
- how social transfers affect the income of certain segments of population
- if population has sufficient income for an adequate living

Generally, the main interest is about changes over time, while differences between countries are second in importance. Income distribution offers information with regard to the overall performance of the whole economy. Also, income distribution statistics shows how this evolves over time, across regions or between subgroups of the population. Moreover, it studies the way that needs of people vary on the basis of composition and age.

The well-being of the population can be expressed in terms of its access to goods and services. Consequently, the more a household can consume, the higher its level of economic well-being. Other approaches have considered other determinants of human well-being reaching beyond the commodities that are available to them. Consumption is also an indicator of economic well-being. However, an individual or household may choose not to consume the maximum amount but to save at least some of the resources it has available. By saving, one can accumulate wealth which will generate income at a later date. In order to have a better understanding, not only income but also wealth and consumption should be considered.

Income is about receipts, whether monetary or in kind, that are received annually or more frequently and are available for current consumption. The mainstream opinion views income as the most important determinant of economic well-being, as it provides a measure of the resources available for consumption and saving. Consumption expenditure can be financed not only by household income but also by savings from previous periods or by incurring debt. For some groups such as farmers, they can average out their consumption over a number of years, while their incomes exhibit wide fluctuations over the same period. In such cases, consumption expenditure represents a better estimate of the individual or of the household sustainable standard of living. In fact, the choice between the income or the consumption expenditure approach to the measurement of economic well-being was made on the basis that income data is more available than data on consumption expenditure.

The income or consumption expenditure data should ideally be accompanied by some assessment of the change in the value of net worth during the accounting period. An increase in the net worth is from savings (the difference between income and consumption), from the receipt of capital transfers, or from other changes in the value of assets.

One way to do that is to annuitise the net worth held and add this annuity to the flow of income and other receipts. Consequently, analysis of economic well-being would benefit to a large extent from the availability of fully articulated survey data covering all aspects: income, expenditure, saving, and the value of wealth held. Where survey data collection is not possible, one may match records or information from different sources in order to allow inferences.

Most users expect the data providers to have undertaken reconciliation between the macro aggregate of household income and the micro income statistics suitably matched to population totals. When this is not possible, the data producer should provide explanations when differences are known to exist. Such reconciliation, with any discrepancies clearly explained, is best practice for national statistical offices. There are several reasons for the maximisation of comparability between income distribution statistics and national accounts. First, it is likely that any datasets collected can be used for multiple purposes. For example, the use of the microdata in compilation or benchmarking of national accounts estimates. Second, statistics compiled under different frameworks are used for mutual checking process, and users can utilise different sets for analytical purposes.

Transfers are receipts for which the recipient does not give any compensation. Transfers can consist of cash (in the monetary sense), of goods or of services, and may be made between households, between households and government, or between households and charities, both within or outside the country. The aim is to redistribute income either by government (e.g. pensions) or privately (e.g. child support).

Longitudinal (or panel) data give a better understanding of income by taking a longer term, dynamic view. It shows the persistence of this state over time, and transitions into and out of it.

The well-being concepts require dealing with income, wealth, and consumption. These concepts are concerned with describing the total economic value of the resources received, owned, or used up by people. However, income only provides a partial view of economic well-being. Income, a flow measure, is volatile sometimes. Wealth, a stock measure, is more stable over time, reflecting accumulated savings and investments over time, which can be drawn on in times of need. Reserves of wealth can be also utilised to generate income. Wealth can be held in assets that are not easily converted into money, its existence may allow people to borrow money to finance expenditures.

Average income, consumption, and wealth are useful statistics, but do not provide the whole picture about living standards. For instance, a rise in average income could be unequally distributed across groups, making some households relatively worse-off than others. Mean measures of income, consumption, and wealth should be accompanied by indicators that reflect their distribution [29].

## 1.5 Literature Review and Theoretical Framework of Statistical Mechanics distributions applied to income, wealth, and expenditure distribution

The investigation of income distribution using statistical physics methods has a long history [30]. Pareto in 1897 suggested that income distribution would obey a universal power law valid for all times and countries [31]. Subsequent authors as Shirras [32] disputed this. Mandelbrot [33] proposed a "weak Pareto law" applicable only asymptotically to the high incomes. In such a form, Pareto distribution is not applicable in describing the majority of population. Many other distributions of income were proposed: Levy, log-normal, Champernowne, Gamma, and two other forms by Pareto himself [34]. The rationale for these was found by two schools: socio-economic and statistical. The former uses economic, political, and demographic grounds to explain the distribution of income [35], whereas the latter invokes stochastic processes. Gibrat [36] proposed log-normal distribution in 1931 considering that income is substantiated on a multiplicative random process. Consequently, these ideas were carried on by Montroll and Shlesinger [37]. However, Kalecki [38] stated that the width of this distribution is not stationary, but increases in time. Levy and Solomon [39] proposed a cut-off at lower incomes, which stabilises the distribution to a power law. Modern econophysicists [39-41] also use various versions of multiplicative random processes in theoretical modelling of wealth and income distributions.

A most exact case analogy consists of thermally isolated system in Physics with a closed macroeconomic system in Economics that was made by Dragulescu and Yakovenko [16]. Thus, energy of particles from a gas can be considered as an analogue for money that each agent possesses. However, nowadays the economy is composed of open free-markets where trading takes place so the number of agents and money vary over time. Given this characteristic of the open-market economies, the best description for such case is *grand canonical ensemble*. Thus, grand-canonical ensemble is characterised by a varying amount of energy and number of particles even though the averages of energy and number of particles of the containing system are fixed. To continue the analogies between the two areas, we use market snapshots as microstates and their averages as macrostates. The procedure to obtain the average from a vast number of market microstates involves getting the average values for characteristic parameters. Given that market is a complex system with a very large number of degrees freedom and with many connections not always fully clarified, long-term market evolution is unpredictable and, therefore, chaotic. Moreover, the composing particles of an economic system are most often chaotic (as majority of transactions occur randomly), so the analogy of this chaotic behaviour in thermodynamic systems is represented by *ergodic hypothesis*. Subsequently, markets pass through all possible states in a time interval given. Consequently, just like in the case of statistical physics, the chaotic nature that characterises the market makes possible to replace time averages for a single system with averages at a fixed time over a certain ensemble trading goods, while each member is identical to the ensemble with regard to its global macroscopic properties [42-43].

The aim of Econophysics so far was to analyse and predict financial markets and financial data. More recently, a new field in Econophysics emerged, which deals with macroeconomic aggregates such as income, wealth, and size of companies [1-2]. This is an area that traditionally belongs to Economics [21]. Thus, three major papers [16], [44-45], dealing with the subject of money and wealth distributions, were published in year 2000.

The quantitative characterisation of income distribution is a persistent problem of Economics. First distribution to tackle this was Pareto distribution which is the most widespread distribution. Yakovenko and Dragulescu [16] introduced exponential distribution which is similar to Boltzmann-Gibbs distribution. Chakraborti et al. [44] came up with an ideal gas model of a closed economic system, which is characterised as fixed with regard to money and number of agents. Chatterjee et al. [46] introduce kinetic exchange models that physicists have developed to understand the reasons and to formulate remedies for income inequalities. Clementi et al. [47-49] used the k-generalized distribution as a descriptive model for the size distribution of income. Lognormal distribution is also another distribution which was applied successfully to low and middle income part of the population. Moura Jr. and Ribeiro [50] showed that the Gompertz curve combined with the Pareto power law provide a good description for whole income distribution. Tsallis [51] is another distribution which claims to fit the entire income range.

Silva and Yakovenko [52] show the very important fact that society is two-class structured, even though this was highlighted for the US case only. Thus, the majority of population is contained in the lower class (97-99%) and is characterised by a very stable exponential in time. The upper class (1-3% of the population) has a power law distribution. Moreover, the parameters change significantly over time depending on the evolutions of stock market.

Yakovenko and Dragulescu use exponential cumulative distribution function and probability density function [16], [40], [53-60] for different measurements of income distribution by taking into account mean disposable income expressed in annual values for different countries (mostly developed) for the recent years. Thus, using log-log scale for lower income part of population and log-linear for the upper one, they show that income is fitted very well by exponential distribution as an analogue for Maxwell-Boltzmann distribution. The countries considered are developed countries such as USA, the UK, Germany, and Australia. The data are very well fitted and they use Lorentz curve and Gini coefficient in order to illustrate the inequality. They show that the upper part class of income data can be fitted well by Pareto distribution using log-linear plotting.

An important contribution is from Kusmartsev and Kurten [42] which, using a numerical simulation based on a trading process of a closed market, show that, in general, money distribution occurs according Bose-Einstein distribution. Also, in [43] Kusmartsev following the same line of reasoning constructs a market model based on statistical mechanics equations and brings additional evidence regarding the applicability of Bose-Einstein distribution by using data about income distribution in the USA.

**1.6 Data description**

The data about income, expenditure, and wealth that we will be using are non-cumulative and cumulative.

In the case of cumulative income and probabilities, we will use deciles of population. Income or wealth deciles are groupings that result from ranking either all households or all individuals in a population in ascending order, according to their income or wealth, and then dividing the population into ten equal groups, each comprising 10% of the estimated population. The first decile contains the bottom 10%, the second decile contains the next 10%, and the tenth decile contains the top 10%. Income which is not stated or not known is excluded from the calculation of deciles [61]. Decile is defined as "one of the values of a variable that divides the distribution of the variable into ten groups having equal frequencies" [62].

Simply defined, let us assume the population of a country be 100 individuals. We rank them increasingly on an axis, according to their income earned or wealth possessed. Thus, the first decile contains the first ten people with the lowest income or wealth. The second decile contains the second group of ten people whose income or wealth are ranked higher than the people from the first group (first decile). Thus, the population can be divided in ten deciles or ten groups of individuals or households, each decile containing ten individuals or households.

Mean income of an income decile is the sum of all income that individuals or households earn in that decile and then divided to the total number of individuals or households contained in that decile.

Upper limit on income is the income of the individual or the household having the highest income from all individuals or the households contained in a decile. The term is used by National Institute of Statistics of Finland [63].

Lower limit on income is income of the individual or the household having the lowest income from all individuals or households contained in a certain decile. The term is used by Office for National Statistics from the UK [64].

Mean value in these cases is calculated as follows: all national data (regardless the fact they are about income, expenditure, or wealth) are ranked in increasing order of their values. All these data are divided in ten equal shares and the mean values (for income, expenditure, or wealth) are calculated as an arithmetic mean of all the values comprised in a decile. The expenditure is calculated based on income deciles. Thus, in each income decile the expenses are summed up and consequently divided to the total number of individuals or households contained in a certain income decile. Mean expenditure (in the case of Uganda) is calculated based on income deciles. Thus, the total amount of expenditure made by people in an income decile is divided to the number of people. The upper limit on income comprises 90% of the population, as for the upper 10 % data were not provided. In the case of mean income, the data cover the entire population as it is possible to calculate the mean income for the richest part of the population.

We present a typical example for mean income in Finland from the year 1987 in Figure 1.1.

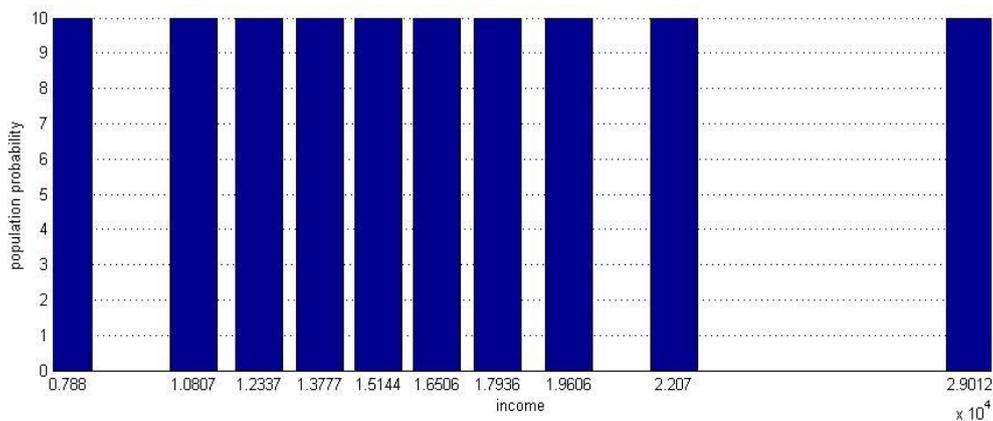

**Figure 1.1 Mean income in Finland in the year 1987.**

On the y-axis, we present the population probability of the ten deciles whereby each decile represents a frequency of 10% of population. On the x-axis, we present deciles of disposable income. Thus, the mean income of the first decile is 7880 euro, for the second decile is 10807 euro, and for the tenth decile is 29012 euro.

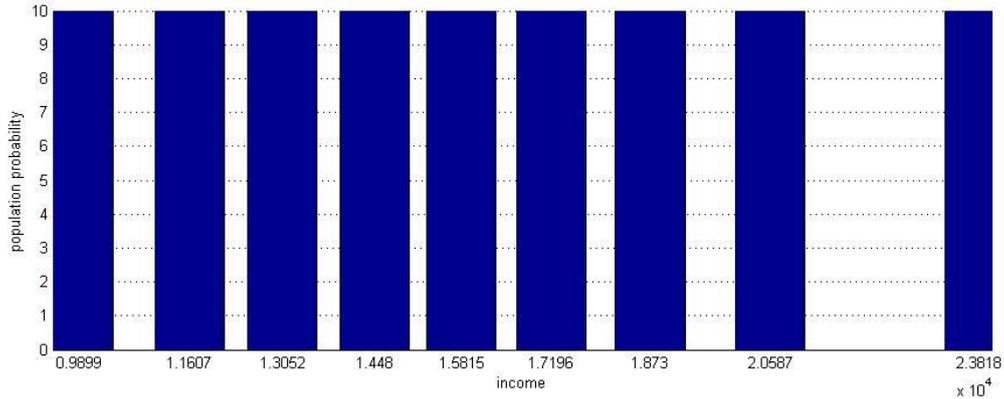

**Figure 1.2 Upper limit on income in Finland in the year 1987.**

On the y-axis, we present the population probability of the deciles, each decile having a frequency of 10% of population. Unfortunately, the data regarding the tenth decile were not provided by the National Institute of Statistics of Finland. On the x-axis, we present the values for deciles of income. Thus, the upper limit on income of the first decile is 9899 euro, for the second decile is 11607 euro, and for the ninth decile is 23818 euro. We can notice that the decile values are higher than in the case of mean income.

Fermi-Dirac and polynomial probability density functions can be also fitted to data which do not consist of deciles of data. The data are expressed using different percentages of population based on thresholds of population arbitrarily chosen. As you can see in the Figure 1.3, the level of income is ranging from below 15000 USD up to higher than 200000 USD per annum.

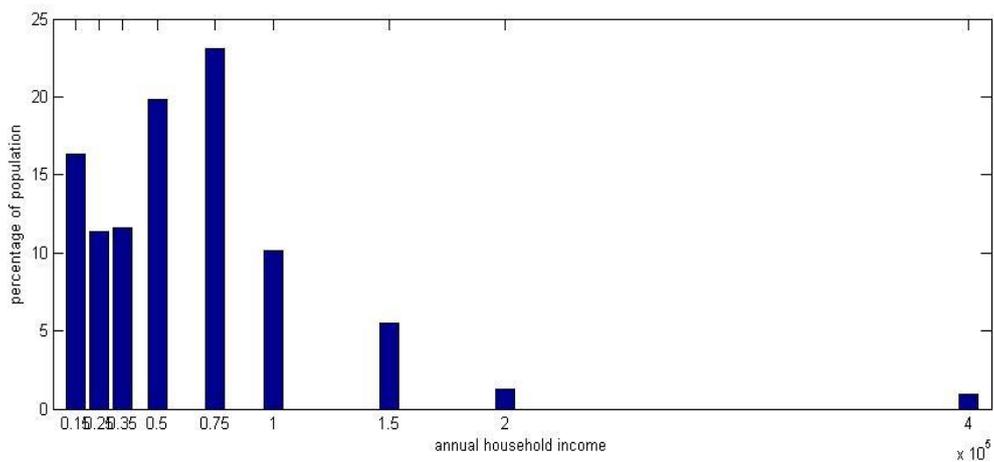

**Figure 1.3 Annual household income in the USA in the year 1967.**

We have obtained the data mostly from national statistics bodies or national bank. In several cases, we were able to get the data from some papers published. We will be using data from Brazil [65], Finland [63, 66], France [67-71], Italy [72-73], Philippine [74], Romania [75], Singapore [76], the UK [64], Uganda [77], and USA [78].

**1.7 Theoretical Framework**

We intend to use this subchapter in order to describe the theoretical notions that we will be using and how the results can be validated based on the theoretical framework.

We will start first with the statistical test which describes how good is fitting function for the data we described above. For the statistical validation of data we will be using correlation coefficient, coefficient of determination, t test, and Durbin Watson test.

**Pearson correlation coefficient (r)** or simply the correlation between two sets of data X and Y shows how strong is the relationship between them and is defined as

$$r = \frac{\sum(X-\bar{X})(Y-\bar{Y})}{\sqrt{\sum(X-\bar{X})^2}\sqrt{\sum(Y-\bar{Y})^2}} \quad (1.1)$$

In the Figure 1.4 we could see how the data are distributed for different values of r.

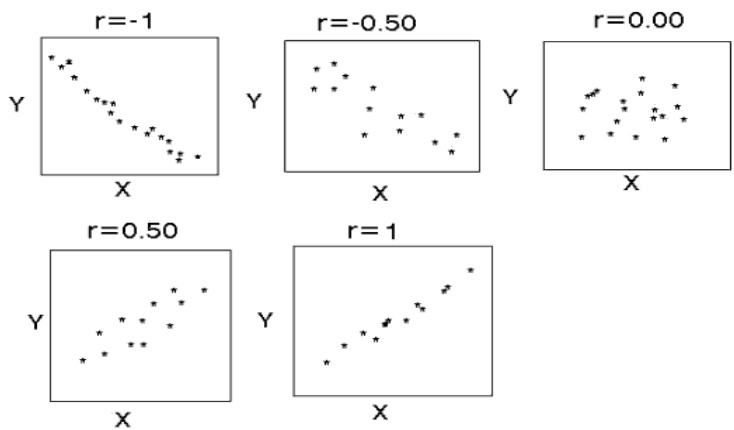

**Figure 1.4 Variance according to different values [79]**

Thus, r lies between 1 and -1. A value for r=0 indicates no relationship between the two variables. While for r=-1 and for r=1 there is a (perfect) strong relationship (negative in the first case and positive for the second)[80].

**Coefficient of determination ($R^2$)** is used to measure the goodness of the fit for two variables. It is actually the square of the correlation coefficient (r). Since the correlation coefficient lies between -1 and 1, the coefficient of determination is between 0 and 1.

$R^2$= (variation attributed to independent variable)/ (total variation of dependent variable). If $R^2$=1, this implies that 100% of the variation in the dependent variable can be explained by the variation in the independent variable [80].

**T-test** can be used for comparing two sets of data in order to test their significance (if they have different means) and the significance of the parameters of a regression function.

When we use it in order to assess the significance of two sets of data, we use the following formula [81]

$$t = \frac{\bar{x}_1 - \bar{x}_2}{\sqrt{\frac{s_1^2}{n_1} + \frac{s_2^2}{n_1}}} \quad (1.2)$$

where $\bar{x}_1$ and $\bar{x}_2$ are the means of the two sets of values, $s_1$ and $s_2$ are the standard deviations of the two sets of data, and $n_1$ and $n_2$ are the number of values in the two sets of data. Standard deviation is calculated as follows

$$s = \sqrt{\frac{\sum(x-\bar{x})^2}{n-1}} \quad (1.3)$$

The values for t test when comparing two sets of data can be determined according to the table of values calculated according to the number of degrees of freedom and a the confidence interval chosen. If the result is higher than the values from the table, we can conclude that the relationship between the two sets of data is significant. If the value from the t-test is below the value described in the table, the relationship is not significant..

In the case of regression, the t test statistic is t = (observed - expected) / (standard error). The expected value for the coefficient is 0 (the assumption is that the null hypothesis is true and the null hypothesis is that the β is 0), the test statistic is found by dividing the coefficient by the standard error of the coefficient [82].

T-test can be calculated then according to [80] as

$$TS = \frac{b}{s_b}$$

where b is a slope

$$s_b{}^2 = \frac{s^2}{\sum x_i}$$

and s is calculated according to the equation (1.3).

**Durbin-Watson test** studies the autocorrelation of the residuals resulted from a regression. Autocorrelation often occurs because of 'momentum' in many economic time-series which lead to self-sustaining upswings and downswings. Thus, a current disturbance is likely to be influenced by the disturbance from the previous period. The most famous model aiming to address autocorrelation is the first–order autoregressive process. This is done by using the Durbin Watson test [80]

$$dw = \frac{\sum_{t=2}^{\infty}(e_t - e_{t-1})^2}{\sum_{t=2}^{\infty}(e_t)^2} \quad (1.4)$$

where $e_t$ are residuals from an OLS regression. The values will be between 0 and 4. Values between 0 and 1 indicate strong positive autocorrelation, between 1 and 2 indicate weak positive autocorrelation. Values equal to 2 indicate no autocorrelation or a perfect model. Values between 2 and 3 indicate weak negative autocorrelation and values between 3 and 4 indicate strong negative correlation.

We calculate probabilities using probability density function, cumulative distribution function, and complementary cumulative distribution function.

**Probability density function** (pdf) is defined as

$$P(a \leq X \leq b) = \int_a^b f(x)dx \quad (1.5)$$

A probability density function must satisfy two requirements:

$$f(x) \geq 0 \text{ for all x} \quad (1.6)$$

and

$$\int_{-\infty}^{\infty} f(x)dx = 1 \quad (1.7)$$

The probability that a random variable X takes on a value at or below a given number *a* is

$$C(a) = P(X \leq a) \quad (1.8)$$

and the function is called **cumulative distribution function** (cdf). A cdf must satisfy certain properties such as:

$$0 \leq C(X) \leq 1 \quad (1.9)$$

$$a < b, C(a) \leq C(b) \quad (1.10)$$

$$\lim_{x \to \infty} C(X) = 1 \text{ and } \lim_{x \to -\infty} C(X) = 0 \quad (1.11)$$

Property (1.9) shows that *C(X)* is a probability [83].

**Complementary cumulative distribution function** unlike cumulative distribution function shows the probability that a variable to be above a certain value [84].

$$\bar{C}(a) = P(X \geq a) \quad (1.12)$$

The complementary cumulative distribution function $\bar{C}$ must also satisfy certain criteria:

$$0 \leq \bar{C}(X) \leq 1 \quad (1.13)$$

$$a < b, \bar{C}(a) \geq \bar{C}(b) \quad (1.14)$$

$$\lim_{x \to \infty} \bar{C}(X) = 0 \text{ and } \lim_{x \to -\infty} \bar{C}(X) = 1 \quad (1.15)$$

# CHAPTER 2: APPLICATIONS OF LOGISTIC FUNCTIONTO THEDISTRIBUTION OF INCOME, WEALTH, AND EXPENDITURE

Fermi-Dirac, Bose-Einstein, and Boltzmann-Gibbs distributions are the most important in statistical physics. Of these statistical physics distributions used so far for modelling socio-economic systems with some degree of success were Bose-Einstein and Maxwell-Boltzmann distributions. The present chapter investigates the applications of logistic distribution to some of the most important economic variables such as income, wealth, and expenditure of the population from nine countries with different economic characteristics. This distribution was used outside economic systems initially and more recently it started being used due to the similarity of economic systems with biological and physical systems.

## 2.1 Methodology

The probability distribution used is cumulative logistic distribution which is applied to cumulated income, expenditure, or wealth on one hand and also to cumulated probabilities on the other hand. Logistic function or sigmoid function is defined as

$$f(x) = \frac{L}{1+exp^{-k(x-x_0)}} \quad (2.1)$$

where L is the curve's maximum value, $x_0$ is the x-value of the sigmoid's midpoint, and $k =$ the steepness of the curve[85]. Logistic map, which is the basis for logistic function, is used to show how complex, chaotic behaviour can arise from very simple non-linear dynamical equations [86].

We use logistic cumulative probability distribution $C(x)$, which is defined as the integral

$$C(x) = \int_{-\infty}^{x} P(x)dx \quad (2.2)$$

It gives the probability that a random variable is below a given value *x*. We present on y-axis the cumulated population probability, which is the share of population with income/wealth/expenditure lower than corresponding level on the x-axis. Cumulated income/wealth/expenditure is contained on the x-axis. According to this type of probability, we calculate the share of population having an income below a certain threshold. Thus, the probability to have an income lower than zero is 0 % (since everyone is assumed to have a certain income).

Cumulated income, wealth, or expenditure is contained on the x-axis. Let us assume X represents the values for cumulated income/wealth/expenditure represented on the x-axis.

$$X_i = \sum x_i$$

where X represents the cumulated income/wealth/expenditure on the x-axis and i=[1,10] for mean values and i=[1,9] for upper limit on income, where i∈N. Thus, the decimal logarithm of probability, which is $\log_{10}(C(x))$, is the dependent probability and decimal logarithm of *X* (cumulated income) is the independent variable. Also, parameters a, b, and c are obtained from fitting the data using logistic distribution as described above in the eq. 2.2.

The results are produced using decimal logarithm values for both axes (i.e. log-log scale). Then, applying the log-log scale, the equation (2.1) becomes

$$log_{10}(C(X)) = \frac{a}{1+exp^{b(\log_{10} X)+c}} \quad (2.3)$$

This is logarithmic form of logistic function. The total cumulated probability is $C_i(x < X_i)$. In the case of mean income, the set which contains the plots representing the probability is S={ (0, 0%), ($X_1$, 10%), ($X_2$, 20%), ($X_3$, 30%), ($X_4$, 40%), ($X_5$, 50%), ($X_6$, 60%), ($X_7$, 70%), ($X_8$, 80%), ($X_9$, 90%), ($X_{10}$, 100%)}. In the case for the upper limit on income data sets, $S_1$= {(0,0%), ($X_1$, 10%), ($X_2$, 20%), ($X_3$, 30%), ($X_4$, 40%), ($X_5$, 50%), ($X_6$, 60%), ($X_7$, 70%), ($X_8$, 80%), ($X_9$, 90%). The fitting was made taking into account the decimal logarithmic values of the probability sets S and $S_1$. The values for the tenth decile, which contains the upper income segment of population, is not comprised in the upper limit on income data set. For the lower limit on income, the set is similar except that each value represents the lowest expenditure value on income decile.

For the first decile (the lowest income decile), *C* represents the population that has an income lower than mean income or upper limit on income or lower limit of the first decile, hence equals 10%. For lower limit on income, the value for the first decile is 0. Subsequently, for the highest income the cumulative distribution function is 100 % (in case of mean income). For the upper limit on income and lower data sets, we do not represent the value for highest decile (tenth) because it was not made available by any of the statistical bodies.

**2.2 Data characteristics**

We will use disposable income data from Brazil [65], Finland [63, 66], France [67-71], Italy [72-73], Philippine [74], Romania [75], Singapore [76], the UK [64], and Uganda [77].
The data were expressed for Brazil, Finland, France, and Singapore with regard to individuals. For Philippine, Romania, Italy, and the UK the data were about households. Also, for France, Finland, and Italy we were able to get data both for mean income and upper limit on income. The UK data were expressed in weekly values for expenditure and in annual values for income.

The data were considered in different monetary units. For example, in the case of France they were expressed in euro for the entire time period considered. Italy was considered both for lire, which was national currency before euro (last year considered for lire in of Italy was in year 1998) and euro starting from year 2000. In the case of Finland, the data were altered such that the numerical value for income from each year was expressed according to last year considered, making the data more reliable and realistic. In Romania, the data were expressed in leu which was the currency until July 2005, when a new currency was introduced called heavy leu. The ratio between 1 heavy leu=10000 leu. In the case of Brazil, we are dealing with different currencies along the years. Thus, in the time interval since 1960 the currencies were cruzeiro, cruzado, new cruzeiro, new cruzado, real cruzeiro and real. The data are expressed in peso for Phillipine, Singapore dollar, and shilling for Uganda.

In the analysis of the parameters of logistic distribution, it is very useful to have a picture of the economic climate in the considered countries. Thus, in the first category are Finland, France, Italy, and the UK, characterised by slow if not negative growth, but they are developed countries and have high income. Brazil is a developed country with middle income. However, in the recent years the economic growth has slowed down considerably due to external factors. Philippine and Singapore, just like most of Far-East economies, were not affected by the recent economic crisis and have high economic growth. Singapore is a developed country, while Philippine is a developing country. Romania is a developing country. Uganda is the country with high economic growth (highest from Sub-Saharan Africa), but still remains by all standards a poor country [77]. It is noteworthy that for the data regarding expenditure from Uganda the data for the 10$^{th}$ decile (upper one) were not made available.

### 2.3 Results

We present the results graphically in Figures 2.1-2.2 and in the tables 2.1-2.14. In the tables 2.1-2.11, we exhibit the results from fitting disposable income, in the table 2.12 we present the results from fitting pensions. Also, in the table 2.13 we show the results for expenditure. The table 2.14 exhibits the results for wealth. We applied the logistic distribution using Matlab.

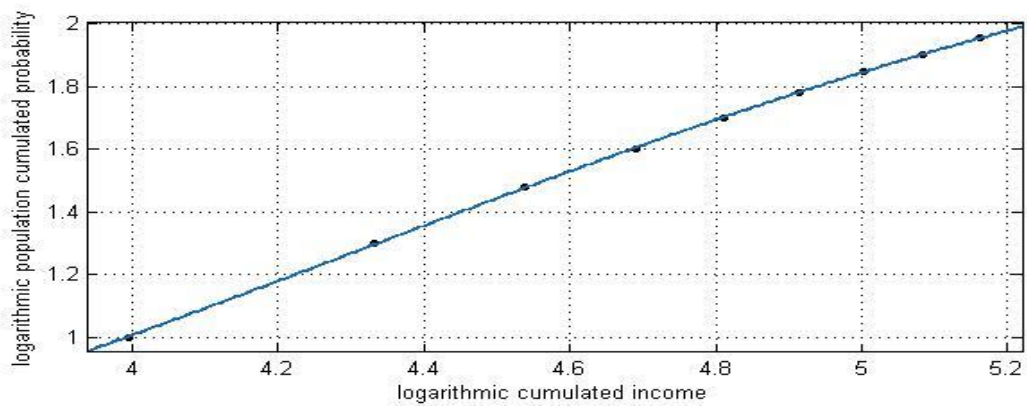

**Figure 2.1 Logistic distribution fitting annual data upper limit on income in Finland for the year 1987.** On the x-axis, we represented decimal logarithmic cumulated income – $\log_{10}(X_i)$ and on the y-axis is decimal logarithm of cumulated probability $\log_{10}(C_i)$.

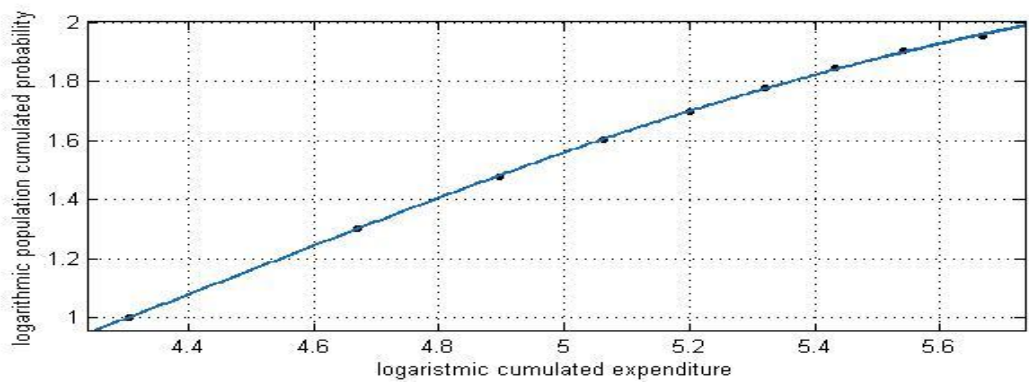

**Figure 2.2 Logistic distribution fitting annual data regarding mean expenditure in the Uganda for the year 2010.** On the x-axis, we represented decimal logarithmic cumulated mean expenditure – $\log_{10}(X_i)$ and on the y-axis is decimal logarithm of cumulated probability $\log_{10}(C_i)$.

The distributions from the other years are very similar to the annual distributions presented in the graphics 2.1-2.2. The annual logistic distributions satisfy the conditions for a cumulated distribution function (1.9-1.11)

Logistic distribution fits very well the data. Thus, the values for coefficient of determination for annual data fitted are above 99% in all cases. However, the data regarding the expenditure and pensions from the UK could not be fitted, even though the data regarding expenditure from Uganda and income for inactive people income from France could be fitted very well.

Logistic distribution can be applied to data having many different characteristics. Thus, it can be applied to income, expenditure and wealth. Also, it is applicable to data calculated as mean value and upper limit. The income, expenditure, and wealth can be expressed in nominal or real values. This shows a wide spectrum of applicability.

Generally, the values for parameters are characterised by small variations when they are compared from one year to another regarding the same variable. On longer time intervals, the variations regarding the same variable between the initial value and the final value are low, however they are bigger compared to the variations from one year to the next. In our opinion, this shows the power of logistic distribution when applied to macroeconomic variables.

For countries such as Finland, France, and Italy we have two sets of data i.e. upper limit on income data set and mean income data set. We can observe from fitting the annual data that the values of $R^2$ are slightly higher in all cases for upper limit on income data set compared to the values for mean income data set in the same year. This is explainable considering that the upper limit on income data set does not contain the tenth decile of income. Namely, this is explainable because the income for people contained in the tenth decile depends mostly on the prices of assets, unlike the rest of the population which depends mainly on wages.

There were countries such as Brazil, Italy, and Romania which had different currencies throughout the years for which data were provided. We can notice that the change of currency did not affect the values for coefficient of determination ($R^2$). However, the values for b parameter changed to a great extent.

We can see that there is no correlation between the economic conditions of a country and the values for coefficient of determination ($R^2$) regarding the income distribution. Developing countries such as Philippine, Romania or a developed country such as Brazil have equal or higher values for coefficient of determination ($R^2$) compared to developed countries with high income such as Finland, France, Italy, and the UK. This is counterintuitive especially since the first group of countries has a higher share of black market.

We can see that logistic distribution is not affected by inflation. The examples showing that are about Italy when the currency was Italian Lira (up to 1998 in the data made available) and Romania during the time interval 2000-2004 when the inflation reached two figures values, up to 40%. Thus, for these countries and time intervals the parameters and coefficient of determination do not behave differently compared with the rest of the data.

Logistic distribution applicability was first suggested by Verhulst in 1838 and derived independently by Pearl and Reed in 1920 in order to model the evolution of animal population. Let N(t) denote population size and N'(t) its derivative. The Verlhust-Pearl model is

$$N'(t) = \lambda N \left(1 - \frac{N}{K}\right) \quad (2.4)$$

where the parameters $\lambda>0$, $K> N(0)$. Thus, $\lambda$ is the intrinsic growth rate, and K is the carrying capacity. The solution according to [87] is

$$N(t) = \frac{K}{1+e^{-\lambda(t-tmax)}} \quad (2.5)$$

Solving the differential equation (2.4), we get the solution (2.5) which is similar to logistic function and Fermi-Dirac function. The equation (2.4) is also called logistic cdf model. The Verlhust-Pearl model shares the Malthusian view, which states that any animal population in nature cannot exceed a certain threshold due to the limitations imposed by resources. According to the model, the number of animals (N) is limited to K (maximum number or carrying capacity). Similarly, the income, expenditure, and wealth distribution have a certain upper limit which is given by productivity and investment limits.

The applicability of logistic function is owed to the fact that the only economic sectors producing added value are agriculture and industry, while services and trade especially redistribute income and wealth. Since most of trade and transactions occur randomly, logistic function, which models chaotic systems, is applicable to chaotic transactions and trade which account for a large part of income and wealth distribution.

## 2.4 Conclusions

Logistic distribution is a very robust distribution. It has the capacity to describe the distribution of several macroeconomic variables, calculated using different methodologies. This can describe the evolution of all groups of income, including the for upper income segment of population which is traditionally described by Pareto distribution.

Logistic distribution is a very powerful statistical tool given the lower variability for coefficients obtained from fitting the annual data. More importantly, pensions, which are not entirely regulated by market but also by state institutions (at least for state pensions), can be fitted very well by logistic distribution.

The utilisation of upper limit or lower limit on income or any other economic variable are preferable to the utilisation of mean income in the analysis of the data comprised in the first nine deciles (for the segment of population with low and medium income).

**2.5 Appendix**

**Table 2.1** Coefficients of the logistic distribution fitting mean disposable income in Brazil

| Year | a | b | c | $R^2$ (%) |
|------|-------|-------|---------|-------|
| 1960 | 2.364 | 3.129 | -0.9353 | 99.93 |
| 1970 | 2.264 | 3.129 | -0.8191 | 99.87 |
| 1980 | 2.239 | 3.409 | -0.7813 | 99.84 |
| 1981 | 2.255 | 1.636 | -0.8598 | 99.9 |
| 1992 | 2.34  | 1.591 | -1.013  | 99.87 |
| 2002 | 2.286 | 1.708 | -0.9223 | 99.88 |

**Table 2.2** Coefficients of the logistic distribution fitting upper limit on disposable income in Finland

| Year | a | b | c | $R^2$ (%) |
|------|-------|-------|---------|-------|
| 1987 | 2.563 | 4.317 | -0.7269 | 99.99 |
| 1988 | 2.571 | 4.329 | -0.7321 | 99.99 |
| 1989 | 2.566 | 4.344 | -0.7321 | 99.99 |
| 1990 | 2.559 | 4.365 | -0.7249 | 99.99 |
| 1991 | 2.563 | 4.377 | -0.7237 | 99.99 |
| 1992 | 2.572 | 4.366 | -0.7231 | 100 |
| 1993 | 2.544 | 4.333 | -0.709  | 100 |
| 1994 | 2.542 | 4.33  | -0.7115 | 99.99 |
| 1995 | 2.543 | 4.335 | -0.7156 | 100 |
| 1996 | 2.534 | 4.33  | -0.7181 | 100 |
| 1997 | 2.5   | 4.319 | -0.7041 | 99.99 |
| 1998 | 2.497 | 4.322 | -0.7114 | 99.99 |
| 1999 | 2.506 | 4.338 | -0.7171 | 99.99 |
| 2000 | 2.498 | 4.332 | -0.719  | 99.99 |
| 2001 | 2.51  | 4.349 | -0.7255 | 99.99 |
| 2002 | 2.512 | 4.362 | -0.728  | 99.99 |
| 2003 | 2.501 | 4.366 | -0.7223 | 99.99 |
| 2004 | 2.514 | 4.386 | -0.7347 | 100 |
| 2005 | 2.512 | 4.396 | -0.7353 | 99.99 |
| 2006 | 2.498 | 4.389 | -0.7296 | 99.99 |
| 2007 | 2.5   | 4.395 | -0.7343 | 99.99 |
| 2008 | 2.521 | 4.413 | -0.7422 | 99.99 |
| 2009 | 2.521 | 4.43  | -0.7411 | 99.99 |

**Table 2.3** Coefficients of the logistic distribution fitting mean disposable income in Finland

| Year | a | b | c | $R^2$ (%) |
|---|---|---|---|---|
| 1987 | 2.678 | 4.321 | -0.8231 | 99.99 |
| 1988 | 2.668 | 4.325 | -0.8184 | 99.99 |
| 1989 | 2.639 | 4.328 | -0.8029 | 99.99 |
| 1990 | 2.642 | 4.355 | -0.8014 | 99.99 |
| 1991 | 2.674 | 4.379 | -0.8188 | 99.99 |
| 1992 | 2.671 | 4.364 | -0.8093 | 99.99 |
| 1993 | 2.635 | 4.324 | -0.7935 | 99.99 |
| 1994 | 2.62 | 4.317 | -0.7852 | 99.99 |
| 1995 | 2.607 | 4.313 | -0.7828 | 99.99 |
| 1996 | 2.522 | 4.287 | -0.7247 | 99.99 |
| 1997 | 2.566 | 4.295 | -0.7758 | 99.98 |
| 1998 | 2.553 | 4.289 | -0.7786 | 99.98 |
| 1999 | 2.539 | 4.294 | -0.7727 | 99.98 |
| 2000 | 2.515 | 4.279 | -0.763 | 99.97 |
| 2001 | 2.554 | 4.307 | -0.7892 | 99.98 |
| 2002 | 2.558 | 4.321 | -0.7928 | 99.98 |
| 2003 | 2.542 | 4.324 | -0.7822 | 99.98 |
| 2004 | 2.54 | 4.335 | -0.7869 | 99.98 |
| 2005 | 2.532 | 4.343 | -0.7827 | 99.98 |
| 2006 | 2.525 | 4.338 | -0.7819 | 99.98 |
| 2007 | 2.521 | 4.338 | -0.7852 | 99.97 |
| 2008 | 2.558 | 4.364 | -0.8057 | 99.98 |
| 2009 | 2.564 | 4.387 | -0.8055 | 99.98 |

**Table 2.4** Coefficients of the logistic distribution fitting upper limit on disposable income in France [61]

| Year | a | b | c | $R^2$ (%) |
|---|---|---|---|---|
| 2002 | 2.466 | 4.272 | -0.7194 | 100 |
| 2003 | 2.478 | 4.278 | -0.7252 | 100 |
| 2004 | 2.483 | 4.282 | -0.7261 | 100 |
| 2005 | 2.484 | 4.284 | -0.7289 | 100 |
| 2006 | 2.478 | 4.286 | -0.7274 | 100 |
| 2007 | 2.475 | 4.293 | -0.7261 | 100 |
| 2008 | 2.494 | 4.315 | -0.7337 | 99.99 |
| 2009 | 2.482 | 4.304 | -0.7323 | 99.99 |

**Table 2.5** Coefficients of the logistic distribution fitting mean disposable income in France

| Year | a | b | c | $R^2$ (%) |
|------|------|------|---------|-------|
| 2003 | 2.519 | 4.23 | -0.7882 | 99.97 |
| 2004 | 2.521 | 4.233 | -0.7879 | 99.97 |
| 2005 | 2.553 | 4.243 | -0.8186 | 99.97 |
| 2006 | 2.533 | 4.239 | -0.8063 | 99.97 |
| 2007 | 2.534 | 4.247 | -0.808 | 99.97 |
| 2008 | 2.548 | 4.266 | -0.8148 | 99.96 |
| 2009 | 2.549 | 4.26 | -0.8211 | 99.97 |

**Table 2.6** Coefficients of the logistic distribution fitting upper limit on disposable income in Italy

| Year | a | b | c | $R^2$ (%) |
|------|------|------|---------|-------|
| 1989 | 2.386 | 7.348 | -0.7389 | 100 |
| 1991 | 2.413 | 7.397 | -0.7592 | 100 |
| 1993 | 2.409 | 7.353 | -0.7949 | 100 |
| 1998 | 2.402 | 7.435 | -0.787 | 99.99 |
| 2000 | 2.4 | 4.195 | -0.7694 | 100 |
| 2002 | 2.377 | 4.214 | -0.7489 | 100 |
| 2004 | 2.392 | 4.267 | -0.7437 | 100 |
| 2006 | 2.383 | 4.306 | -0.7316 | 99.99 |
| 2008 | 2.393 | 4.3 | -0.7509 | 100 |

**Table 2.7** Coefficients of the logistic distribution fitting mean disposable income in Italy

| Year | a | b | c | $R^2$ (%) |
|------|------|------|---------|-------|
| 1989 | 2.466 | 7.294 | -0.8445 | 99.99 |
| 1991 | 2.491 | 7.35 | -0.8593 | 99.99 |
| 1993 | 2.491 | 7.291 | -0.9154 | 99.98 |
| 1995 | 2.479 | 7.334 | -0.8937 | 99.98 |
| 1998 | 2.52 | 7.382 | -0.9471 | 99.97 |
| 2000 | 2.531 | 4.152 | -0.9353 | 99.97 |
| 2002 | 2.473 | 4.158 | -0.8773 | 99.98 |
| 2004 | 2.462 | 4.21 | -0.8426 | 99.97 |
| 2006 | 2.436 | 4.242 | -0.8144 | 99.98 |
| 2008 | 2.463 | 4.24 | -0.8516 | 99.98 |

**Table 2.8** Coefficients of the logistic distribution fitting mean disposable income in Philippine

| Year | a | b | c | $R^2$ (%) |
|---|---|---|---|---|
| 1991 | 2.287 | 4.277 | -0.7644 | 99.93 |
| 1997 | 2.256 | 4.494 | -0.751 | 99.94 |
| 2000 | 2.266 | 4.576 | -0.7624 | 99.95 |
| 2003 | 2.278 | 4.56 | -0.7666 | 99.95 |

**Table 2.9** Coefficients of the logistic distribution fitting mean disposable income in Romania

| Year | a | b | c | $R^2$ (%) |
|---|---|---|---|---|
| 2000 | 2.483 | 6.514 | -0.7032 | 99.98 |
| 2001 | 2.465 | 6.723 | -0.6685 | 99.97 |
| 2002 | 2.52 | 6.828 | -0.7197 | 99.97 |
| 2004 | 2.49 | 6.913 | -0.6878 | 99.96 |
| 2005 | 2.434 | 3.017 | -0.6852 | 99.97 |
| 2006 | 2.408 | 3.037 | -0.683 | 99.98 |
| 2007 | 2.424 | 3.139 | -0.6899 | 99.98 |
| 2008 | 2.465 | 3.273 | -0.7132 | 99.98 |
| 2009 | 2.515 | 3.36 | -0.7285 | 99.98 |
| 2010 | 2.528 | 3.387 | -0.7211 | 99.98 |

**Table 2.10** Coefficients of the logistic distribution fitting mean disposable income in Singapore

| Year | a | b | c | $R^2$ (%) |
|---|---|---|---|---|
| 1980 | 2.741 | 2.733 | -1.301 | 99.74 |
| 1990 | 2.579 | 3.084 | -1.1 | 99.93 |
| 2005 | 2.362 | 2.71 | -0.8827 | 99.95 |
| 2006 | 2.345 | 2.736 | -0.859 | 99.95 |
| 2007 | 2.337 | 2.747 | -0.8611 | 99.94 |

**Table 2.11** Coefficients of the logistic distribution fitting mean disposable income in the UK

| Year | a | b | c | R² (%) |
|------|-------|-------|---------|-------|
| 1977 | 2.388 | 3.408 | -0.7784 | 99.87 |
| 1978 | 2.452 | 3.433 | -0.8044 | 99.93 |
| 1979 | 2.444 | 3.459 | -0.8191 | 99.91 |
| 1980 | 2.427 | 3.518 | -0.8145 | 99.89 |
| 1981 | 2.371 | 3.58  | -0.7373 | 99.94 |
| 1982 | 2.388 | 3.608 | -0.7512 | 99.97 |
| 1983 | 2.376 | 3.654 | -0.7223 | 99.98 |
| 1984 | 2.336 | 3.645 | -0.7013 | 99.96 |
| 1985 | 2.301 | 3.662 | -0.6754 | 99.95 |
| 1986 | 2.315 | 3.666 | -0.7044 | 99.96 |
| 1987 | 2.301 | 3.675 | -0.7077 | 99.94 |
| 1988 | 2.309 | 3.669 | -0.746  | 99.93 |
| 1989 | 2.322 | 3.7   | -0.7631 | 99.94 |
| 1990 | 2.294 | 3.72  | -0.7439 | 99.93 |
| 1991 | 2.316 | 3.773 | -0.759  | 99.95 |
| 1992 | 2.335 | 3.828 | -0.7548 | 99.98 |
| 1993 | 2.302 | 3.84  | -0.7128 | 99.97 |
| 1995 | 2.328 | 3.872 | -0.7338 | 99.98 |
| 1996 | 2.35  | 3.91  | -0.7434 | 99.99 |
| 1997 | 2.354 | 3.918 | -0.7601 | 99.99 |
| 1998 | 2.337 | 3.923 | -0.753  | 99.98 |
| 1999 | 2.359 | 3.952 | -0.7702 | 99.99 |
| 2000 | 2.381 | 3.957 | -0.8056 | 99.99 |
| 2001 | 2.401 | 4.011 | -0.8116 | 99.99 |
| 2002 | 2.393 | 4.024 | -0.8118 | 99.99 |
| 2003 | 2.42  | 4.091 | -0.8116 | 99.99 |
| 2004 | 2.448 | 4.109 | -0.8399 | 99.99 |
| 2005 | 2.456 | 4.154 | -0.8331 | 99.98 |
| 2006 | 2.426 | 4.139 | -0.8137 | 99.99 |
| 2007 | 2.429 | 4.161 | -0.8163 | 99.98 |
| 2008 | 2.456 | 4.176 | -0.8468 | 99.99 |
| 2009 | 2.446 | 4.184 | -0.8338 | 99.98 |
| 2010 | 2.453 | 4.224 | -0.8207 | 99.98 |
| 2011 | 2.461 | 4.243 | -0.803  | 99.96 |
| 2012 | 2.456 | 4.264 | -0.8074 | 99.98 |

**Table 2.12** Coefficients of the logistic distribution fitting income of inactive people in France

| Year | a | b | c | $R^2$ (%) |
|---|---|---|---|---|
| 2003 | 2.506 | 4.177 | -0.8269 | 99.99 |
| 2004 | 2.523 | 4.192 | -0.8332 | 99.99 |
| 2005 | 2.524 | 4.197 | -0.8355 | 100 |
| 2006 | 2.514 | 4.195 | -0.8296 | 99.99 |
| 2007 | 2.52 | 4.204 | -0.8348 | 99.99 |
| 2008 | 2.54 | 4.229 | -0.8438 | 99.99 |
| 2009 | 2.536 | 4.22 | -0.8489 | 99.99 |

**Table 2.13** Coefficients of the logistic function fitting mean expenditure in Uganda

| Year | a | b | c | $R^2$ (%) |
|---|---|---|---|---|
| 2003 | 2.341 | 4.416 | -0.702 | 99.99 |
| 2006 | 2.329 | 4.458 | -0.7013 | 99.99 |
| 2009 | 2.342 | 4.515 | -0.7055 | 99.99 |

**Table 2.14** Coefficients of the logistic function fitting mean wealth in France

| Year | a | b | c | $R^2$ (%) |
|---|---|---|---|---|
| 1998 | 2.633 | 3.541 | -2.221 | 99.63 |
| 2004 | 2.584 | 3.485 | -2.259 | 99.46 |
| 2010 | 2.285 | 3.432 | -1.586 | 99.24 |

# CHAPTER 3: APPLICATIONS OF FERMI-DIRAC FUNCTION TO DISTRIBUTION OF INCOME, WEALTH, AND EXPENDITURE

Statistical mechanics offers a very interesting insight to macroeconomic systems as the particles from thermodynamic systems can be considered as an analogue for macroeconomic systems such as companies, people, and states. Fermi-Dirac and Bose-Einstein distributions are the most important in statistical quantum mechanics. Statistical mechanics distributions used so far for modelling socio-economic systems with some degree of success were Bose-Einstein and Maxwell-Boltzmann distributions. The present chapter investigates the applications of Fermi-Dirac distribution to some of the most important economic variables such as income, wealth, and expenditure for the population of several countries with different economic characteristics.

## 3.1 Methodology

In order to fit the data, we use Fermi-Dirac distribution which calculates the probability distribution for fermions. In Figure 3.1, we present graphically different shapes of Fermi-Dirac function. We can notice the shape of the function changing as the temperature increases or decreases.

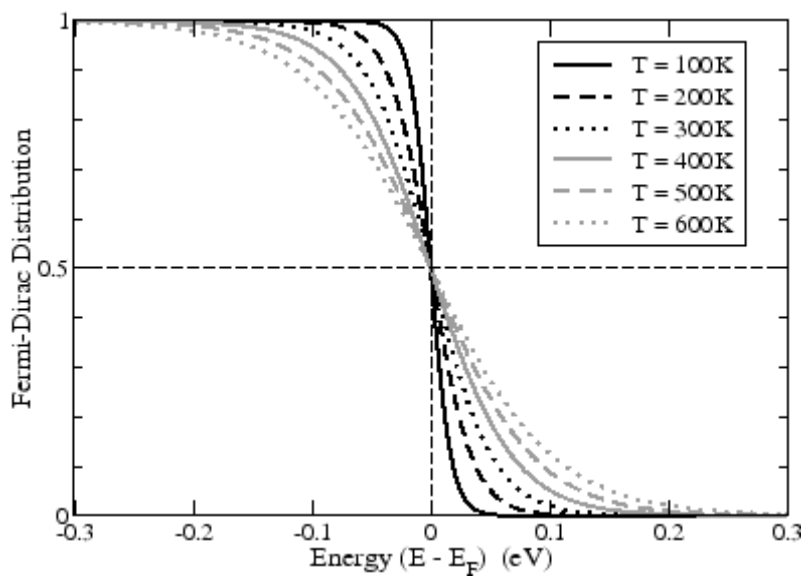

**Figure 3.1** Graphical example of a Fermi-Dirac distribution at different temperatures [88]

Consequently, the parameters of Fermi-Dirac distribution are correlated with certain macroeconomic variables using Matlab software.

The Fermi-Dirac general formula is

$$n(\epsilon_i) = \frac{1}{exp(\frac{\epsilon_i - \mu}{T}) + 1} \quad (3.1)$$

We assume that k=1. Consequently, total number of fermions requires the average number over single-particle states [89]. Therefore, the average number of fermions with the same energy are calculated by multiplying the number of fermions with degeneracy $g_i$ such that

$$N(\epsilon_i) = \frac{g_i}{exp(\frac{\epsilon_i - \mu}{T}) + 1} \quad (3.2)$$

We consider quantity of money (x) to be the statistical thermodynamic counterpart for energy ϵ.

The main parameters which are to be analysed are degeneracy (g), temperature (T), and chemical potential (μ). Temperature (T) has as an analogue notion in the macroeconomic systems. Thus, it is calculated as follows [16]:

$$T = \frac{M}{N} \quad (3.3),$$

where M is total amount of money and N is total number of economic agents.

Chemical potential (μ) is defined in statistical physics as [43]

$$\mu = \frac{\partial U}{\partial N} \quad (3.4)$$

where U is internal energy and N is the number of particles of the system. Another notion that we use is the coefficient of activity or fugacity. The formula is calculated as follows [43]:

$$a = \exp\left(\frac{\mu}{T}\right) \quad (3.5)$$

The probability distribution used is Fermi Dirac. In order to calculate the probability distribution, we use Fermi-Dirac function applied to cumulative values (probability and income) and as probability density function. The complementary cumulative probability distribution $\bar{C}(x)$ is defined as the integral

$$\bar{C}(x) = \int_x^\infty P(x)dx \quad (3.6)$$

It gives the probability that a random variable exceed a given value *x* [21]. We present on y-axis the population probability, which is share of population with income higher than corresponding level on the x-axis. Cumulated income is contained on the x-axis. According to this type of probability, we calculate the share of population having an income above a certain threshold. Thus, the probability to have an income higher than zero is 100% (since everyone is assumed to have a certain income).

Cumulated income is contained on the x-axis. Let us assume X represents the values for cumulated income represented on the x-axis.

$$X_i = \sum x_i$$

where X represents the cumulated income on the x-axis and i=[1,10] for mean income and i=[1,9] for upper limit on income, where i$\epsilon$N.

We applied to these data the formula which describes the probability of occupation of a state by an electron according to Fermi-Dirac distribution, which is

$$\ln(\bar{C}(x)) = \frac{g}{exp\left(\frac{\ln(X)-\mu}{T}\right)+1} \quad (3.7)$$

The equation is the logarithmic Fermi-Dirac function as the results are produced using natural logarithm values for both axes (i.e. log-log scale).

We present on y-axis the cumulated population probability. Income, expenditure, and wealth are displayed on the x-axis. The total cumulated probability is $\bar{C}(x)(a > X_i)$. In the case of mean income, the set which contains the plots representing the probability is *S*={ (0,100%), ($X_1$, 90%), ($X_2$, 80%), ($X_3$, 70%), ($X_4$, 60%), ($X_5$, 50%), ($X_6$, 40%), ($X_7$, 30%), ($X_8$, 20%), ($X_9$, 10%), ($X_{10}$, 0%)}. In the case of the upper limit on income data, the set is $S_1$= {(0,100%), ($X_1$, 90%), ($X_2$, 80%), ($X_3$, 70%), ($X_4$, 60%), ($X_5$, 50%), ($X_6$, 40%), ($X_7$, 30%), ($X_8$, 20%), ($X_9$, 10%). The fitting was made taking into account the natural logarithmic values of the probability sets S and $S_1$.

Thus, for income equal to 0, the cumulative probability is 100 % as it is considered that everyone has an income higher than 0. For the first decile (the lowest income decile), $\bar{C}$ represents the population that has an income higher than mean income or upper limit on income or lower limit of the first decile, hence equals 90%. For lower limit on income, the value for the first decile is 0. Subsequently, for the highest income the cumulative distribution function is 0 % (in case of mean income). For the upper limit on income and lower data set, we do not represent the upper decile as the value corresponding for it was not made available by any of the statistical bodies.

The values for the tenth decile which contains the upper income segment of population are not comprised in the data set. Thus, the natural logarithm of probability, which is $ln(\bar{C}(X))$, is the dependent probability and natural logarithm of $x$ (income) is the independent variable. Also, parameters $g$, $T$, and $\mu$ are obtained from fitting the data using Fermi-Dirac distribution as described above in the eq. 3.7.

Probability density function is calculated on income data provided by the USA using the eq. 3.9. The values on the x-axis are distributed as $x^*_i$, where $i \epsilon N$, $x^* = \log_{10} x$, $x$ being income. Probability density $P_1^*$ is the decimal logarithm of probability density for the population whose income is between $(x_0, x_1)$, $P_2^*$ is the decimal logarithm of probability density for the population whose income is between $(x_1, x_2)$. Similarly, $P_n^*$ is the decimal logarithm of probability density for the population whose income is between $(x_{n-1}, x_n)$, where $i \epsilon N$. According to the probability density function

$$\sum_{i=0}^{n} P_i = 1 \quad (3.8)$$

The set containing the probability plot in decimal logarithmic values is $S_2 = \{(x^*_0, P_0), (x^*_1, P_1), \ldots, (x^*_n, P_n)\}$, where $n \epsilon N$. Thus, $P^* = \log_{10}(P(x^*))$. The equation for Fermi-Dirac density function becomes

$$P^*(x^*) = \frac{g}{exp\left(\frac{x^* - \mu}{T}\right) + 1} \quad (3.9)$$

which is the logarithmic form of Fermi-Dirac pdf.

### 3.2 Data characteristics

The data we analyse are about disposable income from Finland [63, 66], France [67-68], Romania [75], the UK [64], and USA [78]. We chose these countries, which unlike other countries, as their data span a long time interval. Furthermore, these countries maintain the same currency throughout the entire time interval taken into account. The data are expressed using deciles of population, ranked according to their income for all countries except USA. The value for each decile can be calculated as upper limit on income or mean income. The predicted parameters are obtained from fitting Fermi-Dirac distribution expressed in logarithmic form (log-log scale) to the data from the above mentioned countries.

Regarding macroeconomic characteristics of considered countries, Finland, France, the UK, and USA are developed countries with high income. Romania is a developing country. The data from Finland and France are about individuals, while the data about Romania, USA, and the UK are about households. All countries considered were affected by the recent crisis. Finland, France, the UK, and USA were affected seriously by crisis from the beginning of the year 2008, while in the case of Romania crisis started seriously in the year 2009. The impact of the crisis was more severe in the case of Romania than in the cases of the other countries, the economy shrank with 9%. Thus, the developed countries contained in this pool had slow economic growth before the crisis and their economies contract slightly during the crisis. The crisis made Romania to jump from high economic growth to a very severe contraction, phenomenon which occurs in less stable economies.

The time intervals the data span are 1987-2009 in the case of Finland, 2002-2009 in the case of France, 2000-2010 in the case of Romania, 1977-2012 in the case of the UK, and 2003-2013 in the case of USA.

### 3.3 Results

Fermi-Dirac function applied to cumulative values/probabilities satisfies the necessary mathematical conditions for complementary cumulative distribution function (1.12-1.15).

Fermi-Dirac distribution fits very well the data. Thus, the lowest value for coefficient of determination for annual data fitted is 98.31% for income data from the UK in the year 1980. The highest value for coefficient of determination for annual data is 99 % for France. However, most of the values for coefficient of determination are above 98 % in the annual data analysis.

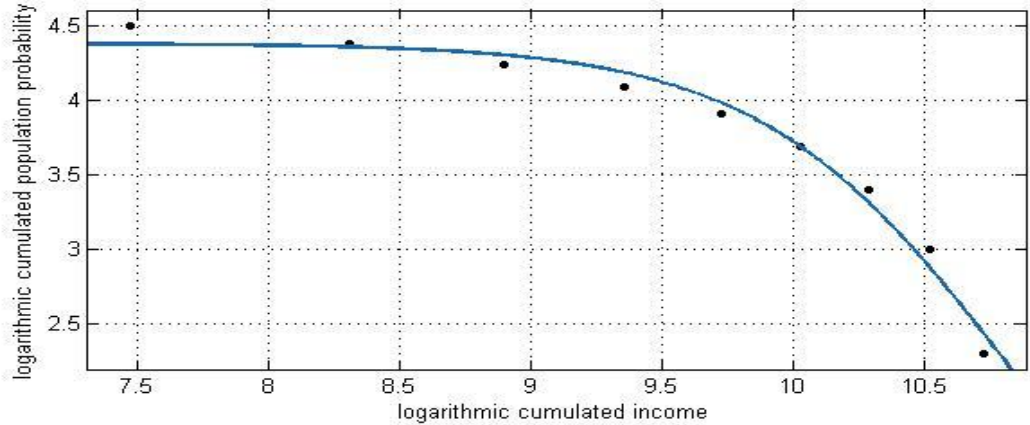

**Figure 3.1 Fermi-Dirac distribution fitting annual data regarding income in the UK for the year 1980.** On the x-axis, we represented natural logarithmic values of cumulated income – ln ($X_i$) and on the y-axis is natural logarithm of cumulated probability- ln($\bar{C}(X)$).

We came across similar findings with the papers using cumulative values when fitting the data using Fermi-Dirac distribution [90-92]. More specifically, they exhibited very good fit to the data when using Fermi-Dirac distribution and have better fit of upper limit on income data compared to mean income data.

We investigated the correlation of Fermi-Dirac parameters obtained from fitting the data with possible macroeconomic parameters. After a very extensive comparison, we could find a significant correlation between chemical potential and exports, degeneracy and Gini coefficient, and between coefficient of activity and inflation using cumulative values/probabilities. Unfortunately, temperature could not be correlated with any economic indicator using cumulative data. We tried to correlate both data sets (mean income and upper limit on income) for Finland and France. We used as a measure for it correlation Pearson correlation coefficient (r).

We tried to fit the cumulative data using Fermi-Dirac complementary cumulative distribution probability by applying the eq. (3.17) from the Appendix 2, but the attempt was unsuccessful.

The macroeconomic data were provided in order to see the analogies with statistical physics and were supplied by several institutions. Finland for exports and inflation were provided by [93] and for Gini coefficient by [94]. For France, exports were provided by [93], inflation by [95], and Gini coefficient by [96]. For Romania all the data were provided by [93]. The UK data were provided by [93] for exports, for inflation by [97] and for Gini coefficient by [98].

In the tables 3.1-3.4, there are displayed the economic indicators and the predicted physical values. For these values, we computed correlation coefficient values which we displayed on the last line of the tables 3.1-3.4 (grey shaded). For France and Finland, there are displayed twice as many coefficients of correlation given that they are calculated both for mean income and upper limit on income. The values from t-test are also presented in the Tables 3.1-3.4 on the last line and shaded in grey. The values obtained show that all correlations are statistically significant. The values for temperature and coefficient of determination from fitting annual data are displayed in the Tables 3.5-3.7.

Chemical potential can be correlated with exports, having very high values for correlation coefficient, the lowest correlation coefficient is 0.8 in the case of France. We applied it for both sets of data (mean and upper limit) and no significant difference was observed. Consequently, there is a positive correlation between chemical potential estimated value and exports.

The applicability of chemical potential to economic systems is due the fact that exports imply an inward flux of money which is very similar to its physical meaning that measures it as the quantity of energy necessary to add a molecule. Thus, chemical potential is higher when the energy (per particle) in physical systems is higher. In the economic systems, the higher are the exports the higher is the amount of money which is considered the analogue for energy (see eq. 7 [21]).

Degeneracy was correlated with Gini coefficient. The lowest correlation coefficient in the case of France was -0.16. For the countries with two data sets, we can notice a significant difference in the correlation coefficient. Thus, in the case of Finland the correlation coefficient for upper limit on income is higher with 0.1 than in the case of mean income. For France, in the case of upper limit on income the correlation coefficient is -0.16, which is negatively correlated. Moreover, this result is highly different from the results for mean income data set.

Degeneracy from quantum systems has some similarities with socio-economic systems in two ways. First, Boltzmann-Gibbs and Bose-Einstein distributions allow that all particles from a system can occupy the same energy level (Bose-Einstein condensation). In the case of Fermi-Dirac function, distribution particles (if in a sufficiently great number) must occupy different energy levels just like in a real economy. Second, degeneracy implies that one or more particles situated on an energy level can have different states i.e. different wave functions, which is similar with different behaviour of individuals having the same amount of money.

Gini coefficient is a measure for inequality. Thus, it plots on the y axis the proportion of the total income of the population which is cumulatively earned by a certain share of the population from a country. Each axis has a maximum value of 1, so the triangle has an area of 0.5. A necessary notion to define Gini coefficient is Lorenz curve. This shows a graph of the proportion of overall income or wealth assumed by the bottom share of the population. The Gini coefficient can then be thought of as the ratio of the area that lies between the line of equality and the Lorenz curve. Gini coefficient can vary according to inequality of income. Thus, when Gini coefficient is zero this means perfect equality (all people have equal income), while a Gini coefficient of 1 (or 100 %) implies maximum inequality i.e. one person has all income and the rest of the people has nothing [99]. A possible explanation for the correlation between these notions is the fact that the two notions evolve similarly regardless if the system is physical or macroeconomic. Thus, the higher the inequality the higher is the number of energy states/ degeneracy. This is true as this index shows the degree inequality and, subsequently, a higher number of income levels (higher income disparity).

In the case coefficient of activity, we can see the lowest value for correlation coefficient is 0.20 in the case of the UK. We can notice then in the cases of France and Finland the values for correlation coefficient are considerably different for the same country (at least 0.1 corresponding to each type of data set). A possible explanation for the analogy between these two notions is the fact that coefficient of activity is correlated with the evolution of the economy. Thus, considering the fact that the pool of countries considered has stable macroeconomic conditions (at least not with hyperinflation), inflation can show an economic boom. Conversely, shrinking economy (or recession) shows a low inflation rate. In the case of Romania, the coefficient of activity is negatively correlated.

The coefficient of activity can be considered an indicator of the overall state of an economy, as it explores the ratio between temperature (money injected in the economic system) and the total energy of the systems to function according to the number of molecules (total activity). Thus, it can be considered as an input/output ratio (energy used given the money flow).

Thus, the highest correlation can be found for chemical potential (correlated with exports). It is followed by degeneracy (correlated with Gini index), and coefficient of activity (correlated with inflation).For mean income and upper limit on income data sets, it can be observed that there is no similar evolution for degeneracy and coefficient of activity predicted values in the same country. Consequently, there is no pattern observed. The correlation for these could be also negative.  There are also differences between correlation values from one country to another. This could be explained by the macroeconomic conditions and particularities of economic model specific for each country. For example, the UK and Romania have high inequality, whereas France and Finland have lower inequality, especially Finland. Exports differ as share of GDP. Thus, the UK and Finland have a high share of exports in GDP, whereas France and Romania have a lower one. Unlike the rest of the countries considered, Romania had a high rate of inflation for few years.

We can notice that graph of Fermi-Dirac distribution applied to income does not observe the symmetry as in the case of fermions in the physical systems by taking into account the mean value of the variables displayed on both axes. This is explained partially in our case by the fact we deal only with positive values on the x-axis, unlike the case of physical systems which can have negative values. However, given the higher values for the coefficient of determination from fitting the annual data in all cases, we conclude that Fermi-Dirac distribution describes successfully the income distribution for population.

We applied also Fermi-Dirac as a probability density function as detailed in the eq. 3.9. The results are displayed in the Figure 3.2 and in the Table 3.8.

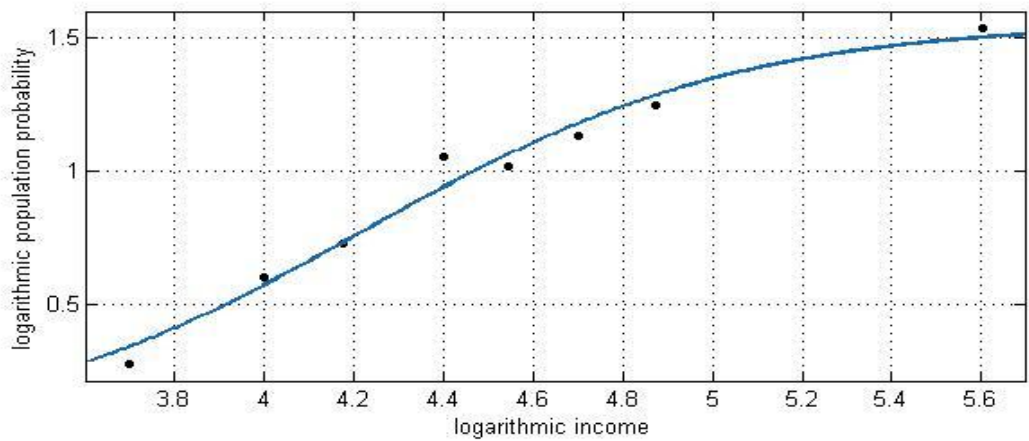

**Figure 3.2 Fermi-Dirac probability density function fitting US annual household income in the year 2013.** On the x-axis, we represented the decimal logarithmic income – $\log_{10}(x_i)$ and on the y-axis are displayed the decimal logarithmic values of probability density- $\log_{10}(P_i)$.

Fermi-Dirac function applied to income values/probabilities satisfies the necessary mathematical conditions for probability density function (1.5-1.7).

We can notice that the distribution fitting the US annual household income for a time interval of ten years shows a good fit to the data. Thus, the lowest annual value for the coefficient of determination is 97.02% in the year 2007. We can notice that the values for parameters from the distribution fitting the data show very little variability from one year to another. The temperature values, unlike those obtained from fitting Fermi-Dirac function to cumulative data/probabilities, have negative values.

We tried to correlate the data we obtained from fitting the US data similarly with the ones from cumulated values/probabilities. Thus, we correlated temperature and income per capita and the correlation coefficient was -0.69. We looked at the correlation between Gini coefficient and degeneracy and the coefficient was -0.80. We also studied the correlation between exports and chemical potential and the correlation coefficient was 0.92. T–test applied to the set of data showed that all the correlations are significant. The data used to perform the analysis were provided by [78].

Starting from [90], we see Fermi-Dirac distribution as being applicable to subatomic particles, which have different characteristics regarding energy from one particle to another from the same system. Rephrased, no two particles from the same system have the same values for the spin numbers. In economy, it means that no money influxes for an economic entity are the same. On the income side, this is true considering that, generally or with very few exceptions, the salaries are not the same to the last sub-unitary division of a currency. On the expenses side, this is true considering that each entity pays different taxes given the nature of their activity and the consumption level, especially due to indirect taxes. Given differences in income level and income sources, there are different taxation levels. Moreover, they can receive different subsidies (in the case of companies) and different social benefits in the case of individuals. Therefore, each entity is characterised by different money fluxes (dynamically). Furthermore, even in the case when two entities have the same amount of money (energy), the composing fluxes remain different.

Fermi-Dirac distribution is more applicable to money/income distribution than Bose-Einstein distribution given that temperature influences the evolution of electrons in a similar way with a real economy. According to Fermi-Dirac distribution, when temperature is zero there are no electrons in the conduction band. As the temperature increases there are more and more electrons in the conduction band where flows are possible (and less in Fermi Sea). Thus, the higher is the temperature the more electrons are in the conduction band. Similarly, the higher the quantity of money per economic agent (higher temperature), the higher are transactions (flows in the conduction band) namely the number of transactions and the value per transaction. The economic explanation is that a reduced level of available money diminishes the number of transactions and the value per transaction. Generally, the higher is the quantity of monetary mass, the number of transactions and the value per transaction increase. However, the analogy is not rigorously equivalent but the relation between energy and monetary mass is true when it refers to trends.

The applicability of Fermi Dirac distribution was tackled by [46]. However, the applicability was tackled more theoretically, while our research is about application to real data from different countries with very different characteristics. Another investigation which deals with the evolution of chemical potential, degeneracy, temperature, and coefficient of activity is [42]. The evolutions over the time of these parameters are analysed. The only analogy drawn is between coefficient of activity and economic activity (loosely defined). Unlike this research, our investigation draws an analogy between coefficient of activity and inflation which is partially related to economic activity. Compared to Yakovenko's research, which is mainly about exponential distribution, we use more statistical tools to substantiate our research.

We consider that Fermi-Dirac distribution is better in describing income than Boltzmann Gibbs distribution due to its similarity with logistic distribution, which shares the Malthusian view on distribution of resources. Boltzmann Gibbs distribution cannot describe accurately increase of economic variables over a certain threshold. This view is illustrated in the function that shows the growth in the number over time such that

$$\frac{dN}{dt} = rN \quad (3.10)$$

Where t is the time and r is a constant growth rate.
The equation becomes

$$\frac{dN}{N} = rdt \quad (3.11)$$

We integrate the equation (3.11)

$$\int \frac{dN}{N} = \int r\, dt$$

$$\ln(N) = rt + C \quad (3.12)$$

as N>0

The equation (3.12) becomes

$$N(t) = e^{rt+C}$$

$$N(t) = e^{rt} * e^{C} = C_1 * e^{rt}$$

where $C_1 = e^C$ and is a constant. This is Boltzmann Gibbs law or the exponential law. This may point to why Fermi-Dirac distribution is applicable to economic phenomena in a similar way to physics, where the electrons cannot exceed a certain number on the same energy level. Thus, in an economy just like in nature the evolution cannot be exponential at all times but eventually income and production cannot exceed an upper limit imposed by investment or demand. Consequently, at least the real income cannot evolve according to an exponential law.

### 3.4 Conclusions

Fermi-Dirac distribution is a very robust distribution. It has the capacity to describe the distribution several macroeconomic variables, calculated using different methodologies, and can describe the evolution of all incomes, including the for upper income segment of population which is traditionally described by Pareto distribution. The data can be fitted very well using this distribution, given the values for coefficient of determination (higher than 98 % when applied to cumulative income/probabilities and 95% when applied as a probability density function).

The parameters obtained from fitting the data show a high correlation between exports and chemical potential and a significant relation for the correlation between degeneracy and Gini index. Temperature can be also correlated (negatively) with income when Fermi-Dirac probability density function fits non-cumulative data.

### 3.5 Appendices

**Appendix 1**

**Table 3.1** Coefficients obtained from fitting the data using Fermi-Dirac distribution and correlation coefficient with their macroeconomic counterparts in Finland

| Year | Exports $\times 10^{11}$ | µ mean | µ upper | Gini index | g mean | g upper | Inflation | a mean $\times 10^{13}$ | a uppper $\times 10^{13}$ |
|---|---|---|---|---|---|---|---|---|---|
| 1987 | 2.27 | 11.9 | 11.96 | 19.7 | 4.39 | 4.396 | 4.08 | 0.9521 | 1.5129 |
| 1988 | 2.57 | 11.93 | 11.99 | 20.2 | 4.392 | 4.398 | 5.09 | 0.9106 | 1.2973 |
| 1989 | 2.73 | 11.97 | 12.04 | 20.5 | 4.392 | 4.398 | 6.63 | 0.9339 | 1.2930 |
| 1990 | 3.13 | 12.02 | 12.08 | 20.2 | 4.392 | 4.399 | 6.1 | 1.1405 | 1.4624 |
| 1991 | 2.71 | 12.03 | 12.1 | 20.1 | 4.392 | 4.398 | 4.11 | 1.3603 | 1.8915 |
| 1992 | 2.86 | 11.99 | 12.05 | 19.9 | 4.392 | 4.399 | 2.6 | 1.5588 | 2.0230 |
| 1993 | 2.78 | 11.95 | 12.01 | 21.1 | 4.395 | 4.402 | 2.1 | 0.9222 | 1.1207 |
| 1994 | 3.5 | 11.95 | 12.01 | 21.1 | 4.395 | 4.402 | 1.08 | 0.9431 | 1.0243 |
| 1995 | 4.77 | 11.97 | 12.03 | 21.7 | 4.395 | 4.402 | 0.98 | 0.7761 | 0.8679 |
| 1996 | 4.78 | 12.00 | 12.06 | 22.3 | 4.397 | 4.402 | 0.61 | 0.7767 | 0.6776 |
| 1997 | 4.77 | 12.03 | 12.09 | 23.7 | 4.397 | 4.404 | 1.19 | 0.3857 | 0.4304 |
| 1998 | 5.01 | 12.06 | 12.13 | 24.8 | 4.396 | 4.403 | 1.39 | 0.3078 | 0.3818 |
| 1999 | 5.06 | 12.09 | 12.16 | 25.9 | 4.396 | 4.403 | 1.15 | 0.3329 | 0.3880 |
| 2000 | 5.31 | 12.1 | 12.17 | 26.7 | 4.397 | 4.404 | 3.36 | 0.2578 | 0.2824 |
| 2001 | 5.17 | 12.12 | 12.19 | 25.8 | 4.396 | 4.403 | 2.56 | 0.2740 | 0.3299 |
| 2002 | 5.48 | 12.15 | 12.22 | 25.6 | 4.395 | 4.403 | 1.56 | 0.3106 | 0.3690 |
| 2003 | 6.36 | 12.18 | 12.25 | 26.0 | 4.396 | 4.403 | 0.87 | 0.2666 | 0.3456 |
| 2004 | 7.54 | 12.22 | 12.29 | 26.6 | 4.396 | 4.404 | 0.18 | 0.2910 | 0.3057 |
| 2005 | 8.18 | 12.25 | 12.33 | 26.7 | 4.396 | 4.403 | 0.86 | 0.2978 | 0.3472 |
| 2006 | 9.46 | 12.26 | 12.33 | 27.3 | 4.397 | 4.404 | 1.56 | 0.2349 | 0.2656 |
| 2007 | 11.3 | 12.28 | 12.36 | 28.0 | 4.398 | 4.405 | 2.51 | 0.2074 | 0.2366 |
| 2008 | 12.7 | 12.28 | 12.36 | 30.7 | 4.395 | 4.403 | 4.06 | 0.3007 | 0.3700 |
| 2009 | 8.94 | 12.32 | 12.39 | 29.7 | 4.394 | 4.402 | 0 | 0.3630 | 0.4457 |
| **Correlation coefficients** | 0.91 | 0.92 | | 0.65 | 0.76 | | 0.48 | 0.58 | |
| **T-values** | 9.14 | 9.14 | | 27.69 | 27.68 | | 7.22 | 6.53 | |

**Table 3.2** Coefficients obtained from fitting the data using Fermi-Dirac distribution and correlation coefficient with their macroeconomic counterparts in France

| Year | Exports $\times 10^{11}$ | µ mean | µ upper | Gini index | g mean | g upper | Infl | a m $\times 10^{11}$ | a up $\times 10^{11}$ |
|---|---|---|---|---|---|---|---|---|---|
| 2002 | 3.99374 | - | 12.14 | 27 | - | 4.412 | 1.8 | - | 5.0199 |
| 2003 | 4.64381 | 12.05 | 12.13 | 27 | 4.401 | 4.411 | 2.1 | 6.5847 | 5.9522 |
| 2004 | 5.37428 | 12.05 | 12.13 | 28.2 | 4.401 | 4.41 | 2.3 | 7.3583 | 6.9318 |
| 2005 | 5.63188 | 12.06 | 12.14 | 27.7 | 4.4 | 4.41 | 1.7 | 6.5716 | 6.7094 |
| 2006 | 6.09571 | 12.08 | 12.16 | 27.3 | 4.401 | 4.411 | 1.5 | 6.1949 | 5.8476 |
| 2007 | 6.93536 | 12.1 | 12.18 | 26.6 | 4.4 | 4.411 | 1.5 | 6.3621 | 6.3789 |
| 2008 | 7.63136 | 12.11 | 12.2 | 29.8 | 4.399 | 4.411 | 2.8 | 8.5364 | 8.0136 |
| 2009 | 6.12193 | 12.12 | 12.2 | 29.9 | 4.4 | 4.411 | 0.1 | 6.2226 | 5.9817 |
| **Correlation coefficients** | **0.80** | **0.80** | | **0.57** | **-0.16** | | **0.75** | **0.52** |
| **T-values** | **16.17** | **13.98** | | **47.58** | **52.14** | | **21.33** | **20.18** |

**Table 3.3** Coefficients obtained from fitting the data using Fermi-Dirac distribution and correlation coefficient with their macroeconomic counterparts in Romania

| Year | Exports $\times 10^{11}$ | µ | Gini index | g mean | Inflation | a $\times 10^{10}$ |
|---|---|---|---|---|---|---|
| 2000 | 1.22 | 17.18 | 30.25 | 4.41 | 45.66 | 0.0165 |
| 2001 | 1.34 | 17.6 | 30.57 | 4.411 | 34.46 | 0.0870 |
| 2002 | 1.62 | 17.83 | 31.46 | 4.407 | 22.53 | 0.1334 |
| 2003 | 2.06 | 18.02 | 31.06 | 4.41 | 15.27 | 0.2641 |
| 2004 | 2.71 | 18.32 | 31.66 | 4.407 | 11.87 | 0.5099 |
| 2005 | 3.26 | 9.219 | 31.57 | 4.411 | 8.98 | 0.1928 |
| 2006 | 3.62 | 9.348 | 32.11 | 4.413 | 6.58 | 0.1261 |
| 2007 | 5.20 | 9.555 | 32.1 | 4.41 | 4.83 | 0.2531 |
| 2008 | 6.22 | 9.798 | 31.15 | 4.404 | 7.84 | 0.9486 |
| 2009 | 5.03 | 9.89 | 30 | 4.402 | 5.58 | 2.3841 |
| 2010 | 5.84 | 9.891 | 24.24 | 4.403 | 6.09 | 4.0650 |
| **Correlation coefficients** | **0.94** | **0.55** | | **-0.42** | |
| **T-values** | **6.20** | **39.27** | | **2.12** | |

**Table 3.4** Coefficients obtained from fitting the data using Fermi-Dirac distribution and correlation coefficient with their macroeconomic counterparts in the UK [91]

| Year | Exports x10$^{11}$ | μ | Gini index | g mean | Inflation | a x10$^9$ |
|---|---|---|---|---|---|---|
| 1977 | 0.76 | 10.61 | 26.7 | 4.386 | 15.80 | 3.5320 |
| 1978 | 0.92 | 10.51 | 25.9 | 4.389 | 8.30 | 4.7490 |
| 1979 | 1.17 | 10.65 | 26.8 | 4.387 | 13.40 | 4.2060 |
| 1980 | 1.47 | 10.83 | 28.0 | 4.385 | 18.00 | 5.9200 |
| 1981 | 1.37 | 10.94 | 28.4 | 4.396 | 11.90 | 6.3430 |
| 1982 | 1.29 | 10.99 | 28.1 | 4.395 | 8.60 | 7.1660 |
| 1983 | 1.23 | 11.03 | 28.2 | 4.399 | 4.60 | 10.1580 |
| 1984 | 1.24 | 11.1 | 27.7 | 4.398 | 5.00 | 7.0820 |
| 1985 | 1.33 | 11.19 | 29.1 | 4.400 | 6.10 | 6.0000 |
| 1986 | 1.46 | 11.25 | 31 | 4.401 | 3.40 | 4.0230 |
| 1987 | 1.78 | 11.35 | 32.8 | 4.399 | 4.20 | 3.2940 |
| 1988 | 1.95 | 11.44 | 34.6 | 4.395 | 4.90 | 2.3190 |
| 1989 | 2.03 | 11.53 | 33.8 | 4.397 | 5.20 | 1.8750 |
| 1990 | 2.47 | 11.63 | 36.5 | 4.396 | 7.00 | 1.8810 |
| 1991 | 2.5 | 11.71 | 35.2 | 4.398 | 7.50 | 2.5140 |
| 1992 | 2.62 | 11.74 | 34.4 | 4.401 | 4.30 | 3.8840 |
| 1993 | 2.55 | 11.74 | 34.5 | 4.406 | 2.50 | 3.9160 |
| 1995 | 3.36 | 11.79 | 33.4 | 4.406 | 2.60 | 4.9420 |
| 1996 | 3.62 | 11.81 | 32.6 | 4.405 | 2.50 | 8.2220 |
| 1997 | 3.92 | 11.88 | 34.0 | 4.408 | 1.80 | 4.9910 |
| 1998 | 3.88 | 11.93 | 34.1 | 4.406 | 1.60 | 5.1690 |
| 1999 | 3.92 | 11.96 | 35.0 | 4.406 | 1.30 | 6.7900 |
| 2000 | 4.08 | 12.01 | 35.3 | 4.404 | 0.80 | 5.4750 |
| 2001 | 4.00 | 12.06 | 34.6 | 4.403 | 1.20 | 10.2040 |
| 2002 | 4.20 | 12.13 | 36.0 | 4.406 | 1.30 | 7.1770 |
| 2003 | 4.78 | 12.18 | 33.5 | 4.411 | 1.40 | 9.9870 |
| 2004 | 5.60 | 12.19 | 33.7 | 4.402 | 1.30 | 20.39030 |
| 2005 | 6.18 | 12.25 | 32.3 | 4.405 | 2.10 | 28.1850 |
| 2006 | 7.13 | 12.27 | 33.7 | 4.41 | 2.30 | 16.2320 |
| 2007 | 7.61 | 12.31 | 34.5 | 4.408 | 2.30 | 22.0160 |
| 2008 | 7.9 | 12.34 | 34.0 | 4.404 | 3.60 | 25.1270 |
| 2009 | 6.27 | 12.36 | - | - | 2.10 | 23.1580 |
| 2010 | 6.88 | 12.38 | - | - | 3.29 | 43.8990 |
| 2011 | 7.95 | 12.33 | - | - | 4.48 | 86.0920 |
| 2012 | 7.8 | 12.42 | - | - | 2.83 | 69.8660 |
| **Correlation coefficients** | **0.9** | | **0.69** | | **0.21** | |
| **T-values** | **9.23** | | **48.76** | | **4.35** | |

**Table 3.5** Coefficients of the Fermi-Dirac distribution fitting disposable income in Finland

|      | Finland mean income | | Upper limit on income | |
|------|--------|---------|--------|---------|
| Year | T | R² (%) | T | R² (%) |
| 1987 | 0.3982 | 98.51 | 0.3941 | 98.64 |
| 1988 | 0.3998 | 98.54 | 0.3971 | 98.68 |
| 1989 | 0.4008 | 98.54 | 0.3988 | 98.68 |
| 1990 | 0.3998 | 98.56 | 0.3985 | 98.71 |
| 1991 | 0.3978 | 98.56 | 0.3958 | 98.70 |
| 1992 | 0.3947 | 98.56 | 0.3933 | 98.70 |
| 1993 | 0.4003 | 98.62 | 0.3997 | 98.76 |
| 1994 | 0.4000 | 98.64 | 0.4009 | 98.80 |
| 1995 | 0.4033 | 98.63 | 0.4038 | 98.79 |
| 1996 | 0.4043 | 98.65 | 0.4082 | 98.78 |
| 1997 | 0.4151 | 98.66 | 0.4156 | 98.80 |
| 1998 | 0.4194 | 98.63 | 0.4187 | 98.78 |
| 1999 | 0.4193 | 98.63 | 0.4195 | 98.78 |
| 2000 | 0.4234 | 98.64 | 0.4245 | 98.81 |
| 2001 | 0.4232 | 98.64 | 0.4229 | 98.79 |
| 2002 | 0.4224 | 98.62 | 0.4223 | 98.79 |
| 2003 | 0.4244 | 98.65 | 0.4243 | 98.81 |
| 2004 | 0.4271 | 98.65 | 0.4275 | 98.80 |
| 2005 | 0.4265 | 98.63 | 0.427 | 98.81 |
| 2006 | 0.4304 | 98.65 | 0.431 | 98.82 |
| 2007 | 0.433 | 98.68 | 0.4338 | 98.84 |
| 2008 | 0.4274 | 98.63 | 0.4271 | 98.80 |
| 2009 | 0.426 | 98.61 | 0.4254 | 98.79 |

**Table 3.6** Coefficients of the Fermi-Dirac distribution fitting disposable income in France

|      | Mean income | | Upper limit on income | |
|------|--------|---------|--------|---------|
| Year | T | R² (%) | T | R² (%) |
| 2002 | - | - | 0.4506 | 98.99 |
| 2003 | 0.4428 | 98.77 | 0.4474 | 98.97 |
| 2004 | 0.441 | 98.76 | 0.4449 | 98.95 |
| 2005 | 0.4432 | 98.76 | 0.4458 | 98.96 |
| 2006 | 0.4449 | 98.77 | 0.4488 | 98.99 |
| 2007 | 0.4452 | 98.76 | 0.4481 | 98.97 |
| 2008 | 0.4408 | 98.77 | 0.4451 | 99.00 |
| 2009 | 0.4463 | 98.77 | 0.4499 | 98.98 |

**Table 3.7** Coefficients of the Fermi-Dirac distribution fitting mean disposable income in the UK and Romania [91]

|      | UK     |           | Romania |           |
|------|--------|-----------|---------|-----------|
| Year | T      | $R^2$ (%) | T       | $R^2$ (%) |
| 1977 | 0.4826 | 98.38     | -       | -         |
| 1978 | 0.4717 | 98.39     | -       | -         |
| 1979 | 0.4806 | 98.33     | -       | -         |
| 1980 | 0.4813 | 98.31     | -       | -         |
| 1981 | 0.4847 | 98.55     | -       | -         |
| 1982 | 0.4843 | 98.48     | -       | -         |
| 1983 | 0.4787 | 98.65     | -       | -         |
| 1984 | 0.4894 | 98.51     | -       | -         |
| 1985 | 0.497  | 98.55     | -       | -         |
| 1986 | 0.5087 | 98.60     | -       | -         |
| 1987 | 0.5179 | 98.57     | -       | -         |
| 1988 | 0.5305 | 98.52     | -       | -         |
| 1989 | 0.54   | 98.47     | -       | -         |
| 1990 | 0.5446 | 98.48     | -       | -         |
| 1991 | 0.541  | 98.53     | -       | -         |
| 1992 | 0.5317 | 98.54     | -       | -         |
| 1993 | 0.5315 | 98.68     | -       | -         |
| 1995 | 0.5282 | 98.61     | -       | -         |
| 1996 | 0.5173 | 98.69     | -       | -         |
| 1997 | 0.532  | 98.70     | -       | -         |
| 1998 | 0.5334 | 98.70     | -       | -         |
| 1999 | 0.5283 | 98.75     | -       | -         |
| 2000 | 0.5356 | 98.64     | 0.4211  | 98.91     |
| 2001 | 0.5233 | 98.68     | 0.4075  | 98.94     |
| 2002 | 0.5345 | 98.73     | 0.4105  | 98.92     |
| 2003 | 0.529  | 98.78     | 0.4058  | 99.03     |
| 2004 | 0.5101 | 98.71     | 0.4077  | 99.03     |
| 2005 | 0.5091 | 98.73     | 0.4312  | 99.00     |
| 2006 | 0.5219 | 98.78     | 0.4461  | 98.95     |
| 2007 | 0.5169 | 98.82     | 0.4411  | 98.86     |
| 2008 | 0.5153 | 98.69     | 0.4265  | 98.87     |
| 2009 | 0.5179 | 98.79     | 0.4139  | 98.85     |
| 2010 | 0.5052 | 98.80     | 0.405   | 98.89     |
| 2011 | 0.4897 | 98.90     | -       | -         |
| 2012 | 0.4974 | 98.82     | -       | -         |

**Table 3.8** Coefficients of the Fermi-Dirac probability density function fitting US annual household income in the time interval 2003-2013

| Year | T | Income /capita (2013 USD) | g | Gini coeff | μ | Exports x $10^9$ (USD) | $R^2$ (%) |
|---|---|---|---|---|---|---|---|
| 2003 | -0.365 | 28829 | 1.396 | 44.8 | 4.01 | 1020418 | 97.34 |
| 2004 | -0.3838 | 28692 | 1.421 | 45.1 | 4.033 | 1161549 | 98.01 |
| 2005 | -0.3775 | 28538 | 1.444 | 45.0 | 4.082 | 1286022 | 97.72 |
| 2006 | -0.37 | 28374 | 1.471 | 44.0 | 4.138 | 1457642 | 97.97 |
| 2007 | -0.3701 | 28812 | 1.488 | 44.3 | 4.167 | 1653548 | 97.02 |
| 2008 | -0.3794 | 29173 | 1.501 | 43.8 | 4.176 | 1841612 | 97.40 |
| 2009 | -0.3943 | 30114 | 1.506 | 43.2 | 4.167 | 1583053 | 97.91 |
| 2010 | -0.4009 | 30446 | 1.501 | 44.4 | 4.156 | 1853606 | 97.53 |
| 2011 | -0.406 | 29874 | 1.518 | 44.0 | 4.177 | 2127021 | 98.00 |
| 2012 | -0.4143 | 29421 | 1.54 | 43.8 | 4.202 | 2216540 | 97.68 |
| 2013 | -0.4175 | 29481 | 1.557 | 43.6 | 4.225 | 2280194 | 97.69 |
| Correlation coefficient | -0.69 | | -0.80 | | 0.92 | | |
| T test | 142.8 | | 234.69 | | 13.05 | | |

**Appendix 2 Fermi-Dirac distribution in continuous approximation**

Similarly with fermions, we apply the calculation for distribution of money using the continuous approximation. We obtain this by integrating the average number of fermions above a certain threshold (in our case $x_0$).

$$N(x_0) = \sum_x \langle n \rangle$$

$$N(x_0) = \int_{x_0}^{\infty} \langle n \rangle \, dx = \int_{x_0}^{\infty} \frac{g}{exp\left(\frac{x-\mu}{T}\right)+1} dx \quad (3.13)$$

$u = exp\left(\frac{x-\mu}{T}\right) + 1$ (3.14), then $u_0 = exp\left(\frac{x_0-\mu}{T}\right) + 1$

$$du = \frac{1}{T} exp\left(\frac{x-\mu}{T}\right) dx$$

$$dx = T \, exp\left(\frac{\mu-x}{T}\right) du \quad (3.15)$$

By substituting (3.14)-(3.15) into (3.13) we get:

$$N(u_0) = \int_{u_0}^{\infty} \frac{gT}{u(u-1)} du \quad (3.16)$$

We integrate

$$\frac{A}{u} + \frac{B}{u-1} = \frac{Bu + A(u-1)}{u(u-1)}$$

$$\frac{A}{u} + \frac{B}{u-1} = \frac{u(A+B) - A}{u(u-1)}$$

$$A + B = 0, A = -gT, B = gT$$

The equation (3.16) becomes

$$N(u_0) = gT\left(-\int_{u_0}^{\infty} \frac{du}{u} + \int_{u_0}^{\infty} \frac{du}{u-1}\right)$$

$$N(u_0) = gT[-\ln(u) + \ln(u-1)]\big|_{u_0}^{\infty}$$

$$N(u_0) = gT\left(\ln\frac{u-1}{u}\right)\Big|_{u_0}^{\infty}$$

$$N(u_0) = gT\left[\ln\left(1 - \frac{1}{u}\right)\right]\Big|_{u_0}^{\infty}$$

The equation becomes

$$N(x_0) = gT\left[\ln\left(1 - \frac{1}{\exp\left(\frac{x-\mu}{T}\right) + 1}\right)\right]\Big|_{x_0}^{\infty}$$

But $\lim_{x \to \infty} \ln\left[1 - \frac{1}{\exp\left(\frac{x-\mu}{T}\right)+1}\right] \to 0$, then

$$N(x_0) = -gT\ln\left[1 - \frac{1}{\exp\left(\frac{x_0-\mu}{T}\right) + 1}\right]$$

$$N(x_0) = gT\ln\left(1 + \exp\frac{\mu - x_0}{T}\right) \quad (3.17)$$

# CHAPTER 4: APPLICATIONS OF POLYNOMIAL FUNCTION TO THEDISTRIBUTION OF INCOME, WEALTH, AND EXPENDITURE

Our approach is based on a totally new distribution, not used so far in the literature regarding income and wealth distribution. Using the complementary cumulative distribution function or probability density function, we find that polynomial functions, regardless of their degree (first, second, third, or higher), can describe with high accuracy income, wealth, and expenditure distribution. Moreover, we find that polynomial functions describe income, expenditure, and wealth distribution for the entire population including upper income segment for which traditionally Pareto distribution is used.

## 4.1 Theoretical framework and literature review

While most of papers claim to cover income and wealth distribution only for low and middle income part of population, there are two exceptions. Fermi-Dirac distribution [90] and Tsallis distribution [51] claim to be robust enough in order to explain income, wealth, and expenditure distribution for the entire population, including for upper income segment of population which traditionally is described by a Pareto distribution.

Polynomial theory uses predictive polynomials as the basic means for the general investigation of complex dynamic systems. The predictive polynomial consists of a regression equation that connects a future value for the output variable with current or past regarding all input and output variables. A more general theory to create an optimising decision algorithm using the information provided in few points of interpolation is not yet available, consequently, polynomial theory creates "hypothesis of selection". Thus, polynomials are handled similarly as seeds in agricultural selection. This allows getting a polynomial description of a component or for the whole complex plant by observing their inputs and outputs over relatively short time. The main advantage of polynomial theory is about finding optimum complexity for polynomial description, which is the one capable to describe the complexity of the plant. Only these descriptions can ensure high prediction accuracy. Polynomial descriptions have some other advantages. For example, it is not necessary to find solutions for equations with finite difference form because the answers to all the interesting questions can be found from the polynomial itself. Consequently, the information regarding initial condition and solution of the equations are not necessary. There is no distinction made between the statics and dynamics of a plant in the polynomial theory. This categorisation of plant regimes is linked with the application of differential equations. There is no need to use identification methods for estimating the coefficients of the differential equations. It is easier to use them directly for the synthesis of polynomial descriptions. Less information is required for this operation. Thus, it is not necessary to know the type of differential equations. Polynomial theory is the only theory allowing us to obtain optimum complexity for a mathematical model of the plant. The most accurate description of extremely complex plants is done by polynomial descriptions of a high degree (up to the sixty-fourth degree). Consequently, finding nonlinear differential equations corresponding to such complex polynomials is impossible. The polynomial theory of complex dynamic systems will cause revolution in the prediction, pattern recognition, identification, optimising control with information storage, and to the other problems of engineering cybernetics. For instance, in order to compare two economic systems would suffice to identify the polynomial descriptions of two systems and to compare their potential possibilities [100].

More importantly, the degree of the polynomials and their shape can be used in order to test different hypothesis regarding the evolution of data. Thus, the authors in the paper [101] test three development hypotheses such as world-systems theory, dependency theory, and modernisation theory using the statistical distribution of Gross National Product in the world's population.

### 4.2 Methodology

In order to calculate the probability distribution using polynomial distribution, we use complementary cumulative probability distribution and probability density function. The complementary cumulative probability distribution $\bar{C}(x)$ is defined as the integral

$$\bar{C}(x) = \int_x^\infty P(x) dx \quad (4.1)$$

It gives the probability that the random variable exceeds a given value $x$ [21]. We present on y-axis the cumulated population probability, which is share of population with income/wealth/expenditure higher than corresponding level on the x-axis. Cumulated income/wealth/expenditure is contained on the x-axis. According to this type of probability, we calculate the share of population having income/wealth/expenditure above a certain threshold. Thus, the probability to have income/wealth/expenditure higher than zero is 100% (since everyone is assumed to have a certain income). Cumulated income, wealth, or expenditure is displayed on the x-axis. Let us assume X represents the values for cumulated income represented on the x-axis.

$$X_i = \sum x_i$$

where X represents the cumulated income on the x-axis and i=[1,10] and i∈N. The total cumulated probability is $\bar{C}_i(a > X_i)$. In the case of mean income, the set which contains the plots representing the probability is $S=\{$ (0,100%), ($X_1$, 90%), ($X_2$, 80%), ($X_3$, 70%), ($X_4$, 60%), ($X_5$, 50%), ($X_6$, 40%), ($X_7$, 30%), ($X_8$, 20%), ($X_9$, 10%), ($X_{10}$, 0%)$\}$. In the case for the upper limit on income data sets, $S_1=\{$(0,100%), ($X_1$, 90%), ($X_2$, 80%), ($X_3$, 70%), ($X_4$, 60%), ($X_5$, 50%), ($X_6$, 40%), ($X_7$, 30%), ($X_8$, 20%), ($X_9$, 10%)$\}$. This is true considering that values for the tenth decile which contains the upper income segment of population is not comprised in the data set. For the lower limit on income, the set is similar upper limit on income, except that each value represents the lowest expenditure value on an income decile.

Thus, for income equal to 0, the cumulative density probability is 100 % as it is considered that everyone has an income higher than 0. For the first decile (the lowest income decile), $\bar{C}_1$ represents the population that has an income higher than mean income or upper limit on income or lower limit of the first decile, hence equals 90%. For lower limit on income, the value for the first decile is 0. Subsequently, for the highest income the complementary cumulative distribution function has probability 0 % (in case of mean income). For the upper limit on income and lower data set, we do not represent the upper decile as the value corresponding for it was not made available by any of the statistical bodies. Thus, complementary cumulative probability distribution addresses the cumulated percentage of population which has income/wealth/expenditure higher than a certain threshold. The distribution we found for income, expenditure, or wealth to fit best the data is the third degree polynomial distribution.

$$\bar{C}(X) = a_0 + a_1 * X + a_2 * X^2 + a_3 * X^3 \quad (4.2)$$

The data that we will be using will be from Brazil [65], Finland [63, 66], France [67-71], Italy [72-73], Philippine [74], Romania [75], Singapore [76], the UK [64], and Uganda [77].

According to the probability density function

$$\sum_{i=0}^{n} P_i = 1 \quad (4.3)$$

Probability density function is calculated based on annual household income data provided by the USA [78]. The values on the x-axis are distributed as $x_i^*$, where i∈N, and $x^* = log_{10}(x)$. Probability density $P^*_1$ is the decimal logarithm of probability density for the population whose income is between $(x_0, x_1)$, $P^*_2$ is the decimal logarithm of probability density for the population whose income is between $(x_1, x_2)$. Similarly, $P^*_n$ is the decimal logarithm of probability density for the population whose income is between $(x_{n-1}, x_n)$, where n∈N.

The set containing the probability plot in decimal logarithmic values is $S_2=\{(x_0^*, P^*_0), (x_1^*, P^*_1),…., (x_n^*, P^*_n)\}$, where n∈N. We fitted the probability plot for income in the USA using as probability density function a third degree polynomial.

$$P^*(x^*) = a_0 + a_1 * x^* + a_2 * x^{*2} + a_3 * x^{*3} \quad (4.4)$$

### 4.3 Results

We applied the polynomial distribution using statistical package R. We present the results from complementary cumulative distribution probability graphically by using the eq. (4.2) in Figures 4.1- 4.2 and in the tables 4.1- 4.20. In the appendices 4.1- 4.11, there are exhibited the results from fitting disposable income. In the tables 4.12- 4.13, we present the results from fitting pensions in the UK and France. Also, in the tables 4.14-4.18 we show the results for expenditure. The table 4.19 exhibits the results for wealth. In the table 4.20, there are displayed the results from fitting the data from USA using Fermi-Dirac probability density function.

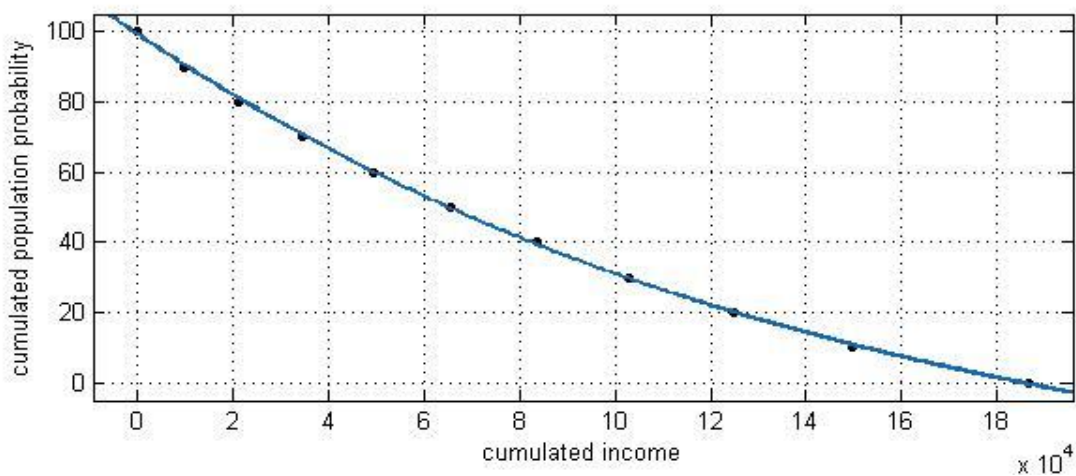

**Figure 4.1 Polynomial distribution fitting annual data regarding mean income in Finland for the year 1996.** On the x-axis, we represented the cumulated income –$X_i$ and on the y-axis the cumulated probability- $\bar{C}_i$.

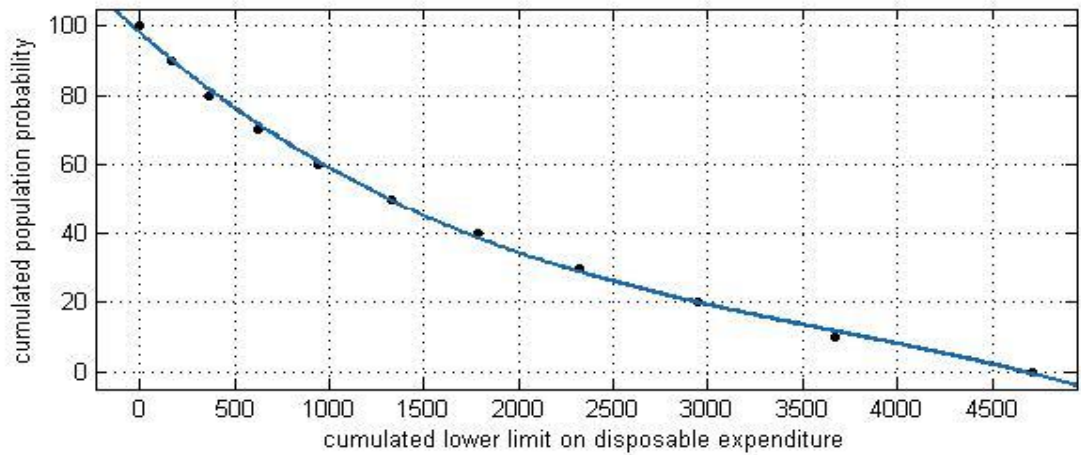

**Figure 4.2 Polynomial distribution fitting annual data regarding lower limit on disposable expenditure in the UK for the year 2008.** On the x-axis, we represented the cumulated disposable expenditure –$X_i$ and on the y-axis the cumulated probability- $\bar{C}_i$.

Polynomial function applied to cumulative values/probabilities satisfies the necessary mathematical conditions for complementary cumulative distribution function (1.12-1.15).

On the x-axis, we represented the cumulated income for Figure 4.1 and the cumulated expenditure for Figure 4.2. On the y-axis we presented the cumulated percentage of the people who have income higher than a specific threshold (Figure 4.1) and a level of expenditure higher than certain level (Figure 4.2)

Polynomial complementary cumulative distribution is applicable to income, expenditure, and wealth. Regarding the goodness of the fitting, the lowest value for coefficient of determination ($R^2$) is 95%. Most of the annual fittings have values for coefficient of determination above 99 %. We can notice that the values for coefficient of determination for wealth are slightly above 92 %.

Regarding the income distribution for countries such as Finland, France, and Italy, upper limit on (disposable) income data sets have similar values with mean income data sets for the values of parameters and the values of the coefficient of determination. Thus, fitting the tenth decile does not decrease the overall goodness of the fit considering that upper income segment of population is more dependent on asset prices than on wages [102].This indicates that the tenth segment which normally is described by Pareto distribution can be described successfully by a polynomial distribution.

Regarding the t-values, which determine the statistical significance for each component of the polynomial, we can observe that t values are between -2 and 2 for some countries and for some time intervals. Thus, in the case of Finland for mean income for the time interval 1998-2009 the first degree component of the polynomial is statistically insignificant. In the case of France, also for the first degree component of the mean income and the income of pensioners data is also statistically insignificant. For Italy, the first degree component of the upper limit on income set for the entire time interval analysed is also insignificant. For Romania also the first degree component is also insignificant for the mean income. As the largest data set is about the UK, we have noticed statistical insignificance for the mean income the second degree component from 1986 to 2012, for gross income data set the second degree component for the time interval 1983-2012 and for the third degree component for the time interval 1987-2012. Also, there is a statistical insignificance for pensions for second and third degree components for the entire time interval and for the UK wages for the third degree component for the time interval 1977-1997. Apparently, there are no correlations that could be drawn as to why these statistical significances occur for some countries and certain components of the polynomial and time interval. One possible exception is that in many cases the first degree component is statistical insignificant in the countries from continental Europe compared to the countries outside of Europe.

Regarding the expenditure, we can say that there are no big differences in the values for coefficient of determination between the UK and Uganda, the countries which made available such data. Also, there is no difference between the mean gross and disposable expenditure data sets. Furthermore, the same phenomenon remains valid in the case of mean and lower limit on disposable expenditure. Regarding the values for coefficient of determination, we can notice that there is no significant difference between the values obtained from fitting expenditure which is expressed in weekly values and disposable income which is expressed in from the UK. Wealth distribution is not similarly distributed as in the case of income because the underlying mechanism is different. Thus, using a third degree polynomial to fit the data yields values for coefficient of determination slightly above 95% [102], which is slightly lower than the overall values for coefficient of determination regarding annual fitting of the income data. Also, the values for parameters are similar over time in the case of all types of income, expenditure, and wealth [102-103], provided that the currency remains the same for the time interval analysed. Generally, for the same country and for the same data set the values of a parameter increase or decrease slightly as result of the fittings from one year to another. The only possible exception is in the case of Romania (where the variations are higher) because of the high inflation at the beginning of the years 2000.

An exception regarding the values of parameters of the polynomials as well as for coefficient of determination was noticed in the case of the countries which changed their currencies. Thus, in case of Brazil it is difficult to determine the evolution of the values for parameters given that in the years considered several different currencies were used. Also, we noticed a slight decrease of about 2% for the annual values of coefficient of determination between the time intervals before the 80s and the time interval after the 80s.

In the cases of Italy and Romania, where a change of currency occurred in 2000 and, respectively, 2005, we noticed no significant changes regarding the values for coefficient of determination. However, a significant and abrupt change was noticed in the years that marked the currency change for the coefficients $a_1$, $a_2$, and $a_3$. The coefficient $a_0$ remains relatively stable from the fitting of the one year to another [102-103].

The results from Durbin Watson test are exhibited in tables 4.1-4.20. We notice that most of the results from Durbin-Watson test fall in the interval 1.4-1.6 for the income from continental countries from Europe. These indicate mild positive autocorrelation of residuals. For the rest of the countries outside Europe and the UK the values are in the interval 0.8-1.2. A possible explanation is the social welfare system in Europe, which influences the income distribution. For expenses, we notice that the values from Durbin-Watson test are in the interval 1.2-1.4, regardless they are from the UK or Uganda. Regarding the wealth data, the DW test indicates a strong positive autocorrelation, which shows that model could be improved.

Since we considered economic systems as having complex system behaviour, the fact that their behaviour can be fitted using polynomials of different degrees from one year to another is evidence that economic systems can be modelled using catastrophe theory. We could see that small variations in the values of parameters lead to large quantitative changes in solutions as the data can be fitted in some years by polynomial having a degree different than the year before or after. So, large quantitative changes in solutions mean that qualitative changes occur [104]. Thus, small variations in the values of parameters leading to a change in the number of roots suggest the occurrence of qualitative changes in the differential equations which are modelled by these polynomials. Thus, the parameters are the analogues of control variables and x is the analogue for internal variables used in catastrophe theory.

The results from using the probability density function as described in the eq. (4.4) in order to fit non-cumulative income/probabilities are displayed in the Table 4.20.

Polynomial pdf function applied to income values/probabilities satisfies the necessary mathematical conditions for a pdf function (1.5-1.7).

The values on both axes are represented in decimal logarithmic values. The annual data span the time interval 1967-2013. The income annual data fitted by the third degree polynomial had lowest value for coefficient of determination ($R^2$) of about 92.7 %. The values from Durbin-Watson test exhibit positive autocorrelation for all the values and most of the values were between (1, 2). For the time interval 1986-1990, the values for DW test are very close to 1.

The model fits best the data towards the year the beginning of the interval and becomes less fitted to the data towards the end of the time interval in the year 2013. The values from t-test show that the parameters are statistically significant, except for the years 1984-1997 when some or all parameters are shown to be statistically insignificant.

To the best of our knowledge, this is the first attempt to describe income distribution using polynomial distribution which models complex systems. The only similar work is Yakovenko's investigation of income and wealth distribution using Boltzmann-Gibbs distribution (exponential distribution). Unlike Yakovenko, we use more statistical tools validate our findings such as t tests and Durbin-Watson test.

### 4.4 Conclusions

Based on the finding, we can draw the conclusion that polynomial distribution is robust. Thus, the polynomial distribution can describe with high accuracy all types of income, wealth, and expenditure. Also, the distribution can fit the entire range of income, wealth, and expenditure including for the upper income segment of population using different methods to calculate them (mean value, upper or lower value) [102-103]. Mean income and upper limit values can be equally used in the analysis of income, wealth and expenditure distribution. However, no data were provided for the tenth decile in the upper limit on income data sets but considering that values for coefficient of determination look similar for the same data/year analysed using upper limit on income or mean income, we can conclude that the data from the tenth does not have a significant impact on the goodness of the fit of the distribution.

Based on the connection with theory of complex systems, catastrophe theory can be used for further research on the income, expenditure, and wealth distribution in order to find out the circumstances when this qualitative changes occur and why this happens.

### 4.5 Appendix

**Table 4.1** Coefficients of the polynomial distribution fitting mean income in Brazil

| Year | $a_0$ | T-value | $a_1$ | T-value | $a_2$ | T-value | $a_3$ | T-value | $R^2$ (%) | DW |
|------|------|------|------|------|------|------|------|------|------|------|
| 1960 | 93.0 | 40.7 | $-5.2*10^{-3}$ | -10.2 | $1.1*10^{-7}$ | 4.7 | $-9.4*10^{-13}$ | -3.2 | 99.1 | 0.9 |
| 1970 | 94.1 | 46.5 | $-4.9*10^{-3}$ | -12.8 | $1*10^{-7}$ | 6.3 | $-6.8*10^{-13}$ | -4.4 | 99.3 | 0.9 |
| 1980 | 94.6 | 50.6 | $-2.6*10^{-3}$ | -14.5 | $2.9*10^{-8}$ | 7.4 | $-1.0*10^{-13}$ | -5.3 | 99.4 | 1.0 |
| 1981 | 92.9 | 38.9 | $-1.2*10^{-1}$ | -11.2 | $6.7*10^{-5}$ | 5.9 | $-1.1*10^{-8}$ | -4.4 | 98.9 | 0.9 |
| 1992 | 91.8 | 35.1 | $-1.3*10^{-1}$ | -9.9 | $7*10^{-5}$ | 5.2 | $-1.2*10^{-8}$ | -3.9 | 98.1 | 0.9 |
| 2002 | 92.3 | 37.0 | $-1.0*10^{-1}$ | -10.6 | $4.3*10^{-5}$ | 5.6 | $-6.0*10^{-9}$ | -4.2 | 98.3 | 0.9 |

**Table 4.2** Coefficients of the polynomial distribution fitting mean income from Finland

| Year | $a_0$ | T-value | $a_1$ | T-value | $a_2$ | T-value | $a_3$ | T-value | $R^2$ (%) | DW |
|------|------|------|------|------|------|------|------|------|------|------|
| 1987 | 98 | 145 | $-1.0*10^{-3}$ | -24 | $3.7*10^{-9}$ | 6 | $-7.1*10^{-15}$ | -2 | 99.9 | 1.4 |
| 1988 | 98 | 144 | $-9.9*10^{-4}$ | -24 | $3.4*10^{-9}$ | 5 | $-6.2*10^{-15}$ | -2 | 99.9 | 1.5 |
| 1989 | 98 | 146 | $-9.4*10^{-4}$ | -25 | $3.1*10^{-9}$ | 6 | $-5.3*10^{-15}$ | -2 | 99.9 | 1.4 |
| 1990 | 98 | 152 | $-8.9*10^{-4}$ | -25 | $2.8*10^{-9}$ | 6 | $-4.4*10^{-15}$ | -2 | 99.9 | 1.4 |
| 1991 | 98 | 145 | $-8.8*10^{-4}$ | -24 | $2.6*10^{-9}$ | 5 | $-4.1*10^{-15}$ | -2.5 | 99.9 | 1.5 |
| 1992 | 98 | 153 | $-9.1*10^{-4}$ | -25 | $2.7*10^{-9}$ | 5 | $-4.1*10^{-15}$ | -2.3 | 99.9 | 1.5 |
| 1993 | 98 | 156 | $-9.4*10^{-4}$ | -26 | $2.9*10^{-9}$ | 5 | $-4.0*10^{-15}$ | -2.1 | 99.9 | 1.5 |
| 1994 | 98 | 157 | $-9.4*10^{-4}$ | -26 | $2.9*10^{-9}$ | 5 | $-4.1*10^{-15}$ | -2.2 | 99.9 | 1.4 |
| 1995 | 98 | 154 | $-9.3*10^{-4}$ | -26 | $2.8*10^{-9}$ | 5 | $-3.9*10^{-15}$ | -2.2 | 99.9 | 1.5 |
| 1996 | 99 | 184 | $-9.1*10^{-4}$ | -31 | $2.7*10^{-9}$ | 7 | $-3.5*10^{-15}$ | -2.6 | 99.9 | 1.4 |
| 1997 | 98 | 145 | $-9.1*10^{-4}$ | -25 | $2.8*10^{-9}$ | 6 | $-3.8*10^{-15}$ | -2.5 | 99.9 | 1.4 |
| 1998 | 98.5 | 132 | $-9.0*10^{-4}$ | -23 | $2.8*10^{-9}$ | 6 | $-3.8*10^{-15}$ | -2.5 | 99.9 | 1.4 |
| 1999 | 98.4 | 127 | $-8.6*10^{-4}$ | -23 | $2.*10^{-9}$ | 5 | $-3.0*10^{-15}$ | -2.2 | 99.9 | 1.4 |
| 2000 | 101.5 | 29 | $2.9*10^{-4}$ | 0.4 | $-2.3*10^{-7}$ | -7 | $3.8*10^{-15}$ | 8 | 99.2 | 1.3 |
| 2001 | 98.4 | 124 | $-8.6*10^{-4}$ | -23 | $2.5*10^{-9}$ | 5 | $-3.1*10^{-15}$ | -2.3 | 99.9 | 1.4 |
| 2002 | 98.3 | 121 | $-8.5*10^{-4}$ | -22 | $2.5*10^{-9}$ | 5 | $-3.2*10^{-15}$ | -2.4 | 99.8 | 1.4 |
| 2003 | 98.3 | 118 | $-8.2*10^{-4}$ | -21 | $2.4*10^{-9}$ | 5 | $-3.1*10^{-15}$ | -2.4 | 99.9 | 1.4 |
| 2004 | 98.3 | 122 | $-8.0*10^{-4}$ | -22 | $2.2*10^{-9}$ | 5 | $-2.8*10^{-15}$ | -2.5 | 99.9 | 1.4 |
| 2005 | 98.3 | 117 | $-7.7*10^{-4}$ | -21 | $2.1*10^{-9}$ | 5 | $-2.5*10^{-15}$ | -2.4 | 99.9 | 1.3 |
| 2006 | 98.2 | 115 | $-7.5*10^{-4}$ | -21 | $2.0*10^{-9}$ | 5 | $-2.3*10^{-15}$ | -2.4 | 99.3 | 1.3 |
| 2007 | 98.2 | 115 | $-7.5*10^{-4}$ | -21 | $2*10^{-9}$ | 5 | $-2.3*10^{-15}$ | -2.5 | 99.8 | 1.3 |
| 2008 | 98.1 | 111 | $-7.4*10^{-4}$ | -20 | $1.9*10^{-9}$ | 5 | $-2.1*10^{-15}$ | -2.4 | 99.9 | 1.3 |
| 2009 | 98.1 | 108 | $-7.3*10^{-4}$ | -20 | $1.9*10^{-9}$ | 5 | $-2.2*10^{-15}$ | -2.3 | 99.9 | 1.3 |

**Table 4.3** Coefficients of the polynomial distribution fitting the upper limit on income from Finland

| Year | $a_0$ | T-value | $a_1$ | T-value | $a_2$ | T-value | $a_3$ | T-value | $R^2$ (%) | DW |
|---|---|---|---|---|---|---|---|---|---|---|
| 1987 | 99.5 | 321 | $-9.6*10^{-4}$ | -46 | $3.5*10^{-9}$ | 10 | $-7.8*10^{-15}$ | -4 | 99.9 | 1.6 |
| 1988 | 99.5 | 312 | $-9.4*10^{-4}$ | -44 | $3.3*10^{-9}$ | 9 | $-7*10^{-15}$ | -4 | 99.9 | 1.6 |
| 1989 | 99.5 | 306 | $-9.0*10^{-3}$ | -44 | $3.0*10^{-9}$ | 9 | $-6.3*10^{-15}$ | -4 | 99.9 | 1.5 |
| 1990 | 99.5 | 325 | $-8.5*10^{-4}$ | -46 | $2.7*10^{-9}$ | 9 | $-5.0*10^{-15}$ | -4 | 99.9 | 1.6 |
| 1991 | 99.5 | 328 | $-8.3*10^{-4}$ | -46 | $2.5*10^{-9}$ | 9 | $-4.6*10^{-15}$ | -4 | 99.9 | 1.5 |
| 1992 | 99.6 | 358 | $-8.6*10^{-4}$ | -50 | $2.6*10^{-9}$ | 10 | $-4.9*10^{-15}$ | -4 | 99.9 | 1.6 |
| 1993 | 99.6 | 412 | $-9.0*10^{-4}$ | -58 | $2.8*10^{-9}$ | 11 | $-4.9*10^{-15}$ | -4 | 99.9 | 1.6 |
| 1994 | 99.6 | 343 | $-9.0*10^{-4}$ | -49 | $2.8*10^{-9}$ | 9 | $-5.0*10^{-15}$ | -3 | 99.9 | 1.6 |
| 1995 | 99.5 | 345 | $-8.9*10^{-4}$ | -49 | $2.8*10^{-9}$ | 10 | $-5.1*10^{-15}$ | -4 | 99.9 | 1.6 |
| 1996 | 99.5 | 319 | $-8.9*10^{-4}$ | -46 | $2.9*10^{-9}$ | 10 | $-5.5*10^{-15}$ | -4 | 99.9 | 1.6 |
| 1997 | 99.5 | 345 | $-8.8*10^{-4}$ | -51 | $2.9*10^{-9}$ | 11 | $-5.3*10^{-15}$ | -5 | 99.8 | 1.6 |
| 1998 | 99.5 | 293 | $-8.7*10^{-4}$ | -44 | $2.9*10^{-9}$ | 10 | $-5.6*10^{-15}$ | -5 | 99.9 | 1.5 |
| 1999 | 99.4 | 291 | $-8.4*10^{-4}$ | -43 | $2.8*10^{-9}$ | 10 | $-5.1*10^{-15}$ | -5 | 99.9 | 1.6 |
| 2000 | 99.4 | 264 | $-8.4*10^{-4}$ | -40 | $2.8*10^{-9}$ | 9 | $-5.2*10^{-15}$ | -4 | 99.9 | 1.6 |
| 2001 | 99.4 | 259 | $-8.2*10^{-4}$ | -39 | $2.7*10^{-9}$ | 9 | $-4.9*10^{-15}$ | -4 | 99.9 | 1.5 |
| 2002 | 99.3 | 243 | $-8.0*10^{-4}$ | -37 | $2.5*10^{-9}$ | 9 | $-4.5*10^{-15}$ | -4 | 99.9 | 1.5 |
| 2003 | 99.4 | 252 | $-7.8*10^{-4}$ | -38 | $2.4*10^{-9}$ | 9 | $-4.2*10^{-15}$ | -4 | 99.9 | 1.5 |
| 2004 | 99.3 | 240 | $-7.5*10^{-4}$ | -36 | $2.3*10^{-9}$ | 9 | $-3.8*10^{-15}$ | -4 | 99.9 | 1.6 |
| 2005 | 99.3 | 221 | $-7.3*10^{-4}$ | -34 | $2.2*10^{-9}$ | 8 | $-3.5*10^{-15}$ | -4 | 99.9 | 1.5 |
| 2006 | 99.3 | 228 | $-7.3*10^{-4}$ | -35 | $2.2*10^{-9}$ | 9 | $-3.5*10^{-15}$ | -4 | 99.9 | 1.5 |
| 2007 | 99.2 | 216 | $-7.2*10^{-4}$ | -33 | $2.1*10^{-9}$ | 8 | $-3.4*10^{-15}$ | -4 | 99.9 | 1.5 |
| 2008 | 99.2 | 215 | $-7.1*10^{-4}$ | -33 | $2.0*10^{-9}$ | 8 | $-3.3*10^{-15}$ | -4 | 99.9 | 1.5 |
| 2009 | 99.2 | 218 | $-6.8*10^{-4}$ | -33 | $1.9*10^{-9}$ | 8 | $-2.9*10^{-15}$ | -4 | 99.9 | 1.6 |

**Table 4.4** Coefficients of the polynomial distribution fitting mean income from France

| Year | $a_0$ | T-value | $a_1$ | T-value | $a_2$ | T-value | $a_3$ | T-value | $R^2$ (%) | DW |
|---|---|---|---|---|---|---|---|---|---|---|
| 2003 | 98.2 | 117 | $-9.5*10^{-4}$ | -22 | $3.2*10^{-9}$ | 6 | $-4.8*10^{-15}$ | -2.9 | 99.9 | 1.4 |
| 2004 | 98.2 | 118 | $-9.4*10^{-4}$ | -22 | $3.2*10^{-9}$ | 6 | $-4.6*10^{-15}$ | -2.8 | 99.9 | 1.4 |
| 2005 | 97.9 | 108 | $-9.4*10^{-4}$ | -20 | $3.2*10^{-9}$ | 5 | $-4.7*10^{-15}$ | -2.6 | 99.9 | 1.4 |
| 2006 | 98.0 | 109 | $-9.2*10^{-4}$ | -21 | $3.1*10^{-9}$ | 6 | $-4.4*10^{-15}$ | -2.7 | 99.9 | 1.4 |
| 2007 | 98.0 | 108 | $-9.1*10^{-4}$ | -21 | $3.0*10^{-9}$ | 6 | $-4.3*10^{-15}$ | -2.7 | 99.9 | 1.4 |
| 2008 | 97.9 | 106 | $-8.8*10^{-4}$ | -20 | $2.8*10^{-9}$ | 5 | $-3.7*10^{-15}$ | -2.4 | 99.9 | 1.4 |
| 2009 | 97.9 | 105 | $-8.9*10^{-4}$ | -20 | $2.9*10^{-9}$ | 5 | $-4.1*10^{-15}$ | -2.7 | 99.9 | 1.4 |

**Table 4.5** Coefficients of the polynomial distribution fitting upper limit on income from France

| Year | $a_0$ | T-value | $a_1$ | T-value | $a_2$ | T-value | $a_3$ | T-value | $R^2$ (%) | DW |
|---|---|---|---|---|---|---|---|---|---|---|
| 2003 | 99.2 | 222 | $-9.1 *10^{-4}$ | -35 | $3.3 *10^{-9}$ | 8 | $-6.2 *10^{-15}$ | -4 | 99.9 | 1.5 |
| 2004 | 99.2 | 228 | $-9.1 *10^{-4}$ | -35 | $3.3 *10^{-9}$ | 9 | $-6.2 *10^{-15}$ | -4 | 99.9 | 1.6 |
| 2005 | 99.2 | 216 | $-9.0 *10^{-4}$ | -34 | $3.3 *10^{-9}$ | 8 | $-6.1 *10^{-15}$ | -4 | 99.9 | 1.6 |
| 2006 | 99.2 | 213 | $-8.9 *10^{-4}$ | -33 | $3.2*10^{-9}$ | 8 | $-5.8 *10^{-15}$ | -4 | 99.9 | 1.5 |
| 2007 | 99.2 | 215 | $-8.7 *10^{-4}$ | -34 | $3.1*10^{-9}$ | 8 | $-5.6 *10^{-15}$ | -4 | 99.9 | 1.6 |
| 2008 | 99.2 | 205 | $-8.5 *10^{-4}$ | -32 | $2.8*10^{-9}$ | 8 | $-4.9 *10^{-15}$ | -3 | 99.9 | 1.5 |
| 2009 | 99.2 | 205 | $-8.6 *10^{-4}$ | -32 | $3.0*10^{-9}$ | 8 | $-5.3 *10^{-15}$ | -4 | 99.9 | 1.5 |

**Table 4.6** Coefficients of the polynomial distribution fitting mean income from Italy

| Year | $a_0$ | T-value | $a_1$ | T-value | $a_2$ | T-value | $a_3$ | T-value | $R^2$ (%) | DW |
|---|---|---|---|---|---|---|---|---|---|---|
| 1989 | 97.0 | 79 | $-6.6 *10^{-7}$ | -17 | $1.8 *10^{-15}$ | 6 | $-2.1 *10^{-24}$ | -3 | 99.8 | 1.3 |
| 1991 | 97.0 | 79 | $-6.1 *10^{-7}$ | -17 | $1.5 *10^{-15}$ | 6 | $-1.7 *10^{-24}$ | -3 | 99.8 | 1.3 |
| 1993 | 96.1 | 64 | $-6.2 *10^{-7}$ | -14 | $1.6 *10^{-15}$ | 5 | $-1.8 *10^{-24}$ | -3 | 99.6 | 1.2 |
| 1995 | 96.3 | 66 | $-5.6 *10^{-7}$ | -14 | $1.3*10^{-15}$ | 5 | $-1.3 *10^{-24}$ | -3 | 99.7 | 1.2 |
| 1998 | 95.8 | 61 | $-5.1 *10^{-7}$ | -13 | $1.1*10^{-15}$ | 5 | $-9.8 *10^{-25}$ | -3 | 99.6 | 1.2 |
| 2000 | 96.2 | 66 | $-9.2 *10^{-5}$ | -14 | $3.6 *10^{-9}$ | 5 | $-5.7 *10^{-15}$ | -3 | 99.7 | 1.3 |
| 2002 | 96.5 | 70 | $-8.6 *10^{-4}$ | -15 | $3.1*10^{-9}$ | 6 | $-4.7 *10^{-15}$ | -3 | 99.7 | 1.2 |
| 2004 | 97.0 | 81 | $-8.0 *10^{-4}$ | -17 | $2.6 *10^{-9}$ | 6 | $-3.5 *10^{-15}$ | -4 | 98.8 | 1.3 |
| 2006 | 97.1 | 81 | $-7.3 *10^{-4}$ | -17 | $2.2*10^{-9}$ | 6 | $-2.7 *10^{-15}$ | -3 | 99.8 | 1.3 |
| 2008 | 96.8 | 76 | $-7.3 *10^{-4}$ | -16 | $2.2*10^{-9}$ | 6 | $-2.8 *10^{-15}$ | -3 | 99.7 | 1.3 |

**Table 4.7** Coefficients of the polynomial distribution fitting upper limit on income from Italy

| Year | $a_0$ | T-value | $a_1$ | T-value | $a_2$ | T-value | $a_3$ | T-value | $R^2$ (%) | DW |
|---|---|---|---|---|---|---|---|---|---|---|
| 1989 | 98.7 | 139 | $-6.4 *10^{-7}$ | -25 | $1.9 *10^{-15}$ | 8 | $-2.7 *10^{-24}$ | -5 | 99.9 | 1.5 |
| 1991 | 98.6 | 133 | $-5.9 *10^{-7}$ | -24 | $1.6 *10^{-15}$ | 8 | $-2.1 *10^{-24}$ | -5 | 99.9 | 1.5 |
| 1993 | 98.1 | 104 | $-6.0 *10^{-7}$ | -19 | $1.7 *10^{-15}$ | 7 | $-2.3 *10^{-24}$ | -4 | 99.8 | 1.5 |
| 1995 | 98.2 | 107 | $-5.5 *10^{-7}$ | -20 | $1.4 *10^{-15}$ | 7 | $-1.7 *10^{-24}$ | -4 | 99.8 | 1.5 |
| 1998 | 98.1 | 101 | $-5.0 *10^{-7}$ | -19 | $1.2 *10^{-15}$ | 7 | $-1.3 *10^{-24}$ | -4 | 99.8 | 1.4 |
| 2000 | 98.3 | 117 | $-8.9 *10^{-3}$ | -21 | $3.8 *10^{-9}$ | 7 | $-7.4 *10^{-15}$ | -4 | 99.9 | 1.5 |
| 2002 | 98.4 | 120 | $-8.4 *10^{-4}$ | -22 | $3.4 *10^{-9}$ | 8 | $-6.1 *10^{-15}$ | -5 | 99.9 | 1.5 |
| 2004 | 98.7 | 142 | $-7.8 *10^{-4}$ | -26 | $2.8 *10^{-9}$ | 9 | $-4.6 *10^{-15}$ | -5 | 99.9 | 1.5 |
| 2006 | 98.7 | 144 | $-7.1 *10^{-4}$ | -25 | $2.3 *10^{-9}$ | 8 | $-3.6 *10^{-15}$ | -5 | 99.9 | 1.5 |
| 2008 | 98.5 | 130 | $-7.1 *10^{-4}$ | -23 | $2.3 *10^{-9}$ | 8 | $-3.5 *10^{-15}$ | -4 | 99.9 | 1.5 |

**Table 4.8** Coefficients of the polynomial distribution fitting mean income in Philippine

| Year | $a_0$ | T-value | $a_1$ | T-value | $a_2$ | T-value | $a_3$ | T-value | $R^2$ (%) | DW |
|---|---|---|---|---|---|---|---|---|---|---|
| 1991 | 96.0 | 64 | $-4.7 \times 10^{-4}$ | -16 | $9.2 \times 10^{-10}$ | 7 | $-6.4 \times 10^{-16}$ | -5 | 99.6 | 1.1 |
| 1994 | 96.0 | 63 | $-3.6 \times 10^{-4}$ | -16 | $5.3 \times 10^{-10}$ | 7 | $-2.9 \times 10^{-16}$ | -4 | 99.6 | 1.1 |
| 1997 | 95.6 | 58 | $-2.6 \times 10^{-4}$ | -15 | $2.9 \times 10^{-10}$ | 7 | $-1.1 \times 10^{-16}$ | -5 | 99.6 | 1.1 |
| 2000 | 95.6 | 57 | $-2.2 \times 10^{-4}$ | -15 | $2.0 \times 10^{-10}$ | 7 | $-6.6 \times 10^{-17}$ | -5 | 99.6 | 1.1 |
| 2003 | 95.8 | 59 | $-2.3 \times 10^{-4}$ | -15 | $2.3 \times 10^{-10}$ | 7 | $-8.3 \times 10^{-17}$ | -5 | 99.6 | 1.1 |

**Table 4.9** Coefficients of the polynomial distribution fitting mean income from Romania

| Year | $a_0$ | T-value | $a_1$ | T-value | $a_2$ | T-value | $a_3$ | T-value | $R^2$ (%) | DW |
|---|---|---|---|---|---|---|---|---|---|---|
| 2000 | 99.2 | 248 | $-5.2 \times 10^{-6}$ | -44 | $8.2 \times 10^{-14}$ | 9 | $-4.5 \times 10^{-22}$ | -2.7 | 99.9 | 1.6 |
| 2001 | 99.5 | 377 | $-3.2 \times 10^{-6}$ | -64 | $2.6 \times 10^{-14}$ | 11 | $-2.1 \times 10^{-23}$ | -0.7 | 99.9 | 1.7 |
| 2002 | 99.1 | 216 | $-2.5 \times 10^{-6}$ | -37 | $1.8 \times 10^{-14}$ | 7 | $-3.5 \times 10^{-23}$ | -1.3 | 99.9 | 1.5 |
| 2003 | 99.3 | 269 | $-2.1 \times 10^{-6}$ | -45 | $1.1 \times 10^{-14}$ | 7 | $-4.2 \times 10^{-24}$ | -0.3 | 99.9 | 1.3 |
| 2004 | 99.0 | 184 | $-1.5 \times 10^{-7}$ | -31 | $6.5 \times 10^{-15}$ | 5 | $-4.6 \times 10^{-24}$ | -0.6 | 99.9 | 1.2 |
| 2005 | 99.1 | 224 | $-1.5 \times 10^{-3}$ | -41 | $7.4 \times 10^{-7}$ | 9 | $-1.3 \times 10^{-11}$ | -3.1 | 99.9 | 1.3 |
| 2006 | 65.4 | 3 | $6.9 \times 10^{-2}$ | 0.4 | $-2.6 \times 10^{-6}$ | -0.8 | $1.2 \times 10^{-10}$ | 0.8 | 55 | 1.5 |
| 2007 | 99.1 | 214 | $-1.1 \times 10^{-3}$ | -41 | $4.4 \times 10^{-7}$ | 10 | $-7.1 \times 10^{-12}$ | -4.3 | 99.9 | 1.5 |
| 2008 | 98.9 | 173 | $-8.6 \times 10^{-4}$ | -31 | $2.5 \times 10^{-7}$ | 7 | $-2.9 \times 10^{-12}$ | -2.9 | 99.9 | 1.3 |
| 2009 | 99.0 | 189 | $-7.5 \times 10^{-3}$ | -33 | $1.8 \times 10^{-7}$ | 7 | $-1.6 \times 10^{-12}$ | -2.3 | 99.9 | 1.4 |
| 2010 | 99.1 | 209 | $-7.3 \times 10^{-2}$ | -35 | $1.5 \times 10^{-7}$ | 7 | $-1.1 \times 10^{-12}$ | -1.7 | 99.9 | 1.4 |

**Table 4.10** Coefficients of the polynomial distribution fitting mean income in Singapore

| Year | $a_0$ | T-value | $a_1$ | T-value | $a_2$ | T-value | $a_3$ | T-value | $R^2$ (%) | DW |
|---|---|---|---|---|---|---|---|---|---|---|
| 1980 | 93.4 | 44 | $-2.4 \times 10^{-2}$ | -11 | $2.5 \times 10^{-6}$ | 5 | $-9.8 \times 10^{-11}$ | -3 | 99.2 | 1.1 |
| 1990 | 94.2 | 48 | $-8.9 \times 10^{-3}$ | -11 | $3.5 \times 10^{-7}$ | 5 | $-5.1 \times 10^{-12}$ | -3 | 99.3 | 1.1 |
| 2000 | 91.0 | 34 | $-5.8 \times 10^{-3}$ | -8 | $1.5 \times 10^{-7}$ | 4 | $-1.4 \times 10^{-11}$ | -2 | 98.6 | 1.0 |
| 2005 | 94.8 | 51 | $-1.6 \times 10^{-2}$ | -13 | $1.1 \times 10^{-6}$ | 6 | $-2.8 \times 10^{-11}$ | -4 | 99.4 | 1.1 |
| 2006 | 95.0 | 52 | $-1.5 \times 10^{-2}$ | -13 | $1.0 \times 10^{-6}$ | 6 | $-2.3 \times 10^{-11}$ | -4 | 99.5 | 1.1 |
| 2007 | 94.8 | 50 | $-1.4 \times 10^{-2}$ | -13 | $8.8 \times 10^{-7}$ | 6 | $-1.9 \times 10^{-11}$ | -4 | 99.4 | 1.1 |

**Table 4.11** Coefficients of the polynomial distribution fitting mean income from the UK

| Year | $a_0$ | T-value | $a_1$ | T-value | $a_2$ | T-value | $a_3$ | T-value | $R^2$ (%) | DW |
|---|---|---|---|---|---|---|---|---|---|---|
| 1977 | 97.0 | 66 | $-4.7*10^{-3}$ | -13 | $1.0*10^{-7}$ | 5 | $-9.8*10^{-13}$ | -3 | 99.7 | 1.1 |
| 1978 | 97.4 | 78 | $-5.1*10^{-3}$ | -15 | $1.1*10^{-7}$ | 5 | $-1.2*10^{-12}$ | -3 | 99.8 | 1.2 |
| 1979 | 97.0 | 69 | $-4.5*10^{-3}$ | -14 | $9.4*10^{-8}$ | 5 | $-8.6*10^{-13}$ | -3 | 99.8 | 1.2 |
| 1980 | 96.8 | 64 | $-3.8*10^{-3}$ | -13 | $6.5*10^{-8}$ | 5 | $-4.9*10^{-13}$ | -3 | 99.7 | 1.2 |
| 1981 | 97.7 | 83 | $-3.4*10^{-3}$ | -17 | $5.0*10^{-8}$ | 6 | $-3.2*10^{-13}$ | -4 | 99.8 | 1.2 |
| 1982 | 97.7 | 84 | $-3.2*10^{-3}$ | -17 | $4.5*10^{-8}$ | 6 | $-2.8*10^{-13}$ | -4 | 99.8 | 1.2 |
| 1983 | 98.1 | 99 | $-3.0*10^{-3}$ | -20 | $3.8*10^{-8}$ | 7 | $-2.1*10^{-13}$ | -4 | 99.9 | 1.2 |
| 1984 | 98.1 | 91 | $-2.9*10^{-3}$ | -19 | $3.7*10^{-8}$ | 7 | $-2.0*10^{-13}$ | -4 | 99.9 | 1.2 |
| 1985 | 98.1 | 86 | $-2.7*10^{-3}$ | -18 | $3.1*10^{-8}$ | 7 | $-1.5*10^{-13}$ | -4 | 99.8 | 1.2 |
| 1986 | 97.6 | 78 | $-2.5*10^{-3}$ | -17 | $2.8*10^{-8}$ | 7 | $-1.3*10^{-13}$ | -4 | 99.8 | 1.2 |
| 1987 | 97.3 | 71 | $-2.4*10^{-3}$ | -15 | $2.5*10^{-8}$ | 6 | $-1.0*10^{-13}$ | -4 | 99.8 | 1.1 |
| 1988 | 96.5 | 58 | $-2.2*10^{-3}$ | -13 | $2.1*10^{-8}$ | 5 | $-8.4*10^{-14}$ | -3 | 99.6 | 1.1 |
| 1989 | 96.4 | 59 | $-2.0*10^{-3}$ | -13 | $1.9*10^{-8}$ | 5 | $-7.2*10^{-14}$ | -4 | 99.6 | 1.1 |
| 1990 | 96.2 | 55 | $1.8*10^{-3}$ | -12 | $1.5*10^{-8}$ | 5 | $-5.2*10^{-14}$ | -3 | 99.6 | 1.1 |
| 1991 | 96.4 | 58 | $-1.7*10^{-3}$ | -13 | $1.3*10^{-8}$ | 5 | $-4.0*10^{-14}$ | -3 | 99.6 | 1.1 |
| 1992 | 96.9 | 67 | $-1.6*10^{-3}$ | -15 | $1.2*10^{-8}$ | 6 | $-3.5*10^{-14}$ | -4 | 99.7 | 1.2 |
| 1993 | 97.3 | 73 | $-1.6*10^{-3}$ | -16 | $1.1*10^{-8}$ | 6 | $-3.2*10^{-14}$ | -4 | 99.8 | 1.2 |
| 1995 | 97.3 | 76 | $-1.5*10^{-3}$ | -17 | $1.0*10^{-8}$ | 7 | $-2.8*10^{-14}$ | -4 | 99.8 | 1.2 |
| 1996 | 97.4 | 80 | $-1.4*10^{-3}$ | -17 | $9.3*10^{-9}$ | 7 | $-2.4*10^{-14}$ | -4 | 99.8 | 1.2 |
| 1997 | 97.2 | 77 | $-1.4*10^{-3}$ | -17 | $8.5*10^{-9}$ | 7 | $-2.1*10^{-14}$ | -4 | 99.8 | 1.2 |
| 1998 | 97.0 | 71 | $-1.3*10^{-3}$ | -16 | $7.7*10^{-9}$ | 6 | $-1.8*10^{-14}$ | -4 | 99.8 | 1.2 |
| 1999 | 97.0 | 72 | $-1.2*10^{-3}$ | -16 | $6.9*10^{-9}$ | 6 | $-1.5*10^{-14}$ | -4 | 99.8 | 1.2 |
| 2000 | 96.6 | 66 | $-1.2*10^{-3}$ | -15 | $6.6*10^{-9}$ | 6 | $-1.4*10^{-14}$ | -4 | 99.7 | 1.2 |
| 2001 | 96.7 | 70 | $-1.1*10^{-3}$ | -15 | $5.5*10^{-9}$ | 6 | $-1.0*10^{-14}$ | -3 | 99.7 | 1.2 |
| 2002 | 96.7 | 69 | $-1.0*10^{-3}$ | -15 | $5.0*10^{-9}$ | 6 | $-9.3*10^{-15}$ | -4 | 99.7 | 1.2 |
| 2003 | 97.2 | 83 | $-1.0*10^{-3}$ | -18 | $4.4*10^{-9}$ | 7 | $-8*10^{-15}$ | -4 | 99.8 | 1.3 |
| 2004 | 96.8 | 72 | $-9.7*10^{-3}$ | -15 | $4.0*10^{-9}$ | 6 | $-6.8*10^{-15}$ | -3 | 99.8 | 1.2 |
| 2005 | 97.1 | 80 | $-9.2*10^{-3}$ | -17 | $3.5*10^{-9}$ | 6 | $-5.6*10^{-15}$ | -4 | 99.8 | 1.3 |
| 2006 | 97.2 | 82 | $-9.1*10^{-3}$ | -18 | $3.5*10^{-9}$ | 7 | $-5.5*10^{-15}$ | -4 | 99.8 | 1.3 |
| 2007 | 97.1 | 80 | $-8.6*10^{-3}$ | -17 | $3.1*10^{-9}$ | 6 | $-4.6*10^{-15}$ | -4 | 99.8 | 1.3 |
| 2008 | 96.8 | 74 | $-8.4*10^{-3}$ | -16 | $3.0*10^{-9}$ | 6 | $-4.5*10^{-15}$ | -3 | 99.8 | 1.3 |
| 2009 | 97.0 | 78 | $-8.3*10^{-3}$ | -17 | $2.9*10^{-9}$ | 6 | $-4.2*10^{-15}$ | -4 | 99.8 | 1.3 |
| 2010 | 97.3 | 86 | $-7.9*10^{-3}$ | -18 | $2.5*10^{-9}$ | 6 | $-3.4*10^{-15}$ | -4 | 99.8 | 1.3 |
| 2011 | 97.6 | 99 | $-7.9*10^{-3}$ | -21 | $2.4*10^{-9}$ | 7 | $-3.1*10^{-15}$ | -3 | 99.9 | 1.4 |
| 2012 | 97.6 | 95 | $-7.5*10^{-3}$ | -20 | $2.3*10^{-9}$ | 7 | $-2.9*10^{-15}$ | -4 | 99.9 | 1.4 |

**Table 4.12** Coefficients of the polynomial distribution fitting income of inactive people in France

| Year | $a_0$ | T-value | $a_1$ | T-value | $a_2$ | T-value | $a_3$ | T-value | $R^2$ (%) | DW |
|---|---|---|---|---|---|---|---|---|---|---|
| 2003 | 98.2 | 106 | $-1.0*10^{-4}$ | -18 | $5.4*10^{-7}$ | 6 | $-1.3*10^{-14}$ | -3 | 99.8 | 1.5 |
| 2004 | 98.2 | 108 | $-1.0*10^{-5}$ | -18 | $5.3*10^{-7}$ | 6 | $-1.2*10^{-14}$ | -3 | 99.9 | 1.5 |
| 2005 | 98.2 | 106 | $-1.0*10^{-4}$ | -18 | $5.2*10^{-7}$ | 6 | $-1.2*10^{-14}$ | -3 | 99.8 | 1.5 |
| 2006 | 98.2 | 106 | $-1.0*10^{-4}$ | -18 | $5.0*10^{-7}$ | 6 | $-1.1*10^{-14}$ | -3 | 99.8 | 1.5 |
| 2007 | 98.2 | 106 | $-1.0*10^{-4}$ | -18 | $4.9*10^{-7}$ | 6 | $-1.1*10^{-14}$ | -3 | 99.8 | 1.5 |
| 2008 | 98.2 | 105 | $-9.9*10^{-6}$ | -18 | $4.6*10^{-7}$ | 5 | $-1.0*10^{-14}$ | -3 | 99.8 | 1.5 |
| 2009 | 98.1 | 101 | $-1.0*10^{-4}$ | -17 | $4.7*10^{-7}$ | 5 | $-1.0*10^{-14}$ | -3 | 99.8 | 1.5 |

**Table 4.13** Coefficients of the polynomial distribution fitting pensions in the UK

| Year | $a_0$ | T-value | $a_1$ | T-value | $a_2$ | T-value | $a_3$ | T-value | $R^2$ (%) | DW |
|---|---|---|---|---|---|---|---|---|---|---|
| 1977 | 98.5 | 59.0 | $-1.1*10^{-2}$ | -3.0 | $-2.2*10^{-6}$ | -1.0 | $-2.0*10^{-10}$ | -0.6 | 99.7 | 1.8 |
| 1978 | 98.5 | 60.6 | $-7.5*10^{-3}$ | -2.4 | $-3.1*10^{-6}$ | -2.1 | $1.1*10^{-10}$ | 0.5 | 99.8 | 2.0 |
| 1979 | 98.7 | 60.7 | $-8.2*10^{-3}$ | -3.0 | $-7.4*10^{-7}$ | -0.6 | $-1.8*10^{-10}$ | -1.4 | 99.8 | 1.6 |
| 1980 | 98.5 | 53.1 | $-7.2*10^{-3}$ | -2.7 | $-5.1*10^{-7}$ | -0.5 | $-1.1*10^{-10}$ | -1.1 | 99.7 | 1.3 |
| 1981 | 98.4 | 66.2 | $-6.9*10^{-3}$ | -3.9 | $-6.0*10^{-6}$ | -1.1 | $-1.0*10^{-11}$ | -0.2 | 99.8 | 1.8 |
| 1982 | 98.7 | 78.5 | $-8.4*10^{-3}$ | -5.9 | $9.0*10^{-8}$ | 0.2 | $-6.0*10^{-11}$ | -1.9 | 99.8 | 1.4 |
| 1983 | 97.8 | 62.4 | $-7.7*10^{-3}$ | -5.1 | $-2.6*10^{-7}$ | -0.7 | $7.0*10^{-12}$ | 0.3 | 99.7 | 1.8 |
| 1984 | 97.9 | 60.0 | $-6.7*10^{-3}$ | -4.4 | $-2.2*10^{-6}$ | -0.6 | $-2.4*10^{-12}$ | -0.1 | 99.7 | 1.5 |
| 1985 | 97.6 | 60.5 | $-6.2*10^{-3}$ | -4.6 | $-2.1*10^{-6}$ | -0.7 | $3.4*10^{-12}$ | 0.2 | 99.7 | 1.5 |
| 1986 | 98.0 | 72.0 | $-5.9*10^{-3}$ | -5.7 | $-1.7e*10^{-6}$ | -0.8 | $2.3*10^{-12}$ | 0.2 | 99.8 | 1.7 |
| 1987 | 98.0 | 58.8 | $-5.4*10^{-3}$ | -4.7 | $-1.8*10^{-6}$ | -0.9 | $4.3*10^{-12}$ | 0.4 | 99.7 | 1.6 |
| 1988 | 97.9 | 58.2 | $-4.9*10^{-3}$ | -4.6 | $-1.6*10^{-7}$ | -1.0 | $4.0*10^{-12}$ | 0.5 | 99.7 | 1.6 |
| 1989 | 98.1 | 61.9 | $-4.7*10^{-3}$ | -4.9 | $-8.3*10^{-8}$ | -0.5 | $-6.1*10^{-13}$ | -0.1 | 99.7 | 1.3 |
| 1990 | 98.4 | 86.9 | $-5.0*10^{-3}$ | -8.0 | $-1.1*10^{-8}$ | -0.1 | $-1.5*10^{-12}$ | -0.4 | 99.8 | 1.8 |
| 1991 | 98.2 | 81.7 | $-4.8*10^{-3}$ | -7.9 | $-6.3*10^{-9}$ | -0.0 | $-6.3*10^{-13}$ | -0.2 | 99.8 | 1.5 |
| 1992 | 97.9 | 73.9 | $-4.8*10^{-3}$ | -7.8 | $9.7*10^{-9}$ | 0.1 | $-1.9*10^{-13}$ | 0.0 | 99.8 | 1.5 |
| 1993 | 97.8 | 72.8 | $-5.3*10^{-3}$ | -8.8 | $8.3*10^{-8}$ | 1.2 | $-1.9*10^{-12}$ | -1.0 | 99.8 | 1.5 |
| 1995 | 97.8 | 73.0 | $-4.7*10^{-3}$ | -8.3 | $5.1*10^{-8}$ | 0.8 | $-1.1*10^{-12}$ | -0.7 | 99.8 | 1.5 |
| 1996 | 97.7 | 69.4 | $-4.5*10^{-3}$ | -8.2 | $4.6*10^{-8}$ | 0.8 | $-7.9*10^{-13}$ | -0.5 | 99.8 | 1.5 |
| 1997 | 98.0 | 85.3 | $-4.7*10^{-3}$ | -10.7 | $6.2*10^{-8}$ | 1.5 | $-1.0*10^{-12}$ | -1.0 | 99.8 | 1.7 |
| 1998 | 97.8 | 78.0 | $-4.5*10^{-3}$ | -10.0 | $6.7*10^{-8}$ | 1.7 | $-1.1*10^{-12}$ | -1.2 | 99.8 | 1.5 |
| 1999 | 97.8 | 73.3 | $-4.0*10^{-3}$ | -8.9 | $5.1*10^{-8}$ | 1.3 | $-9.8*10^{-13}$ | -1.1 | 99.8 | 1.4 |
| 2000 | 97.9 | 80.9 | $-4.1*10^{-3}$ | -10.4 | $6.6*10^{-8}$ | 2.1 | $-1.1*10^{-12}$ | -1.7 | 99.8 | 1.5 |
| 2001 | 98.1 | 86.6 | $-3.6*10^{-3}$ | -10.6 | $4.6*10^{-8}$ | 1.8 | $-8.4*10^{-13}$ | -1.6 | 99.8 | 1.4 |
| 2002 | 97.9 | 78.8 | $-3.5*10^{-3}$ | -9.7 | $5.7*10^{-8}$ | 2.2 | $-1.1*10^{-12}$ | -2.3 | 99.8 | 1.8 |
| 2003 | 98.0 | 81.3 | $-3.3*10^{-3}$ | -10.4 | $4.1*10^{-8}$ | 2.0 | $-6.1*10^{-13}$ | -1.7 | 99.8 | 1.4 |
| 2004 | 98.3 | 106.6 | $-3.2*10^{-3}$ | -13.8 | $4.9*10^{-8}$ | 3.4 | $-7.8*10^{-13}$ | -3.2 | 99.9 | 1.6 |
| 2005 | 98.6 | 95.8 | $-2.9*10^{-3}$ | -12.1 | $4.0*10^{-8}$ | 2.7 | $-6.6*10^{-13}$ | -2.8 | 99.9 | 1.6 |
| 2006 | 97.62 | 75.1 | $-2.8*10^{-3}$ | -9.2 | $2.5*10^{-8}$ | 1.4 | $-2.9*10^{-13}$ | -1.1 | 99.8 | 1.4 |
| 2007 | 97.76 | 79.5 | $-2.7*10^{-3}$ | -10.0 | $2.9*10^{-8}$ | 1.9 | $-3.8*10^{-13}$ | -1.7 | 99.8 | 1.6 |
| 2008 | 97.89 | 76.7 | $-2.7*10^{-3}$ | -9.8 | $3.5*10^{-8}$ | 2.3 | $-4.8*10^{-13}$ | -2.1 | 99.8 | 1.4 |
| 2009 | 98.03 | 83.2 | $-2.7*10^{-3}$ | -11.4 | $3.4*10^{-8}$ | 2.8 | $-4.0*10^{-13}$ | -2.4 | 99.8 | 1.5 |
| 2010 | 98.05 | 99.4 | $-2.8*10^{-3}$ | -14.7 | $4.5*10^{-8}$ | 4.9 | $-5.2*10^{-13}$ | -4.4 | 99.9 | 1.5 |
| 2011 | 97.80 | 86.6 | $-2.6*10^{-3}$ | -13.0 | $3.5*10^{-8}$ | 3.9 | $-3.3*10^{-13}$ | -3.1 | 99.8 | 1.5 |
| 2012 | 97.93 | 79.8 | $-2.6*10^{-3}$ | -12.5 | $3.4*10^{-8}$ | 3.8 | $-3.0*10^{-13}$ | -2.9 | 99.8 | 1.5 |

**Table 4.14** Coefficients of the polynomial distribution fitting mean disposable expenditure in the UK

| Year | $a_0$ | T-value | $a_1$ | T-value | $a_2$ | T-value | $a_3$ | T-value | $R^2$ (%) | DW |
|---|---|---|---|---|---|---|---|---|---|---|
| 2000/2001 | 97.6 | 83 | $-5.8*10^{-2}$ | -17 | $1.5*10^{-5}$ | 7 | $-1.6*10^{-7}$ | -4 | 99.8 | 1.2 |
| 2001/2002 | 97.8 | 88 | $-5.6*10^{-2}$ | -18 | $1.4*10^{-5}$ | 7 | $-1.4*10^{-7}$ | -4 | 99.8 | 1.2 |
| 2002/2003 | 97.8 | 86 | $-5.5*10^{-2}$ | -18 | $1.3*10^{-5}$ | 7 | $-1.3*10^{-7}$ | -4 | 99.8 | 1.2 |
| 2003/2004 | 94.8 | 54 | $-4.8*10^{-2}$ | -10 | $1.0*10^{-5}$ | 3 | $-9.7*10^{-7}$ | -2.1 | 99.5 | 2.2 |
| 2004/2005 | 97.8 | 87 | $-5.0*10^{-2}$ | -18 | $1.1*10^{-5}$ | 6 | $-1.0*10^{-7}$ | -4 | 99.7 | 1.2 |
| 2005/2006 | 97.8 | 92 | $-5.0*10^{-2}$ | -19 | $1.0*10^{-5}$ | 7 | $-9.8*10^{-7}$ | -4 | 99.8 | 1.2 |
| 2006 | 97.8 | 97 | $-4.7*10^{-2}$ | -20 | $9.7*10^{-5}$ | 7 | $-8.7*10^{-7}$ | -4 | 99.8 | 1.2 |
| 2007 | 98.2 | 94 | $-4.7*10^{-2}$ | -19 | $9.8*10^{-5}$ | 7 | $-8.8*10^{-7}$ | -4 | 99.8 | 1.2 |
| 2008 | 103.4 | 6.4 | $-7.5*10^{-2}$ | -2.2 | $2.8*10^{-5}$ | 1.6 | $-3.6*10^{-7}$ | -1.4 | 71.2 | 1.5 |
| 2009 | 38.1 | 3.2 | $4.5*10^{-2}$ | 0.9 | $-1.2*10^{-3}$ | -0.5 | $9.5*10^{-7}$ | 0.4 | 32.7 | 0.6 |
| 2010 | 98.5 | 111 | $-4.5*10^{-2}$ | -23 | $8.6*10^{-6}$ | 8 | $-7.3*10^{-7}$ | -4 | 99.9 | 1.2 |
| 2011 | 98.1 | 102 | $-4.3*10^{-2}$ | -21 | $8.1*10^{-6}$ | 7 | $-6.7*10^{-7}$ | -4 | 99.8 | 1.2 |
| 2012 | 98.1 | 95 | $-4.3*10^{-2}$ | -19 | $7.8*10^{-6}$ | 6 | $-6.3*10^{-7}$ | -4 | 99.8 | 1.2 |

**Table 4.15** Coefficients of the polynomial distribution fitting mean gross expenditure in the UK

| Year | $a_0$ | T-value | $a_1$ | T-value | $a_2$ | T-value | $a_3$ | T-value | $R^2$ (%) | DW |
|---|---|---|---|---|---|---|---|---|---|---|
| 2000/2001 | 94.1 | 42 | $-45.3*10^{-1}$ | -14 | $5.7*10^{-4}$ | 7 | $-0.04$ | -6 | 99.3 | 0.8 |
| 2001/2002 | 97.5 | 82 | $-5.6*10^{-1}$ | -17 | $1.4*10^{-4}$ | 6 | $-1.4*10^{-9}$ | -4 | 99.8 | 1.2 |
| 2002/2003 | 97.5 | 79 | $-5.5*10^{-2}$ | -16 | $1.3*10^{-4}$ | 6 | $-1.4*10^{-9}$ | -4 | 99.8 | 1.1 |
| 2003/2004 | 97.4 | 75 | $-5.3*10^{-2}$ | -16 | $1.2*10^{-4}$ | 6 | $-1.3*10^{-9}$ | -4 | 99.7 | 1.2 |
| 2004/2005 | 97.4 | 75 | $-5.3*10^{-2}$ | -16 | $1.2*10^{-4}$ | 6 | $-1.3*10^{-9}$ | -4 | 99.7 | 1.2 |
| 2005/2006 | 98.1 | 100 | $-5.7*10^{-2}$ | -20 | $1.4*10^{-4}$ | 7 | $-1.5*10^{-9}$ | -4 | 99.8 | 1.2 |
| 2006 | 97.6 | 89 | $-4.7*10^{-2}$ | -18 | $9.7*10^{-4}$ | 6 | $-8.7*10^{-10}$ | -4 | 99.8 | 1.2 |
| 2007 | 97.9 | 87 | $-4.7*10^{-2}$ | -18 | $9.7*10^{-4}$ | 6 | $-8.8*10^{-10}$ | -4 | 99.8 | 1.2 |
| 2008 | 97.4 | 80 | $-4.7*10^{-2}$ | -17 | $9.9*10^{-4}$ | 6 | $-9.0*10^{-10}$ | -4 | 99.8 | 1.2 |
| 2009 | 97.8 | 94 | $-4.5*10^{-2}$ | -18 | $8.7*10^{-4}$ | 6 | $-7.4*10^{-10}$ | -3 | 99.8 | 1.3 |
| 2010 | 98.2 | 99 | $-4.5*10^{-2}$ | -20 | $8.5*10^{-4}$ | 7 | $-7.2*10^{-10}$ | -4 | 99.8 | 1.2 |
| 2011 | 97.8 | 92 | $-4.3*10^{-2}$ | -18 | $8.1*10^{-4}$ | 6 | $-6.8*10^{-10}$ | -4 | 99.8 | 1.2 |
| 2012 | 98.0 | 91 | $-4.3*10^{-2}$ | -18 | $7.7*10^{-4}$ | 6 | $-6.2*10^{-10}$ | -3 | 99.8 | 1.2 |

**Table 4.16** Coefficients of the polynomial distribution fitting lower limit on disposable expenditure in the UK

| Year | $a_0$ | T-value | $a_1$ | T-value | $a_2$ | T-value | $a_3$ | T-value | $R^2$ (%) | DW |
|---|---|---|---|---|---|---|---|---|---|---|
| 2000/2001 | 97.3 | 74 | $-6.3*10^{-1}$ | -14 | $2.0*10^{-5}$ | 6 | $-2.7*10^{-9}$ | -4 | 99.7 | 1.3 |
| 2001/2002 | 97.2 | 72 | $-5.9*10^{-1}$ | -14 | $1.7*10^{-5}$ | 6 | $-2.2*10^{-9}$ | -4 | 99.7 | 1.3 |
| 2002/2003 | 97.4 | 77 | $-5.6*10^{-1}$ | -15 | $1.6*10^{-5}$ | 6 | $-1.9*10^{-9}$ | -4 | 99.7 | 1.3 |
| 2003/2004 | 97.4 | 77 | $-5.5*10^{-1}$ | -15 | $1.5*10^{-5}$ | 6 | $-1.8*10^{-9}$ | -4 | 99.8 | 1.4 |
| 2004/2005 | 97.4 | 78 | $-5.2*10^{-2}$ | -15 | $1.3*10^{-5}$ | 6 | $-1.5*10^{-9}$ | -4 | 99.8 | 1.3 |
| 2005/2006 | 97.5 | 80 | $-5.2*10^{-1}$ | -16 | $1.3*10^{-5}$ | 6 | $-1.5*10^{-9}$ | -4 | 99.8 | 1.4 |
| 2006 | 97.5 | 80 | $-5.0*10^{-2}$ | -15 | $1.2*10^{-5}$ | 6 | $-1.3*10^{-9}$ | -4 | 99.8 | 1.4 |
| 2007 | 97.6 | 82 | $-4.7*10^{-2}$ | -16 | $1.1*10^{-5}$ | 6 | $-1.1*10^{-9}$ | -4 | 99.8 | 1.3 |
| 2008 | 96.7 | 63 | $-4.2*10^{-2}$ | -13 | $9.1*10^{-6}$ | 6 | $-7.8*10^{-9}$ | -4 | 99.7 | 1.3 |
| 2009 | 97.7 | 84 | $-4.6*10^{-2}$ | -16 | $1.0*10^{-5}$ | 6 | $-1.0*10^{-9}$ | -4 | 99.8 | 1.4 |
| 2010 | 97.7 | 85 | $-4.5*10^{-2}$ | -16 | $1.0*10^{-5}$ | 6 | $-9.9*10^{-10}$ | -4 | 99.8 | 1.4 |
| 2011 | 97.8 | 89 | $-4.2*10^{-2}$ | -17 | $8.9*10^{-6}$ | 6 | $-8.1*10^{-10}$ | -4 | 99.8 | 1.4 |
| 2012 | 97.8 | 87 | $-4.2*10^{-2}$ | -16 | $9.0*10^{-6}$ | 6 | $-8.1*10^{-10}$ | -4 | 99.8 | 1.4 |

**Table 4.17** Coefficients of the polynomial distribution fitting lower limit on gross expenditure in the UK

| Year | $a_0$ | T-value | $a_1$ | T-value | $a_2$ | T-value | $a_3$ | T-value | $R^2$ (%) | DW |
|---|---|---|---|---|---|---|---|---|---|---|
| 2000/2001 | 96.5 | 60 | $-5.6*10^{-2}$ | -12 | $1.6*10^{-5}$ | 5 | $-1.8*10^{-9}$ | -3 | 99.6 | 1.3 |
| 2001/2002 | 96.5 | 60 | $-5.3*10^{-2}$ | -12 | $1.4*10^{-5}$ | 5 | $-1.5*10^{-9}$ | -3 | 99.6 | 1.3 |
| 2002/2003 | 96.8 | 64 | $-5.0*10^{-2}$ | -13 | $1.3*10^{-5}$ | 5 | $-1.3*10^{-9}$ | -4 | 99.7 | 1.3 |
| 2003/2004 | 96.7 | 63 | $-4.9*10^{-2}$ | -13 | $-1.0*10^{-5}$ | 5 | $-1.2*10^{-9}$ | -4 | 99.6 | 1.3 |
| 2004/2005 | 96.7 | 64 | $-4.6*10^{-2}$ | -13 | $1.0*10^{-5}$ | 5 | $-1.0*10^{-9}$ | -3 | 99.6 | 1.3 |
| 2005/2006 | 96.8 | 65 | $-4.6*10^{-2}$ | -13 | $1.1*10^{-5}$ | 6 | $-1.0*10^{-9}$ | -4 | 99.7 | 1.3 |
| 2006 | 96.9 | 65 | $-4.4*10^{-2}$ | -13 | $1.0*10^{-5}$ | 6 | $-9.0*10^{-10}$ | -4 | 99.7 | 1.3 |
| 2007 | 96.9 | 64 | $-4.2*10^{-2}$ | -13 | $9.1*10^{-6}$ | 5 | $-8.0*10^{-10}$ | -4 | 99.7 | 1.3 |
| 2008 | 96.7 | 63 | $-4.2*10^{-2}$ | -13 | $9.1*10^{-6}$ | 6 | $-7.8*10^{-10}$ | -4 | 99.6 | 1.3 |
| 2009 | 97.1 | 69 | $-4.1*10^{-2}$ | -14 | $8.6*10^{-6}$ | 6 | $-7.3*10^{-10}$ | -4 | 99.6 | 1.3 |
| 2010 | 97.1 | 70 | $-4.1*10^{-2}$ | -14 | $8.5*10^{-6}$ | 6 | $-7.1*10^{-10}$ | -4 | 99.6 | 1.3 |
| 2011 | 97.2 | 73 | $-3.8*10^{-2}$ | -15 | $7.5*10^{-6}$ | 6 | $-5.9*10^{-10}$ | -4 | 99.8 | 1.3 |
| 2012 | 97.2 | 73 | $-3.8*10^{-2}$ | -14 | $7.5*10^{-6}$ | 6 | $-5.9*10^{-10}$ | -4 | 99.8 | 1.3 |

**Table 4.18** Coefficients of the polynomial distribution fitting mean expenditure in Uganda

| Year | $a_0$ | T-value | $a_1$ | T-value | $a_2$ | T-value | $a_3$ | T-value | $R^2$ (%) | DW |
|---|---|---|---|---|---|---|---|---|---|---|
| 2003 | 98.8 | 151 | $-5.2*10^{-4}$ | -27 | $1.2*10^{-9}$ | 9 | $-1.2*10^{-13}$ | -5 | 99.9 | 1.4 |
| 2006 | 98.7 | 150 | $-4.6*10^{-4}$ | -28 | $9.7*10^{-10}$ | 9 | $-8.7*10^{-13}$ | -5 | 99.9 | 1.4 |
| 2010 | 98.7 | 15 | $-4.1*10^{-4}$ | -28 | $7.5*10^{-10}$ | 9 | $-5.9*10^{-13}$ | -5 | 99.9 | 1.5 |

**Table 4.19** Coefficients of polynomial distribution fitting mean wealth from France

| Year | $a_0$ | T-value | $a_1$ | T-value | $a_2$ | T-value | $a_3$ | T-value | $R^2$ (%) | DW |
|---|---|---|---|---|---|---|---|---|---|---|
| 1998 | 82.6 | 17 | $-3.1*10^{-4}$ | -4 | $4.6*10^{-9}$ | 2.5 | $-2.1*10^{-16}$ | -2.0 | 93.5 | 0.6 |
| 2004 | 81.9 | 16 | $-2.2*10^{-4}$ | -4 | $2.4*10^{-9}$ | 2.4 | $-8.3*10^{-17}$ | -1.9 | 92.9 | 0.6 |
| 2010 | 81.9 | 16 | $-1.5*10^{-4}$ | -4 | $1.0*10^{-10}$ | 2.4 | $-2.3*10^{-17}$ | -1.9 | 92.8 | 0.6 |

**Table 4.20** Coefficients of polynomial probability density function fitting income in USA

| Year | $a_0$ | T-value | $a_1$ | T-value | $a_2$ | T-value | $a_3$ | T-value | $R^2$ (%) | DW |
|---|---|---|---|---|---|---|---|---|---|---|
| 1967 | $-2.8*10^{-1}$ | -6.8 | $2.6*10^{-1}$ | 13.7 | $-1.6*10^{-2}$ | -8.0 | $3.5*10^{-3}$ | 6.2 | 99.6 | 2.1 |
| 1968 | $-2.2*10^{-1}$ | -5.4 | $2.2*10^{-1}$ | 11.6 | $-1.2*10^{-2}$ | -6.2 | $2.5*10^{-4}$ | 4.5 | 99.5 | 2.1 |
| 1969 | $-2.7*10^{-1}$ | -7.9 | $2.0*10^{-1}$ | 13.6 | $-9.5*10^{-3}$ | -6.2 | $1.5*10^{-4}$ | 3.9 | 99.7 | 1.6 |
| 1970 | $-2.7*10^{-1}$ | -9.2 | $2.0*10^{-1}$ | 14.9 | $-9.3*10^{-3}$ | -6.6 | $1.5*10^{-4}$ | 4.1 | 99.8 | 1.6 |
| 1971 | $-2.5*10^{-1}$ | -8.2 | 0.1 | 13.2 | $-8*10^{-3}$ | -5.5 | $1*10^{-4}$ | 3.2 | 99.8 | 1.7 |
| 1972 | $-3.2*10^{-1}$ | -4.5 | $1.8*10^{-1}$ | 5.6 | $-7.1*10^{-3}$ | -2.0 | $9.3*10^{-5}$ | 0.9 | 99.3 | 1.2 |
| 1973 | $-3.3*10^{-1}$ | -2.7 | $1.8*10^{-1}$ | 3.1 | $-6.4*10^{-3}$ | -1.0 | $7.3*10^{-5}$ | 0.4 | 98.2 | 1.4 |
| 1974 | $-2.7*10^{-1}$ | -4.8 | $1.6*10^{-1}$ | 5.8 | $-4.7*10^{-3}$ | -1.5 | $3.0*10^{-5}$ | 0.3 | 99.5 | 1.3 |
| 1975 | $-2.6*10^{-1}$ | -6.4 | $1.7*10^{-1}$ | 8.6 | $-6.5*10^{-3}$ | -2.9 | $8.5*10^{-5}$ | 1.3 | 99.7 | 2.3 |
| 1976 | $-2.5*10^{-1}$ | -4.6 | $1.5*10^{-1}$ | 5.3 | $-3.8*10^{-3}$ | -1.2 | $7.6*10^{-6}$ | 0.1 | 99.5 | 1.7 |
| 1977 | $-2.7*10^{-1}$ | -3.1 | $1.5*10^{-1}$ | 3.4 | $-3.6*10^{-3}$ | -0.7 | $2.8*10^{-6}$ | 0.02 | 99.1 | 1.7 |
| 1978 | $-2.8*10^{-1}$ | -2.1 | $1.3*10^{-1}$ | 2.0 | $-1.2*10^{-3}$ | -0.1 | $-6.8*10^{-5}$ | -0.3 | 98.4 | 1.0 |
| 1979 | $-3.1*10^{-1}$ | -2.2 | $1.4*10^{-1}$ | 2.1 | $-2.4*10^{-3}$ | -0.3 | $-4.0*10^{-5}$ | -0.1 | 98.3 | 0.8 |
| 1980 | $-2.3*10^{-1}$ | -2.9 | 0.1 | 2.6 | $1*10^{-3}$ | 0.2 | $1*10^{-4}$ | -1.0 | 99.3 | 1.1 |
| 1981 | $-2.1*10^{-1}$ | -3.4 | 0.1 | 2.8 | $2*10^{-3}$ | 0.5 | $1*10^{-4}$ | -1.4 | 99.6 | 1.8 |
| 1982 | $-2.8*10^{-1}$ | -4.3 | $1.3*10^{-1}$ | 4.0 | $-8.5*10^{-4}$ | -0.2 | $-8.2*10^{-5}$ | -0.7 | 99.6 | 1.9 |
| 1983 | $-2.8*10^{-1}$ | -3.1 | 0.1 | 2.6 | $8*10^{-4}$ | 0.1 | -0.0 | -0.9 | 99.4 | 1.9 |
| 1984 | $-2.5*10^{-1}$ | -1.6 | 0.1 | 1.1 | $4*10^{-3}$ | 0.5 | -0.0 | -1.0 | 98.8 | 1.4 |
| 1985 | $-3.9*10^{-1}$ | -1.5 | $1.5*10^{-1}$ | 1.2 | $-2.2*10^{-3}$ | -0.1 | $-5.1*10^{-5}$ | -0.1 | 97.3 | 1.4 |
| 1986 | $-5*10^{-1}$ | -1.3 | 0.1 | 1.0 | $-4*10^{-3}$ | -0.2 | $-8*10^{-6}$ | 0.0 | 96.0 | 1.0 |
| 1987 | $-6.9*10^{-1}$ | -1.5 | 0.2 | 1.3 | $-13*10^{-3}$ | -0.6 | $2*10^{-4}$ | 0.4 | 94.8 | 1.0 |
| 1988 | $-7.7*10^{-1}$ | -1.4 | 0.2 | 1.2 | $-14*10^{-3}$ | -0.6 | $2*10^{-4}$ | 0.4 | 94.5 | 1.1 |
| 1989 | $-1.0*10^{-1}$ | -1.7 | 0.3 | 1.5 | $-23*10^{-3}$ | -0.9 | $5*10^{-4}$ | 0.7 | 94.0 | 1.1 |
| 1990 | $-7.5*10^{-1}$ | -1.4 | 0.2 | 1.2 | $-11*10^{-3}$ | -0.5 | $1*10^{-4}$ | 0.3 | 94.8 | 1.0 |
| 1991 | $-5.5*10^{-1}$ | -1.1 | $1.8*10^{-1}$ | 0.9 | $-4.4*10^{-3}$ | -0.2 | $-5.0*10^{-6}$ | 0.0 | 95.5 | 1.1 |
| 1992 | $-5.2*10^{-1}$ | -1.1 | $1.6*10^{-1}$ | 0.8 | $-2.5*10^{-3}$ | -0.1 | $-5.3*10^{-5}$ | -0.1 | 96.4 | 1.2 |
| 1993 | $-6.1*10^{-1}$ | -1.0 | $1.8*10^{-1}$ | 0.7 | $-3.6*10^{-3}$ | -0.1 | $-2.8*10^{-5}$ | -0.03 | 96.3 | 1.4 |
| 1995 | $-8.7*10^{-1}$ | -1.1 | 0.2 | 0.8 | $-12*10^{-3}$ | -0.3 | $2*10^{-4}$ | 0.2 | 95.5 | 1.6 |
| 1996 | $-1.1*10^{-1}$ | -1.3 | 0.3 | 1.1 | $-24*10^{-3}$ | -0.6 | $5*10^{-4}$ | 0.5 | 94.2 | 1.3 |
| 1997 | $-2.4*10^{-1}$ | -2.1 | 0.8 | 1.9 | $-61*10^{-3}$ | -1.5 | $15*10^{-4}$ | 1.3 | 93.2 | 1.4 |
| 1998 | -3.4 | -3.2 | 1.09 | 3.0 | $-88*10^{-3}$ | -2.5 | $23*10^{-4}$ | 2.2 | 93.6 | 1.7 |
| 1999 | -4.9 | -3.9 | 1.5 | 3.7 | $-123*10^{-3}$ | -3.2 | $32*10^{-4}$ | 2.9 | 94.4 | 2.1 |
| 2000 | -4.6 | -3.8 | 1.3 | 3.6 | $-111*10^{-3}$ | -3.1 | $28*10^{-4}$ | 2.8 | 93.7 | 1.7 |
| 2001 | -4.5 | -3.4 | 1.4 | 3.2 | $-114*10^{-3}$ | -2.7 | $30*10^{-4}$ | 2.5 | 93.9 | 2.0 |
| 2002 | -4.2 | -3.1 | 1.3 | 2.8 | $-111*10^{-3}$ | -2.4 | $29*10^{-4}$ | 2.2 | 93.9 | 1.9 |
| 2003 | -4.4 | -2.9 | 1.3 | 2.7 | $-112*10^{-3}$ | -2.3 | $29*10^{-4}$ | 2.1 | 93.5 | 1.6 |
| 2004 | -3.5 | -2.3 | 1.0 | 2.0 | $-86*10^{-3}$ | -1.7 | $22*10^{-4}$ | 1.5 | 92.7 | 1.3 |
| 2005 | -4.6 | -3.1 | 1.4 | 2.9 | $-113*10^{-3}$ | -2.5 | $29*10^{-4}$ | 2.2 | 93.8 | 1.9 |
| 2006 | -5.3 | -3.1 | 1.6 | 2.8 | $-131*10^{-3}$ | -2.4 | $35*10^{-4}$ | 2.2 | 93.8 | 1.8 |
| 2007 | -5.2 | -3.3 | 1.5 | 3.1 | $-128*10^{-3}$ | -2.7 | $34*10^{-4}$ | 2.4 | 93.9 | 1.7 |
| 2008 | -4.2 | -2.2 | 1.3 | 2.0 | $-106*10^{-3}$ | -1.7 | $28*10^{-4}$ | 1.5 | 93.3 | 1.5 |
| 2009 | -3.8 | -2.0 | 1.1 | 1.8 | $-93*10^{-3}$ | -1.5 | $24*10^{-4}$ | 1.3 | 93.3 | 1.5 |
| 2010 | -3.3 | -1.7 | 1.0 | 1.5 | $-78*10^{-3}$ | -1.2 | $20*10^{-4}$ | 1.1 | 93.5 | 1.6 |
| 2011 | -2.2 | -1.4 | 0.6 | 1.1 | $-42*10^{-3}$ | -0.8 | $10*10^{-4}$ | 0.6 | 94.3 | 1.6 |
| 2012 | -2.8 | -1.4 | 0.8 | 1.2 | $-62*10^{-3}$ | -0.9 | $15*10^{-4}$ | 0.8 | 93.8 | 1.6 |
| 2013 | -2.9 | -1.5 | 0.8 | 1.3 | $-61*10^{-3}$ | -1.0 | $15*10^{-4}$ | 0.8 | 93.9 | 1.5 |

# CHAPTER 5: APPLICATIONS OF POLYNOMIAL FUNCTION TO DYNAMICAL DISTRIBUTION OF INCOME, WEALTH, AND EXPENDITURE

### 5.1 Literature review

The topic of income and wealth distribution is very important considering its influence and the close link to economic development and social stability. One of the recent trends in Econophysics has developed towards the explanation of income distribution by using distributions specific to statistical mechanics and thermodynamics.

Most researchers dealt with income and wealth distribution by considering static models/methods, whereby the analyses were performed for a determined time interval (most usually for one year). This was as a result of the grand canonical distribution, which assumes that the average quantity of money per capita remains unchanged for longer time intervals even though the quantity of money and number of economic agents (i.e. molecules) change as time goes by [42-43]. This is a limitation at least in the case of the applications from statistical mechanics to social systems and in particular to income distribution.

For example, most of the papers analyse national distribution of income by using Pareto distribution for upper income which consists of 1-10% of the population and for the rest (low and middle income) mostly lognormal distribution, Maxwell-Boltzmann, and Bose-Einstein for a certain time interval most often for one year.

Therefore, we try to construct a method which analyses the income, wealth, and/or expenditure distribution in a dynamic manner, where time is taken into account for longer time intervals (more than one year). Thus, we try do that ideally for as long as the data allow that, which enables to have a more robust and reliable distribution. Consequently, the approach that we propose is a more general one, while the traditional (static) ones may be considered as an analogue of a snapshot.

Also, most of the papers that tackle income distribution from the perspective of nominal income, which is represented by the quantity of money per capita or the amount of money that an individual receives in a time interval. However, this is not the most accurate representation of the income when we try to capture the right picture. Instead, real income is a more correct notion used to describe income as inflation can make nominal income to increase without any real improvement in the purchasing power.

## 5.2 Theoretical considerations

Dynamical systems theory tries to describe the changes that occur over time in physical and artificial "systems" [105]. A dynamical system appears when various kinds of conflicting forces interact resulting into some kind of equilibrium partly stable, partly unstable. The relationships created as consequence of interaction of these forces and substances generate a range of possible states for the system. Mathematically, this set of possibilities is called the state space of the system. Consequently, the dimensions of the state space are the variables of the system. The state space of dynamical system have many dimensions each measuring variations in a relevant variable: air pressure, temperature, concentration of a certain chemical, or position in physical space. But Mathematics is the same regardless number of variables the space contains, or the physical or biological process that each dimension is tracking [106].

Branches of Biology, Physics, Economics, and applied mathematics require a detailed analysis of systems using particular laws governing their change which are derived from a suitable theory (Newtonian mechanics, fluid dynamics, mathematical economics, etc.). All these types of models called dynamical systems have two parts the phase space and the dynamics. The phase space of a dynamical system is the collection of all possible states of the system. Each state represents a complete snapshot of the system at some moment in time. The dynamics is a rule that transforms one point in the phase space representing a state of system now into another point representing the state of the system one time unit "later". Mathematically, the dynamics is a function mapping state into other state [105].

Almost all the papers that tackle dynamically income and wealth distribution deal with this by getting the annual values in order to assess their temporal evolution in support of some mathematical models [107-110].

A landmark in this area can be considered the paper of Clementi and Gallegati [111], which is the most illustrative contribution to the temporal evolution of income. This paper studies the evolution of real mean income per capita, real GDP growth (economic cycle) in the case of Italy by using the data which span from 1987 to 2002. Among the main findings of this paper is that they come across the finding that lognormal distribution describes low and middle income of the population, while Pareto law describes the upper income. However, the curvature of lognormal distribution and the slope of Pareto law differ from one year to another. Also, the authors show that personal income and GDP growth are well fitted by a Laplace function. Empirical results regarding growth of Personal Income and GDP in Italy show that they are similar, which indicate a possible common mechanism which may characterise the growth dynamics for both economic indicators and points to the existence of a partial similarity.

### 5.3 Methodology

Let |X| denote the cardinality of a state set X. Assuming |X| is a power of a prime, then X can be a finite field with usual modular arithmetic. Let k denote a state set satisfying the primality condition in order to distinguish it as a finite field. Primality allows the utilisation of the theorem below.

THEOREM (Generalised Lagrange Interpolation). Let k be a finite field. Then any function $f: k^n \to k$ can be represented as a polynomial of degree at most n. In fact, each transition function of a Finite Dynamical System as an element of a polynomial ring $k[x_1, \ldots \ldots x_n]$.
Definition: Let k be a finite field. A Finite Dynamical System $F = (f_1, \ldots \ldots f_n): k^n \to k$ over k is a polynomial dynamical system [112].

Most papers analysing income and wealth distribution make little or no distinction regarding income. Thus, the authors use nominal income i.e. the (physical) quantity of money an individual/household earns and do not consider inflation, which erodes the quantity of goods and services that can be bought with the income earned. Given that study of dynamics of income and wealth is the purpose of this chapter, real income is more useful as it is a dynamic notion that considers income growth over a certain time interval by taking into account the inflation rate for the same time interval (i.e. prices increase). This notion expresses the purchasing power which is a compounded notion of nominal income and inflation [113]. Generally, the data about real income are expressed in values moderated with Consumer Price Index (CPI). There are several definitions regarding CPI. The most accurate are the ones provided by National Institute of Statistics from France [114] and US Department of Labour [115].

Most papers tackle income and wealth distribution in a static manner, most often for a time interval of one year. We aim to analyse the increase or decrease of each income and wealth decile throughout the time for the entirety of the time interval over which the data span.

In our case, as the annual income/ wealth is expressed in income deciles, we consider the annual growth of real/nominal income and/or wealth of a decile of population (individuals or households). In order to calculate this, we use the following formula [116]:

$$growth\ in\ real\ income = \frac{\Delta r_i}{\Delta p_i} (5.1),$$

where $\Delta r_i$ is the annual change in nominal income of decile i and $\Delta p_i$ is annual change in price level in the economy.

We start from the idea that, in order to compare two economic systems, would suffice to identify the polynomial descriptions of two systems and to compare their potential possibilities [100]. Subsequently, on the x axis we represented cumulated growth of real income, nominal income, expenditure, and wealth calculated as the difference from the same decile but from different years. Growth is ranked in increasing order and the values can be negative or positive. On the y axis, we represent the cumulative probability for each decile of population that has a certain growth of income or wealth greater than a certain threshold chosen arbitrarily on the x-axis. The graphical representations will be made by using normal values and not logarithmic values (i.e. not a log-log scale).

We will use the complementary cumulative distribution function. Mathematically, the methodology described for mean income data is as follows: let $a_1, \ldots \ldots, a_n$ be the values assigned for mean income for each decile from the base year (first year analysed) and let $b_1, \ldots \ldots, b_n$ the values assigned for mean income of each decile from the second year (analysed). Second year is chronologically following the first year regardless if this is consecutive or not. Let $\Delta c_1, \Delta c_2, \ldots \ldots, \Delta c_n$ be real numbers such that $\Delta c_1 = b_1 - a_1$ for the first decile and $\Delta c_n = b_n - a_n$ for the n-th decile. Let S be a set such that $S = \{\Delta c_1, \Delta c_2, \ldots \ldots, \Delta c_n\}$. Let $\Delta x_1, \ldots \ldots, \Delta x_n \in S$ be such that $\Delta x_1 < \Delta x_2 < \Delta x_3 \ldots < \Delta x_n$, where $\Delta x_1, \Delta x_2, \ldots \Delta x_n$ are real numbers.

$\Delta c_1, \ldots \ldots, \Delta c_n$ are the values which represent the differences of the deciles in no order regarding their values (can be negative or positive). We introduced $\Delta x_1 < \Delta x_2 \ldots \Delta x_n$, which are exactly the same values as $\Delta_{c1}, \ldots \ldots \Delta_{cn}$, but ranked increasingly according to their values.

Thus, S becomes such that $S = \{\Delta x_1, \ldots \ldots, \Delta x_n\}$. Assuming that $X$ represents the values for cumulated income represented on the x-axis.

$$X_i = \sum \Delta x_i$$

Then $X$ represents the cumulated difference of income, wealth, and expenditure on the x-axis and i=[1,10] and i$\epsilon$N for mean income/expenditure/wealth.

The formula of the complementary cumulative distribution is:

$$\bar{C}(x) = \int_x^\infty P(x)dx \quad (5.2)$$

Where x is the income, wealth, or expenditure difference from the corresponding deciles from two different years considered. In the case of mean income/expenditure/wealth, $\Delta X_1$ is represented in the complementary cdf with a probability of 90% and $\Delta X_2$ is represented in the cdf with a probability of 80%. Similarly, all the other values are assigned decreasing probabilities such that the value for the highest decile is represented in the cdf with a probability of 0%. Let H the set of plots of the graphic. So, H={ $(X_1,90\%)$, $(X_2,80\%)$, $(X_3,70\%)$, $(X_4,60\%)$, $(X_5,50\%)$, $(X_6,40\%)$, $(X_7,30\%)$, $(X_8,20\%)$, $(X_9,10\%)$, $(X_{10}, 0\%)$}. H is the set representing the plots of the complementary cumulative distribution function for mean income increase of a decile of population.

In the case of upper limit on income, the values for the highest decile values for income and wealth were not made available. Thus, n= [1,9] and consequently, in our analysis we omitted the values for this decile and the lowest probability assigned to the probabilistic values for the increase of income/wealth of a decile is 10%. So, in this case I={ $(X_1,90\%)$, $(X_2,80\%)$, $(X_3,70\%)$, $(X_4,60\%)$, $(X_5,50\%)$, $(X_6,40\%)$, $(X_7,30\%)$, $(X_8,20\%)$, $(X_9,10\%)$}.

In the case of lower limit on expenditure, the value corresponding for the first decile is zero. Also, $n = [1,9]$ as the values for the tenth decile were not made available. Therefore, J = {$(X_1, 90\%)$, $(X_2, 80\%)$, $(X_3, 70\%)$, $(X_4, 60\%)$, $(X_5, 50\%)$, $(X_6, 40\%)$, $(X_7, 30\%)$, $(X_8, 20\%)$, $(X_9, 10\%)$}.

The function fitting the data is a polynomial of third degree such that

$$\bar{C}(X) = a_0 + a_1 * X + a_2 * X^2 + a_3 * X^3 \quad (5.3)$$

We illustrate this by giving an actual numerical example. Let us take the case for mean income difference for the years 1988-1987 for Finland.

Income (1987) = (7880, 10807, 12337, 13777, 15144, 16506, 17936, 19606, 22070, 29012)

Income (1988) = (8068, 11030, 12648, 14066, 15407, 16766, 18341, 20110, 22710, 30609)

We calculate the differences between the same deciles from the time interval 1987-1988.

Income (1988-1987) = (188, 223, 311, 289, 263, 260, 405, 504, 640, 1597)

This is the equivalent of $\Delta c_1 \ldots, \Delta c_n$ in the explanatory model.

Now we order the values increasingly according to $\Delta x_1, \ldots \Delta x_n$.

Income.ordered (1988-1987) = (188, 223, 260, 263, 289, 311, 405, 504, 640, 1597)

And now we cumulate them, becoming cumulated income difference

Cumulated.income.ordered (1988-1987) = (188, 411, 671, 934, 1223, 1534, 1939, 2443, 3083, 4680), which is the equivalent of $X_1, \ldots \ldots X_n$.

We used data without inflation since the data provided for Finland, France, and the UK are about real income and expenditure. Only in the case of Romania the data are expressed in nominal terms.

We considered this degree of polynomial distribution to be optimal of all others in describing the data we used, since a higher degree polynomial used does not alter significantly the goodness of the fit and the number of parameters is kept to a minimum.

### 5.4 Data characteristics

We chose countries such as Finland, France, Romania, and the UK which provided data that cover longer time intervals for consecutive years and do not have significant changes that would alter the stability of macroeconomic frame. Namely, we chose data from the countries that have macroeconomic stability and the same currency for the entire time interval considered. Also, we chose countries that have a high credibility of data. However, in the case of Romania the data should be regarded cautiously given the high degree of fiscal evasion.

We use data regarding mean income and upper limit on decile income from Finland [63, 66], France [67-71], Romania [75], and the UK [64]. For these countries the data were available for consecutive years. The time intervals used for the analyses of the data were for one year (when we used data from consecutive years) and for the entire time interval which is the time interval from the first year of the data to the final year of the data analysed. Thus, in the case of Finland the time interval considered is 23 years, for Romania is 10 years, for the UK is for 34 years, and France is 7-8 years (according to each kind of income) and 3 non-consecutive years for mean wealth.

We plan to analyse the evolution of income by taking into account the annual values for income, wealth, and expenditure from consecutive years when the data allow that. We will consider for the calculation of real income in the cases of France and Finland the CPI index. In order to calculate net income in the case of Romania, we use the inflation rate [117]. Also, the wealth data set, which is entirely about France, will be analysed for larger intervals (6 years). Finally, for each country and for each set of data we analyse the annual values from the first and the last year of the data row in order to see how the distribution works in the case of the largest time interval, which is meaningful to our analysis.

### 5.5 Results

The results were obtained using statistical software R. We present some results graphically in Figure 5.1. Income results are shown in the tables 5.1-5.6. In the tables 5.7-5.10, we show the results for expenditure. The table 5.11 exhibits the results for wealth.

Polynomial function applied to cumulative values/probabilities satisfies the necessary mathematical conditions for complementary cumulative distribution function (1.12-1.15).

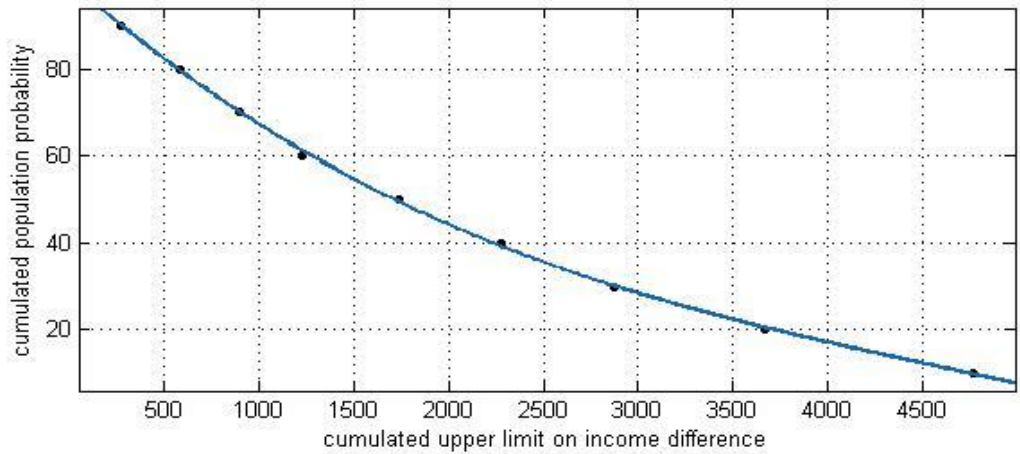

**Figure 5.1 Polynomial dynamical distribution fitting upper disposable income evolution in Finland for the time interval 2003/2002.** On the x-axis, we represented cumulated income difference– $X_i$ and on the y-axis is cumulated probability- $\bar{C}_i$.

We see that in the most cases the distribution has values for coefficient of determination higher than 99%, for all range of income, expenditure or distribution. This shows the power of this approach for several reasons. First, this is true even if the years considered are consecutive or not. Secondly, the countries we considered have different economic characteristics such as level of development, size, geographic location, and macroeconomic stability. Thirdly, the data provided used different methods of calculation such as mean values, lower value, or upper limit. Furthermore, we used in our analysis different types of the same variable (income) such as nominal or real income [118].

In the analysis of income and expenditure, different methodologies for calculation of data apart from mean values were introduced such as upper limit on income for Finland and lower limit on expenditure in the case of the UK. We can notice that the values for coefficient of determination are relatively similar when comparing income and expenditure data, with no significant discrepancies for the same time interval. This is an additional proof regarding the robustness of polynomial dynamical distribution. Also, we notice that the values for parameters are not as stable as in the case of polynomial distribution applied to annual time intervals. This is caused by the differences in the values for the same decile from different years, which can be very different from one year to another (regardless if it is consecutive or not)

Regarding the income and expenditure, we notice that the values for coefficient of determination increase when the time interval between the years analysed is longer or when the degree of the polynomial used is higher. Moreover, there is a trade-off up to a certain extent between the length of time interval and the degree of the polynomial used. Thus, the longer the time interval, the lower can be the degree of the polynomial. This is caused by a general increase of nominal income on the long run (and subsequently for expenditure). Drops in income and expenditure for longer time intervals occur only in cases of severe economic recession. Looking on the small amount of data that we have regarding wealth from France, we see that this characteristic mentioned above for income and expenditure is not applicable to wealth [118-119].

The t values and the results from Durbin Watson test are also exhibited in tables 5.1-5.11. We can notice that for some years, t values are outside the interval from −2 to 2. Also, most of the values from the Durbin-Watson test are in the interval 1.5-2. We see also that in the years characterised by economic contraction or crisis(2008-2012 depending on each country), most parameters have t-values in the interval -2 to 2 the goodness of the fit is lower (lower $R^2$) and also the values from the Durbin-Watson test get away from 2 further than usual. This can be compensated by taking longer time intervals over which we can analyse the evolution of income, expenditure, and wealth. Thus, from the analysis of the time interval which contains the first year and the last year from the data we can see that t values for parameters are far outside the interval -2 to 2, the values for coefficient of determination are high (higher than 99%), and most the values from Durbin Watson test are in the interval 1.5-2, which indicates positive autocorrelation of residuals, inherent to all models. Exceptions for Durbin Watson test are from fitting France wealth data, which shows there is room for improvement regarding wealth.

The overall economic situation affects the shape of the distribution and the values of the parameters. Thus, generally the shape of the distribution looks similar to the static approach of the polynomial distribution for time intervals of one year. In the years with good macroeconomic conditions, the evolution for income or expenditure has similar growth for all deciles and, therefore, we found the similarity with the polynomial distribution applied to annual values static analysis. The values of parameters stay generally the same. Economic or financial crisis distorts to a greater extent the evolution for one or more deciles. Thus, the shape of the distribution becomes quite different and the values for coefficient of determination decrease sometimes dramatically. Also, the values of parameter $a_0$ change significantly and sometimes becomes negative. Moreover, the t values for some parameters become statistically insignificant (between -2 and 2). We can observe on the table that most of these changes occur in the time interval 2008-2012.

A similar approach from analysis of income distribution is that of Clementi and Gallegati [111]. They analyse the shift of the distributions and the evolution of parameters from fitting the annual data using several distributions. They succeed to use a Laplace distribution in order to investigate the time evolution of GDP and personal income growth probability. Moreover, they find that the evolution of these two macroeconomic indicators is linked, pointing to a similar underlying mechanism. Though we analyse the evolution over time of income using a different method, it has applicability to distribution of wealth and expenditure as well. Considering that the same distribution is applicable to expenditure and wealth, this shows that they may have also a similar underlying mechanism as well, similarly with the case described by Clementi and Gallegati for other macroeconomic indicators [111]. This indicates a similarity or partial resemblance regarding underlying mechanism for these economic variables, as pointed out by Clementi and Gallegati [111].

Just like in the case of polynomial distribution applied to annual changes, t-values in the interval (-2, 2) indicate that polynomials with different degrees can be used for the analysis of different time intervals data. Changing number of roots implies qualitative changes in the differential equations modelled by these polynomials as catastrophe theory states [104].

### 5.6 Conclusions

The dynamic approach using polynomial distribution is robust and it can describe successfully the entire range of income, expenditure, and wealth using different methodologies and notions in most of the cases. Even though the results suggest a partial resemblance in the underlying mechanism for the distribution of income, expenditure, and wealth, it is difficult to point the determinants that account for differences in their distribution. The success of the analysis can be increased by extending the time interval we used more often (one year). Thus, the values for goodness of fit can be higher by increasing it in the analysis.

Given that polynomial distribution is generally successful in describing the dynamic evolution of income and wealth distribution, it can be utilised in many applications. First, it can be used in the assessment of income and wealth distribution mechanism in order to see possible theoretical correlations with other economic indicators and theoretical models that regarding income and wealth distribution. Second, polynomial distribution can be used in a variety of economic circumstances. Thus, it can be applied to different types of income (income before distribution, net income), different methodologies to assess income (mean income, upper limit on income, or median income for individuals or households), different currencies (using conversion), and for different segments of population (entire population, active population, pensioners). Third, in macroeconomic analysis it can be used to test different hypothesis regarding the net income and wealth fluctuations for various taxation levels and fiscal regimes. Subsequently, it can be dealt with analysis of inequality. Fourth, in business practice, this distribution allows the dynamic analysis of income and wealth of the groups targeted by a company in the marketing and investment policies [113-114].

## 5.7 Appendix

**Table 5.1** Coefficients of the polynomial fitting dynamical mean income distribution from France [118]

| Time interval | $a_0$ | T-value | $a_1$ | T-value | $a_2$ | T-value | $a_3$ | T-value | $R^2$ (%) | DW |
|---|---|---|---|---|---|---|---|---|---|---|
| 2004/2003 | -685.6 | -2.7 | -5.354 | -2.7 | $-1.1*10^{-2}$ | -2.4 | $-7.8*10^{-6}$ | -2.1 | 70.3 | 0.8 |
| 2005/2004 | 57.0 | 53.4 | $-6.663*10^{-2}$ | -23.3 | $3.1*10^{-5}$ | 5.8 | $-5.5*10^{-9}$ | -3 | 99.2 | 1.3 |
| 2006/2005 | 96.3 | 126.8 | $-5.551*10^{-2}$ | -33.9 | $1.1*10^{-5}$ | 13.6 | $-8.4*10^{-10}$ | -7 | 99.5 | 1.4 |
| 200720/06 | 95.6 | 146.8 | $-5.288*10^{-2}$ | -30.5 | $1.3*10^{-5}$ | 11.4 | $-1.7*10^{-9}$ | -8 | 99.9 | 2.2 |
| 2008/2007 | 95.7 | 60.4 | $-4.589*10^{-2}$ | -12.7 | $6.1*10^{-6}$ | 3.0 | $-1.4*10^{-10}$ | -0.5 | 99.8 | 1.6 |
| 2009/2008 | -2.6 | -0.1 | $-6.767*10^{-2}$ | -1.5 | $1.7*10^{-4}$ | 0.8 | $2.01*10^{-7}$ | 1.1 | 53.8 | 0.5 |
| 2009/2003 | 83.5 | 116.9 | $-1.242*10^{-2}$ | -23.3 | $5.7*10^{-7}$ | 6.3 | $-8.8*10^{-12}$ | -2.2 | 99.9 | 1.3 |

**Table 5.2** Coefficients of the polynomial fitting dynamical upper limit on income distribution from France [119]

| Time interval | $a_0$ | T-value | $a_1$ | T-value | $a_2$ | T-value | $a_3$ | T-value | $R^2$ (%) | DW |
|---|---|---|---|---|---|---|---|---|---|---|
| 2004/2003 | -1997 | -0.7 | -11.04 | -0.7 | $-1.8*10^{-2}$ | -0.7 | $-9.8*10^{-6}$ | -0.7 | 49.11 | 0.4 |
| 2005/2004 | 89.57 | 68 | $-9.4*10^{-2}$ | -12 | $4.4*10^{-5}$ | 4 | $-8.8*10^{-9}$ | -2.5 | 99.23 | 2.2 |
| 2006/2005 | 96.96 | 142 | $-6.0*10^{-2}$ | -30 | $1.4*10^{-5}$ | 10 | $-1.4*10^{-9}$ | -5.2 | 99.35 | 2.0 |
| 2007/2006 | 99.01 | 101 | $-5.0*10^{-2}$ | -19 | $1.3*10^{-5}$ | 7 | $-1.8*10^{-9}$ | -5.5 | 99.56 | 2.3 |
| 2008/2007 | 98.65 | 150 | $-4.9*10^{-2}$ | -27 | $9.1*10^{-6}$ | 7 | $-7.3*10^{-9}$ | -2.9 | 99.67 | 2.4 |
| 2009/2008 | 56.12 | 4.7 | $-6.2*10^{-2}$ | -2.1 | $-5.8*10^{-5}$ | -0.4 | $7.0*10^{-8}$ | 0.6 | 71.65 | 0.4 |
| 2009/2003 | 95.32 | 137 | $-1.2*10^{-2}$ | -23 | $6.4*10^{-7}$ | 6 | $-1.5*10^{-11}$ | -2.9 | 99.24 | 1.8 |

**Table 5.3** Coefficients of the polynomial fitting dynamical mean income distribution from Finland [118]

| Time interval | $a_0$ | T-value | $a_1$ | T-value | $a_2$ | T-value | $a_3$ | T-value | $R^2$ (%) | DW |
|---|---|---|---|---|---|---|---|---|---|---|
| 1988/1987 | 99.07 | 234 | $-4.894*10^{-2}$ | -56 | $7.6*10^{-6}$ | 17 | $-3.5*10^{-10}$ | -5 | 99 | 1.7 |
| 1989/1988 | 97.32 | 131 | $-1.950*10^{-2}$ | -26 | $1.1*10^{-6}$ | 6 | $-2.5*10^{-11}$ | -1.9 | 99 | 1.7 |
| 1990/1989 | 98.20 | 207 | $-1.611*10^{-2}$ | -35 | $5.4*10^{-7}$ | 4 | $7.0*10^{-13}$ | 0.0 | 99 | 2.8 |
| 1991/1990 | 87.04 | 97 | $-6.883*10^{-2}$ | -16 | $2.2*10^{-5}$ | 4 | $-4.0*10^{-9}$ | -2.8 | 99 | 1.3 |
| 1992/1991 | 110.1 | 13 | $1.642*10^{-2}$ | 2.6 | $2.4*10^{-6}$ | 1.8 | $2.2*10^{-10}$ | 2.6 | 99 | 1.3 |
| 1993/1992 | 84.90 | 2.4 | $-1.166*10^{-2}$ | -0.2 | $-9.9*10^{-6}$ | -0.6 | $-9.6*10^{-10}$ | -0.5 | 89 | 0.8 |
| 1994/1993 | 30.03 | 2.1 | $-3.323*10^{-1}$ | -2.0 | $-2.2*10^{-4}$ | -0.6 | $3.4*10^{-6}$ | 1.3 | 59 | 0.4 |
| 1995/1994 | 91.29 | 51 | $-5.526*10^{-2}$ | -12 | $1.3*10^{-5}$ | 5 | $-1.1*10^{-9}$ | -3.3 | 99 | 1.2 |
| 1996/1995 | 88.20 | 67 | $-4.195*10^{-2}$ | -15 | $8.5*10^{-6}$ | 6 | $-7.2*10^{-10}$ | -4 | 99 | 1.3 |
| 1997/1996 | 59.92 | 51 | $-2.710*10^{-2}$ | -17 | $5.2*10^{-6}$ | 6 | $-3.5*10^{-10}$ | -4 | 99 | 1.2 |
| 1998/1997 | 83.37 | 33 | $-3.084*10^{-2}$ | -7 | $4.5*10^{-6}$ | 2.7 | $-2.5*10^{-10}$ | -1.6 | 98 | 1.1 |
| 1999/1998 | 95.82 | 105 | $-2.544*10^{-2}$ | -23 | $1.6*10^{-6}$ | 5 | $-5.8*10^{-12}$ | -0.2 | 99 | 0.7 |
| 2000/1999 | 71.97 | 12 | $-7.739*10^{-2}$ | -2.8 | $3.1*10^{-6}$ | 1.2 | $-4.1*10^{-9}$ | -0.8 | 88 | 0.8 |
| 2001/2000 | 96.3 | 165 | $-2.790*10^{-2}$ | -24 | $1.7*10^{-6}$ | 2.9 | $-3.7*10^{-12}$ | -0.0 | 99 | 2.2 |
| 2002/2001 | 94.17 | 66 | $-2.581*10^{-2}$ | -12 | $2.6*10^{-6}$ | 3.2 | $-1.6*10^{-10}$ | -1.9 | 99 | 1.4 |
| 2003/2002 | 102.4 | 193 | $-3.671*10^{-2}$ | -51 | $4.6*10^{-6}$ | 18 | $-2.1*10^{-10}$ | -9 | 9 | 1.7 |
| 2004/2003 | 92.15 | 89 | $-1.772*10^{-2}$ | -19 | $1.1*10^{-6}$ | 5 | $-2.8*10^{-11}$ | -2.5 | 99 | 1.5 |
| 2005/2004 | 94.33 | 54 | $-2.160*10^{-2}$ | -10 | $1.3*10^{-6}$ | 2.1 | $-2.2*10^{-11}$ | -0.4 | 99 | 1.4 |
| 2006/2005 | 65.77 | 14 | $-1.127*10^{-1}$ | -3.6 | $7.4*10^{-5}$ | 2.0 | $-1.4*10^{-8}$ | -1.6 | 91 | 0.7 |
| 2007/2006 | 85.42 | 38 | $-4.179*10^{-2}$ | -9 | $7.7*10^{-6}$ | 4 | $-5.0*10^{-10}$ | -2.7 | 99 | 1.0 |
| 2008/2007 | 151.9 | 0.2 | 1.530 | 0.2 | $4.9*10^{-4}$ | 0.3 | $5.1*10^{-8}$ | 0.3 | 44 | 0.3 |
| 2009/2008 | 76.47 | 409 | $-1.750*10^{-2}$ | -70 | $1.4*10^{-6}$ | 13 | $-8.7*10^{-11}$ | -7 | 99 | 1.7 |
| 2009/1987 | 94.23 | 94 | $-2.089*10^{-3}$ | -20 | $1.6*10^{-6}$ | 6 | $-4.7*10^{-14}$ | -2.9 | 99 | 1.4 |

**Table 5.4** Coefficients of the polynomial fitting dynamical upper limit on income distribution from Finland [118]

| Time interval | $a_0$ | T-value | $a_1$ | T-value | $a_2$ | T-value | $a_3$ | T-value | $R^2$ (%) | DW |
|---|---|---|---|---|---|---|---|---|---|---|
| 1988/1987 | 96.5 | 256 | $-5.0*10^{-2}$ | -48 | $1.0*10^{-5}$ | 13 | $-7.8*10^{-10}$ | -5 | 99 | 2.6 |
| 1989/1988 | 97.7 | -29 | $-2.0*10^{-2}$ | 6.8 | $1.7*10^{-6}$ | 8 | $-8.1*10^{-11}$ | -4 | 99 | 1.6 |
| 1990/1989 | 100.4 | 190 | $-1.5*10^{-2}$ | -29 | $2.7*10^{-7}$ | 1.9 | $2.9*10^{-11}$ | 2.7 | 099 | 2.5 |
| 1991/1990 | 93.9 | 51 | $-5.5*10^{-2}$ | -7 | $9.1*10^{-6}$ | 1.0 | $-9.0*10^{-10}$ | -0.3 | 99 | 2.0 |
| 1992/1991 | 109.0 | 21 | $1.4*10^{-2}$ | 3.7 | $2.0*10^{-6}$ | 2.2 | $2.0*10^{-10}$ | 3 | 99 | 1.6 |
| 1993/1992 | 103.0 | 57 | $1.7*10^{-2}$ | 7 | $2.5*10^{-6}$ | 2.9 | $4.5*10^{-10}$ | 5 | 99 | 2.0 |
| 1994/1993 | 39.92 | 0.8 | $-1.5*10^{-2}$ | -0.4 | $-1.2*10^{-4}$ | -0.6 | $7.7*10^{-7}$ | 0.3 | 32 | 0.4 |
| 1995/1994 | 92.9 | 300 | $-5.4*10^{-2}$ | -58 | $1.5*10^{-5}$ | 21 | $-1.9*10^{-9}$ | -14 | 99 | 3.0 |
| 1996/1995 | 83.7 | 40 | $-5.4*10^{-2}$ | -7 | $1.8*10^{-5}$ | 3.1 | $-2.7*10^{-9}$ | -2.3 | 99 | 1.4 |
| 1997/1996 | 95.0 | 74 | $-4.2*10^{-2}$ | -18 | $8.9*10^{-6}$ | 8.8 | $-7.4*10^{-10}$ | -6 | 99 | 2.1 |
| 1998/1997 | 87.3 | 46 | $-2.9*10^{-2}$ | -7 | $5.4*10^{-6}$ | 3.0 | $-4.7*10^{-10}$ | -2.1 | 99 | 1.6 |
| 1999/1998 | 96.3 | 254 | $-2.6*10^{-2}$ | -41 | $2.3*10^{-6}$ | 8 | $-8.3*10^{-11}$ | -2.4 | 99 | 3.1 |
| 2000/1999 | 70.4 | 25 | $-8.2*10^{-2}$ | -5 | $4.8*10^{-6}$ | 2.1 | $-1.0*10^{-8}$ | -1.4 | 96 | 1.3 |
| 2001/2000 | 96.3 | -24 | $-2.7*10^{-2}$ | 0.001 | $1.7*10^{-6}$ | 2.9 | $-3.7*10^{-12}$ | -0.04 | 99 | 2.2 |
| 2002/2001 | 95.3 | 64 | $-2.2*10^{-2}$ | -9 | $1.5*10^{-6}$ | 1.7 | $-4.5*10^{-11}$ | -0.4 | 99 | 1.9 |
| 2003/2002 | 99.9 | 94 | $-3.8*10^{-2}$ | -19 | $6.3*10^{-6}$ | 6.7 | $-4.5*10^{-10}$ | -3 | 99 | 2.0 |
| 2004/2003 | 93.7 | 247 | $-1.7*10^{-2}$ | -42 | $1.3*10^{-6}$ | 12 | $-5.3*10^{-11}$ | -6 | 99 | 2.3 |
| 2005/2004 | 93.9 | 50 | $-1.9*10^{-2}$ | -7 | $1.3*10^{-6}$ | 1.4 | $-4.5*10^{-11}$ | -0.5 | 99 | 1.9 |
| 2006/2005 | 59.7 | 9 | $-1.1*10^{-1}$ | -2.28 | $1.1*10^{-4}$ | 1.0 | $-4.2*10^{-8}$ | -0.7 | 85 | 0.7 |
| 2007/2006 | 91.0 | 66 | $-3.2*10^{-2}$ | -11 | $4.6*10^{-6}$ | 3 | $-2.8*10^{-10}$ | -1.5 | 99 | 1.5 |
| 2008/2007 | 0.1 | 0.4 | $2.9*10^{-1}$ | 0.3 | $3.0*10^{-4}$ | 0.5 | $1.5*10^{-7}$ | 0.6 | 66 | 0.4 |
| 2009/2008 | 98.8 | 161 | $-1.7*10^{-2}$ | -26 | $1.0*10^{-6}$ | 5 | $-3.8*10^{-11}$ | -2.6 | 99 | 2.9 |
| 2009/1987 | 96.2 | 161 | $-2.0*10^{-3}$ | -29 | $1.8*10^{-8}$ | 8 | $-8.0*10^{-14}$ | -4 | 99 | 1.7 |

**Table 5.5** Coefficients of the polynomial fitting dynamical mean income distribution from Romania [118]

| Time interval | $a_0$ | T-value | $a_1$ | T-value | $a_2$ | T-value | $a_3$ | T-value | $R^2$ (%) | DW |
|---|---|---|---|---|---|---|---|---|---|---|
| 2001/2000 | 99.8 | -58 | -2.8 | 0.04 | 0.01 | 5.1 | $1.5*10^{-5}$ | 6.5 | 1 | 2.8 |
| 2002/2001 | 94.3 | 187 | -2.6 | -34.3 | $2.0*10^{-2}$ | 6.9 | $-4.5*10^{-5}$ | -1.4 | 99.9 | 1.7 |
| 2003/2002 | 100.2 | 183 | -1.7 | -33.9 | $3.6*10^{-3}$ | 2.8 | $3.5*10^{-5}$ | 4.0 | 99.9 | 2.5 |
| 2004/2003 | 95.9 | 86 | -0.6 | -16.3 | $1.1*10^{-3}$ | 2.9 | $2.8*10^{-8}$ | 0.02 | 99.9 | 1.4 |
| 2005/2004 | 86.2 | 31 | -1.8 | -7.7 | $1.7*10^{-2}$ | 3.8 | $-6.1*10^{-5}$ | -2.8 | 98.8 | 1.1 |
| 2006/2005 | 94.8 | 61 | -0.9 | -15.7 | $3.8*10^{-3}$ | 6.9 | $-6.2*10^{-6}$ | -4.5 | 99.8 | 1.7 |
| 2007/2006 | 98.2 | 110 | -0.2 | -22.6 | $2.8*10^{-4}$ | 6.2 | $-1.3*10^{-7}$ | -2.8 | 99.9 | 1.7 |
| 2008/2007 | 95.8 | 87 | -0.2 | -14.5 | $1.2*10^{-4}$ | 1.7 | $3.2*10^{-8}$ | 0.4 | 99.9 | 1.6 |
| 2009/2008 | 99.1 | 264 | -0.4 | -44.4 | $5.3*10^{-4}$ | 8.7 | $-5.9*10^{-7}$ | -5.2 | 100 | 2.0 |
| 2010/2009 | -255.3 | -0.4 | -21.3 | -0.4 | -0.5 | -0.3 | $-4.1*10^{-3}$ | -0.2 | 31.7 | 0.3 |
| 2010/2000 | 97.74 | 216 | -0.007 | -40.0 | $1.5*10^{-7}$ | 6.7 | $-6.8*10^{-14}$ | -0.09 | 99.9 | 1.2 |

**Table 5.6** Coefficients of the polynomial fitting dynamical mean income distribution from the UK [119]

| Time interval | $a_0$ | T-value | $a_1$ | T-value | $a_2$ | T-value | $a_3$ | T-value | $R^2$ (%) | DW |
|---|---|---|---|---|---|---|---|---|---|---|
| 1978/1977 | 159.6 | 2.4 | $8.3*10^{-2}$ | 1.0 | $2.8*10^{-5}$ | 1.0 | $3.7*10^{-9}$ | 1.3 | 95.5 | 0.8 |
| 1979/1978 | 89.0 | 41.9 | $-3.3*10^{-2}$ | -9.0 | $5.4*10^{-6}$ | 3.8 | $-3.9*10^{-10}$ | -2.6 | 99.4 | 1.3 |
| 1980/1979 | 92.3 | 46.4 | $-2.0*10^{-2}$ | -9.9 | $1.8*10^{-6}$ | 3.5 | $-7.1*10^{-11}$ | -2.1 | 99.6 | 1.4 |
| 1981/1980 | 99.0 | 104.1 | $-2.9*10^{-2}$ | -24.1 | $2.9*10^{-6}$ | 7.4 | $-1.1*10^{-10}$ | -3.1 | 99.9 | 1.8 |
| 1982/1981 | 89.2 | 38.5 | $-7.0*10^{-2}$ | -8.9 | $2.4*10^{-5}$ | 4.0 | $-3.3*10^{-9}$ | -2.8 | 99.3 | 1.3 |
| 1983/1982 | 97.0 | 74.2 | $-5.1*10^{-2}$ | -17.1 | $8.2*10^{-6}$ | 4.6 | $-3.6*10^{-10}$ | -1.3 | 99.8 | 2.2 |
| 1984/1983 | 84.9 | 34.2 | $-5.3*10^{-2}$ | -7.4 | $1.5*10^{-5}$ | 3.5 | $-1.8*10^{-9}$ | -2.6 | 98.9 | 1.2 |
| 1985/1984 | 92.4 | 55.3 | $-3.2*10^{-2}$ | -14.3 | $4.5*10^{-6}$ | 6.3 | $-2.3*10^{-10}$ | -4.0 | 99.7 | 1.3 |
| 1986/1985 | 82.9 | 59.9 | $-4.0*10^{-2}$ | -15.3 | $7.7*10^{-6}$ | 7.2 | $-5.3*10^{-10}$ | -4.9 | 99.6 | 1.2 |
| 1987/1986 | 86.9 | 36.2 | $-2.5*10^{-2}$ | -8.3 | $3.3*10^{-6}$ | 3.8 | $-1.6*10^{-10}$ | -2.7 | 99.1 | 1.3 |
| 1988/1987 | 78.9 | 29.6 | $-1.9*10^{-2}$ | -6.4 | $1.9*10^{-6}$ | 2.6 | $-7.1*10^{-11}$ | -1.7 | 98.2 | 1.1 |
| 1989/1988 | 92.8 | 93.5 | $-3.3*10^{-2}$ | -24.5 | $5.2*10^{-6}$ | 11.6 | $-3.2*10^{-10}$ | -8.1 | 99.9 | 2.0 |
| 1990/1989 | 90.3 | 30.7 | $-1.7*10^{-2}$ | -7.7 | $1.3*10^{-6}$ | 3.5 | $-3.8*10^{-11}$ | -2.3 | 99.0 | 1.0 |
| 1991/1990 | 94.8 | 200.4 | $-1.8*10^{-2}$ | -42.3 | $1.3*10^{-6}$ | 13.2 | $-4.4*10^{-11}$ | -6.7 | 99.9 | 2.8 |
| 1992/1991 | 66.2 | 112.8 | $-3.5*10^{-2}$ | -33.9 | $6.7*10^{-6}$ | 5.1 | $-8.4*10^{-10}$ | -2.2 | 99.9 | 1.6 |
| 1993/1992 | 49.2 | 13.0 | $5.5*10^{-2}$ | -6.1 | $1.6*10^{-5}$ | 3.4 | $-9.8*10^{-9}$ | -2.7 | 93.9 | 1.0 |
| 1995-1994/1993 | 94.8 | 130.3 | $-4.4*10^{-2}$ | -30.4 | $8.4*10^{-6}$ | 11.5 | $-6.8*10^{-10}$ | -6.9 | 99.9 | 2.7 |
| 1996-1995/1995-1994 | 82.3 | 46.3 | $-4.4*10^{-2}$ | -8.7 | $1.0*10^{-5}$ | 3.1 | $-1.0*10^{-9}$ | -2.0 | 99.3 | 1.4 |
| 1997-1996/1996-1995 | 85.8 | 40.9 | $-2.4*10^{-2}$ | -11.3 | $2.9*10^{-6}$ | 6.0 | $-1.1*10^{-10}$ | -4.5 | 99.2 | 1.3 |
| 1998-1997/1997-1996 | 86.8 | 33.6 | $-2.3*10^{-2}$ | -7.6 | $2.5*10^{-6}$ | 3.3 | $-1.1*10^{-10}$ | -2.2 | 99.0 | 1.1 |
| 1999-1988/1998-1997 | 92.0 | 80.6 | $-3.1*10^{-2}$ | -19.0 | $3.7*10^{-6}$ | 6.6 | $-1.4*10^{-10}$ | -3.2 | 99.8 | 1.3 |
| 2000-1999/1999-1988 | 74.5 | 28.8 | $-2.8*10^{-2}$ | -6.9 | $4.4*10^{-6}$ | 3.6 | $-2.4*10^{-10}$ | -2.8 | 97.8 | 1.1 |
| 2001-2000/2000-1999 | 95.2 | 84.7 | $-1.7*10^{-2}$ | -16.5 | $1.3*10^{-6}$ | 5.6 | $-5.7*10^{-10}$ | -3.8 | 99.9 | 1.4 |
| 2002-2001/2001-2000 | 88.7 | 44.1 | $-1.7*10^{-2}$ | -12.0 | $1.3*10^{-6}$ | 6.0 | $-3.8*10^{-11}$ | -4.2 | 99.4 | 1.4 |
| 2003-2002/2002-2001 | 38.6 | 72.7 | $-9.0*10^{-3}$ | -33.2 | $8.4*10^{-7}$ | 13.9 | $-8.4*10^{-11}$ | -4.9 | 99.9 | 2.1 |
| 2004-2003/2003-2002 | 46.1 | 3.6 | $-2.0*10^{-2}$ | -2.4 | $-4.8*10^{-6}$ | -0.6 | $1.1*10^{-9}$ | 0.8 | 65.1 | 0.4 |
| 2005-2004/2004-2003 | 98.4 | 102.9 | $-1.6*10^{-2}$ | -24.3 | $1.1*10^{-6}$ | 9.5 | $-3.6*10^{-11}$ | -6.1 | 99.9 | 2.0 |
| 2006-2005/2005-2004 | 57.3 | 12.4 | $-3.0*10^{-2}$ | -3.8 | $6.9*10^{-6}$ | 1.7 | $-5.5*10^{-10}$ | -1.2 | 90.8 | 0.6 |
| 2007-2006/2006-2005 | 93.2 | 66.3 | $-1.7*10^{-2}$ | -15.8 | $1.2*10^{-6}$ | 5.9 | $-3.4*10^{-11}$ | -3.2 | 99.8 | 1.6 |
| 2008-2007/2007-2006 | 44.3 | 4.3 | $-2.0*10^{-2}$ | -3.4 | $3.8*10^{-7}$ | 0.0 | $5.1*10^{-10}$ | 0.3 | 79.3 | 0.4 |
| 2009-2008/2008-2007 | 53.3 | 9.0 | $-3.2*10^{-2}$ | -3.1 | $8.3*10^{-6}$ | 1.447 | $-7.1*10^{-10}$ | -1.0 | 86.2 | 0.6 |
| 2010-2009/2009-2008 | 79.5 | 44.0 | $-2.1*10^{-2}$ | -7.6 | $1.8*10^{-6}$ | 1.807 | $-4.7*10^{-11}$ | -0.5 | 99.1 | 1.3 |
| 2011-2010/2010-2009 | 78.1 | 0.6 | $-7.4*10^{-3}$ | -0.1 | $-1.3*10^{-6}$ | -0.146 | $-1.9*10^{-11}$ | -0.0 | 77.9 | 0.4 |
| 2012-2011/2011-2010 | 92.4 | 32.3 | $-1.5*10^{-2}$ | -8.9 | $1.1*10^{-6}$ | 5.039 | $-3.2*10^{-11}$ | -3.9 | 99.1 | 1.3 |
| 2012- | 95.4 | 161.5 | $-8.1*10^{-4}$ | -36.5 | $2.5*10^{-6}$ | 12.248 | $-3.3*10^{-15}$ | -6.1 | 99.9 | 1.5 |

| 2011/1977 | | | | | | | | | |

**Table 5.7** Coefficients of the polynomial fitting dynamical mean disposable expenditure distribution in the UK [119]

| Time interval | $a_0$ | T-value | $a_1$ | T-value | $a_2$ | T-value | $a_3$ | T-value | $R^2$ (%) | DW |
|---|---|---|---|---|---|---|---|---|---|---|
| 2002-2001 /2000-2001 | 85.8 | 56.7 | -1.8 | -12.6 | 0.01 | 5.0 | $-5.2*10^{-5}$ | -3.1 | 99.4 | 1.2 |
| 2003-2002 /2001-2000 | 85.9 | 30.7 | -3.6 | -8.2 | 0.06 | 4.7 | $-4.4*10^{-4}$ | -3.7 | 98.7 | 1.5 |
| 2004-2003 /2003-2002 | 83.5 | 28.6 | -2.1 | -7.0 | 0.02 | 3.7 | $-1.0*10^{-4}$ | -2.8 | 98.4 | 1.1 |
| 2005-2004 /2004-2003 | 91.2 | 90.5 | -1.1 | -17.9 | 0.006 | 6.6 | $-1.7*10^{-5}$ | -4.5 | 99.9 | 1.4 |
| 2006-2005 /2005-2004 | 33.6 | 2.6 | -1.3 | -2.9 | 0.01 | 0.5 | $-7.5*10^{-5}$ | -0.2 | 72.5 | 0.5 |
| 2006/2006-2005 | 29.6 | 4.4 | -0.5 | -4.0 | 0.006 | 1.8 | $-2.9*10^{-5}$ | -0.9 | 91.5 | 0.7 |
| 2007/2006 | 2.9 | 0.1 | -0.3 | -0.7 | 0.0 | 0.3 | $3.3*10^{-5}$ | 0.2 | 54.4 | 0.3 |
| 2008/2007 | 45.7 | 4.1 | -0.7 | -2.5 | 0.001 | 0.1 | $1.4*10^{-5}$ | 0.1 | 72.6 | 0.4 |
| 2009/2008 | -12.7 | 0.0 | -3.4 | -0.6 | -0.03 | -0.8 | $-8.8*10^{-5}$ | -0.8 | 48.7 | 0.4 |
| 2010/2009 | 71.6 | 23.9 | -1.0 | -4.9 | 0.006 | 2.0 | $-1.7*10^{-5}$ | -1.3 | 96.6 | 1.0 |
| 2011/2010 | 44.2 | 4.6 | -0.9 | -3.7 | 0.001 | 0.1 | $3.1*10^{-5}$ | 0.2 | 80.6 | 0.4 |
| 2012/2011 | 12.9 | 0.6 | -1.1 | -1.7 | 0.006 | 0.7 | $1.9*10^{-4}$ | 0.9 | 59.6 | 0.3 |
| 2012/2000-2001 | 98.1 | 82.5 | -0.1 | -15.3 | $8.5*10^{-5}$ | 3.0 | $-1.6*10^{-8}$ | -0.8 | 99.9 | 1.6 |

**Table 5.8** Coefficients of the polynomial fitting dynamical mean gross expenditure distribution in the UK [119]

| Time interval | $a_0$ | T-value | $a_1$ | T-value | $a_2$ | T-value | $a_3$ | T-value | $R^2$ (%) | DW |
|---|---|---|---|---|---|---|---|---|---|---|
| 2002-2001 /2000-2001 | 66.7 | 23.3 | -1.6 | -5.5 | $1.8*10^{-2}$ | 2.5 | $-7.8*10^{-5}$ | -1.8 | 96.5 | 0.9 |
| 2003-2002 /2001-2000 | 64.3 | 22.5 | -2.2 | -5.7 | 0.03 | 2.2 | $-2.2*10^{-4}$ | -1.5 | 96.6 | 1.0 |
| 2004-2003 /2003-2002 | 81.6 | 34.1 | -2.3 | -8.7 | $2.9*10^{-2}$ | 5.1 | $-1.3*10^{-4}$ | -4.1 | 98.7 | 1.5 |
| 2005-2004 /2004-2003 | 95.7 | 126.3 | -1.2 | -28.8 | $6.8*10^{-3}$ | 10.8 | $-1.6*10^{-5}$ | -6.6 | 99.9 | 2.6 |
| 2006-2005 /2005-2004 | 132 | 5.6 | 0.4 | 1.9 | $1.0*10^{-3}$ | 1.8 | $1.0*10^{-6}$ | 2.2 | 98.2 | 1.1 |
| 2006/2006-2005 | 91.6 | 78.3 | -0.2 | -16.7 | $2.3*10^{-4}$ | 6.2 | $-1.0*10^{-7}$ | -3.8 | 99.8 | 1.6 |
| 2007/2006 | 20.8 | 1.2 | -1.7 | -2.0 | $9.8*10^{-3}$ | 0.4 | $7.7*10^{-4}$ | 1.0 | 60.0 | 0.4 |
| 2008/2007 | 33.5 | 3.4 | -1.2 | -4.1 | $2.1*10^{-2}$ | 1.5 | $-1.1*10^{-4}$ | -1.2 | 85.1 | 0.6 |
| 2009/2008 | -18.6 | -0.1 | -3.6 | -0.8 | $-3.6*10^{-2}$ | -1.0 | $-9.4*10^{-5}$ | -1.0 | 58.2 | 0.5 |
| 2010/2009 | 75.3 | 42.2 | -0.9 | -8.3 | $5.8*10^{-3}$ | 3.1 | $-1.4*10^{-5}$ | -2.1 | 98.9 | 1.4 |
| 2011/2010 | 60.4 | 13.9 | -1.1 | -4.2 | $9.3*10^{-3}$ | 0.9 | $-3.9*10^{-5}$ | -0.5 | 94.2 | 1.0 |
| 2012/2011 | 2.99 | 0.1 | -0.9 | -0.9 | $1.1*10^{-2}$ | 1.3 | $1.1*10^{-4}$ | 0.3 | 69.4 | 0.4 |
| 2012/2000-2001 | 96.7 | 90.7 | -0.1 | -15.0 | $4*10^{-5}$ | 1.6 | $9.1*10^{-9}$ | 0.5 | 99.9 | 1.9 |

**Table 5.9** Coefficients of the polynomial dynamical fitting lower limit on disposable expenditure distribution in the UK [119]

| Time interval | $a_0$ | T-value | $a_1$ | T-value | $a_2$ | T-value | $a_3$ | T-value | $R^2$ (%) | DW |
|---|---|---|---|---|---|---|---|---|---|---|
| 2002-2001 /2000-2001 | 86.2 | 45.0 | -0.9 | -10.9 | 0.004 | 5.2 | $-8.2*10^{-6}$ | -3.7 | 99.4 | 1.2 |
| 2003-2002 /2001-2000 | 89.0 | 117.0 | -0.9 | -17.0 | 0.003 | 3.0 | $-7.8*10^{-6}$ | -1.4 | 99.9 | 1.5 |
| 2004-2003 /2003-2002 | 42.0 | 5.2 | -2.9 | -3.0 | -0.002 | -0.0 | $2*10^{-3}$ | 0.5 | 85.4 | 0.5 |
| 2005-2004 /2004-2003 | 88.1 | 84.6 | -0.8 | -17.8 | 0.003 | 5.8 | $-4.4*10^{-6}$ | -2.9 | 99.8 | 1.2 |
| 2006-2005 /2005-2004 | 47.4 | 5.9 | -1.9 | -2.8 | 0.03 | 1.1 | $-2*10^{-4}$ | -0.8 | 83.3 | 0.6 |
| 2006/2006-2005 | 87.5 | 60.2 | -1.0 | -11.6 | 0.005 | 4.0 | $-1.2*10^{-5}$ | -2.4 | 99.7 | 1.3 |
| 2007/2006 | 88.7 | 89.2 | -1.0 | -14.9 | 0.004 | 3.5 | $-9.2*10^{-6}$ | -1.6 | 99.8 | 1.5 |
| 2008/2007 | 76.7 | 19.5 | -0.2 | -4.9 | 0.0004 | 2.8 | $-2.0*10^{-7}$ | -2.2 | 95.7 | 0.9 |
| 2009/2008 | 23.1 | 1.4 | 0.8 | 0.9 | 0.001 | 1.0 | $1.0*10^{-6}$ | 1.2 | 86.9 | 0.5 |
| 2010/2009 | 77.5 | 18.9 | -4.9 | -5.6 | 0.1 | 3.6 | $-9*10^{-4}$ | -3.0 | 95.5 | 1.2 |
| 2011/2010 | 88.6 | 117.1 | -0.6 | -18.2 | 0.002 | 4.8 | $-4.9*10^{-6}$ | -3.3 | 99.9 | 2.0 |
| 2012/2011 | 250.8 | 0.1 | 11.9 | 0.0 | -0.8 | -0.0 | $-5*10^{-2}$ | -0.0 | 33.3 | 0.3 |
| 2012/2000-2001 | 88.8 | 141.0 | -0.1 | -26.8 | $9.2*10^{-5}$ | 8.1 | $-2.7*10^{-8}$ | -4.3 | 99.9 | 1.6 |

**Table 5.10** Coefficients of the polynomial dynamical fitting lower limit on gross expenditure distribution in the UK [119]

| Time interval | $a_0$ | T-value | $a_1$ | T-value | $a_2$ | T-value | $a_3$ | T-value | $R^2$ (%) | DW |
|---|---|---|---|---|---|---|---|---|---|---|
| 2002-2001 /2000-2001 | 86.3 | 48.3 | -0.9 | -13.1 | $4.1*10^{-3}$ | 6.8 | $-6.8*10^{-6}$ | -5.1 | 99.4 | 1.3 |
| 2003-2002 /2001-2000 | 89.3 | 128.1 | -1.0 | -19.6 | $5.5*10^{-3}$ | 5.0 | $-2.0*10^{-5}$ | -3.5 | 99.9 | 1.5 |
| 2004-2003 /2003-2002 | 85.0 | 43.7 | -2.6 | -8.6 | 0.03 | 3.3 | $-2.2*10^{-4}$ | -2.2 | 99.3 | 1.5 |
| 2005-2004 /2004-2003 | 87.1 | 70.9 | -0.7 | -16.6 | $2.8*10^{-3}$ | 6.6 | $-3.9*10^{-6}$ | -4.0 | 99.7 | 1.2 |
| 2006-2005 /2005-2004 | 44.0 | 5.9 | -1.8 | -3.4 | 0.03 | 1.7 | $-2.1*10^{-4}$ | -1.3 | 86.2 | 0.7 |
| 2006/2006-2005 | 87.3 | 50.1 | -0.9 | -10.3 | $4.6*10^{-3}$ | 4.0 | $-9.9*10^{-6}$ | -2.7 | 99.5 | 1.2 |
| 2007/2006 | 86.8 | 46.4 | -0.9 | -8.9 | $4.5*10^{-3}$ | 3.2 | $-9.8*10^{-6}$ | -2.1 | 99.5 | 1.3 |
| 2008/2007 | 55.4 | 8.0 | -1.6 | -1.9 | $2.1*10^{-2}$ | 1.0 | $-8.4*10^{-5}$ | -0.8 | 77.8 | 0.4 |
| 2009/2008 | 1223. | 1.2 | 53.7 | 1.1 | 0.7 | 1.0 | $3.7*10^{-3}$ | 0.9 | 29.7 | 0.6 |
| 2010/2009 | 83.0 | 40.7 | -5.5 | -13.5 | 0.1 | 7.9 | $-1*10^{-3}$ | -6.0 | 99.1 | 1.4 |
| 2011/2010 | 88.6 | 108.7 | -0.6 | -17.9 | $1.8*10^{-3}$ | 5.3 | $-3.4*10^{-6}$ | -3.6 | 99.9 | 1.5 |
| 2012/2011 | 100.1 | 0.5 | 0.9 | 0.0 | $3.1*10^{-2}$ | 0.0 | $5*10^{-4}$ | 0.1 | 74.9 | 0.3 |
| 2012/2000-2001 | 88.5 | 122.2 | -0.1 | -24 | $7.9*10^{-5}$ | 7.3 | $-2*10^{-8}$ | -3.7 | 99.9 | 1.6 |

**Table 5.11** Coefficients of the polynomial dynamical fitting mean wealth distribution in France [118]

| Time interval | $a_0$ | T-value | $a_1$ | T-value | $a_2$ | T-value | $a_3$ | T-value | $R^2$ (%) | DW |
|---|---|---|---|---|---|---|---|---|---|---|
| 2004/1998 | 75.0 | 16.8 | $-7*10^{-4}$ | -4.1 | $2.5*10^{-9}$ | 2.2 | $-3.1*10^{-15}$ | -1.7 | 94.0 | 0.7 |
| 2010/2004 | 77.1 | 17.9 | $-3.8*10^{-4}$ | -4.5 | $7.2*10^{-10}$ | 2.5 | $-4.3*10^{-16}$ | -1.9 | 95.1 | 0.7 |
| 2010/1998 | 76.4 | 17.5 | $-2.4*10^{-4}$ | -4.4 | $3.0*10^{-10}$ | 2.4 | $-1.2*10^{-16}$ | -1.8 | 94.7 | 0.7 |

# CHAPTER 6: MATHEMATICAL APPLICATIONS

We try to find a connection between economic theory and dynamical evolution of income, expenditure, and wealth by using standard widespread macroeconomic model for the distributions we introduced. Thus, we use a Hamiltonian based on a polynomial and Fermi-Dirac utility functions. The model upon which our work is basedon is the Ramsey growth model.

### 6.1 Theoretical considerations

Ramsey model is a sophisticated model of optimal saving in a society, which was created by Frank Ramsey [120].The Ramsey growth model is a neoclassical model of economic growth that explains the fundamentals of consumption and capital accumulation in a dynamic real equilibrium setting. It develops the standard Solow growth model by taking into account an endogenous determination of the level of savings [121]. It is one of the most important models in macroeconomics. Time is continuous. The model does not contain any stochastic elements. Households understand exactly how the economy functions and are able to forecast the future path of wages and interest rates. This implies rational expectations, which in non-stochastic setting is equivalent with perfect foresight. This characteristic makes aggregation very simple: the overall behaviour is multiplication of the behaviour of a single household with the number of households. Thus, the household is considered as an infinitely-lived family, whose members act in unity and are concerned about the utility from own and future consumption. Births and population growth are considered as an expansion of the size of already existing infinitely-lived households.

We consider a single household where its preferences can be represented by an additive inter-temporal utility function with a constant rate of time preference.

$$U = \int_0^\infty u(c)\, e^{-\beta t} dt$$

where β is the effective rate of time preference and c is the consumption. The household chooses a consumption-saving plan which maximises U subject to its budget constraint. The decision problem is: choose a plan $(c_t)_0^\infty$ so as to achieve a maximum of U subject to non-negativity of the control variable, $c_t$ and the constraints. In order to solve the problem one shall apply the Maximisation Principle [120]. Assume a life time consumption problem with fixed assets in continuous time. The interest rate in the economy is r. The budget constraint in continuous time can be written as a constraint on the change of the assets and the initial condition [122-123], where $a$ represents the assets:

$$\dot{a} = ra - c \text{ and } a(0) = a_0$$

$$H = u(c)e^{-\beta t} + \lambda(ra - c), \text{ where } \lambda \text{ is the Lagrange multiplier}$$

$$\dot{\lambda} = -\frac{\partial H}{\partial a} = -\lambda r \quad (6.1)$$

$$\frac{\partial H}{\partial c} = e^{-\beta t}u'(c) - \lambda = 0$$

$$\lambda = e^{-\beta t}u'(c) \quad (6.2)$$

$$\dot{\lambda} = -\beta e^{-\beta t}u'(c) + e^{-\beta t}u''(c)\dot{c} \quad (6.3)$$

From (6.1) and (6.3)

$$-\beta e^{-\beta t}u'(c) + e^{-\beta t}u''(c)\dot{c} = -\lambda r \quad (6.4)$$

From (6.2) and (6.4)

$$e^{-\beta t}u''(c)\dot{c} = e^{-\beta t}u'(c)(\beta - r)$$

$$\dot{c} = \frac{u'(c)}{u''(c)}(\beta - r) \quad (6.5)$$

**6.2 The model using polynomial consumption function**

Since in the assessment of income, wealth, and expenditure distribution we used polynomial distribution, we will use utility function a second degree polynomial function.

$$u(c) = p_1 c^2 + p_2 c + p_3$$

$$u'(c) = 2p_1 c + p_2$$

$$u''(c) = 2p_1$$

But we know from the equation (6.5) that

$$\dot{c} = \frac{2p_1 c + p_2}{2p_1}(\beta - r)$$

$$\dot{c} = \left(c + \frac{p_2}{2p_1}\right)(\beta - r)$$

$$\frac{dc}{dt} = \left(c + \frac{p_2}{2p_1}\right)(\beta - r)$$

$$\frac{dc}{c + \frac{p_2}{2p_1}} = dt(\beta - r) \text{ we integrate}$$

$$\int \frac{dc}{c + \frac{p_2}{2p_1}} = \int (\beta - r)dt$$

$$\ln\left(c + \frac{p_2}{2p_1}\right) = (\beta - r)t + \text{const}$$

$$c + \frac{p_2}{2p_1} = e^{(\beta - r)t + const}$$

$$c = e^{(\beta - r)t + const} - \frac{p_2}{2p_1} \quad (6.6)$$

We must note that c>0. Therefore,

$$e^{(\beta - r)t + const} - \frac{p_2}{2p_1} > 0$$

This is in line with theories regarding the consumption and its connection with time preference and interest rate. Namely, this is theoretically linked with the line of thought of the Austrian School starting with Karl Menger in the 19th century.

The whole expression is linked and depends on the values of β and r, as all the other symbols represent constants. Consequently, their difference could be negative or positive. Thus, if their difference is positive, the function is (monotonically) increasing and if negative, the expression is (monotonically) decreasing. Over the time, the value of the exponential can be both negative and positive and, consequently, the graphic of the parameter is similar to the evolution of GDP upwards/downwards. This makes sense as consumption and GDP are positively correlated and consumption is the main component taken into account in calculating GDP. Therefore, when β-r>0, then the whole expression is increasing corresponding to economic boom, while β-r<0, the expression shows economic recession or crisis.

According to the theory of time preference of the Austrian School, the time preference rate is equal to interest rate [124]. According to the equation (6.6), this can occur only as an exception but cannot occur for most of the times. First, β=r implies that the consumption is kept constant (since all the other symbols in the equation are constant). This is clearly not true as consumption is positively correlated with the cyclic evolution of GDP. Second, a constant quantity of a good consumed cannot yield a maximum utility as this contradicts the law of diminishing marginal utility which states that total utility cannot become and stay constant as the first unit of consumption of a good or service yields more utility than the subsequent units [125].

Moreover, the eq. 6 shows the relation between time preference, interest rate, and economic cycle.

### 6.3 The model using Fermi-Dirac consumption function

$$u(c) = \frac{g}{e^{\frac{(c-\mu)}{(kT)}} + 1}$$

We substitute $y = \frac{(c-\mu)}{kT}$ (6.7)

$$u(c) = \frac{g}{e^y + 1}$$

$$u'(c) = -\frac{ge^y}{kT(e^y+1)^2} \quad (6.8)$$

$$u'(c) = -\left(\frac{g}{kT}\right)\left[\frac{e^y + 1 - 1}{(e^y + 1)^2}\right]$$

$$u'(c) = -\left(\frac{g}{kT}\right)\left[\frac{1}{(e^y + 1)} - \frac{1}{(e^y + 1)^2}\right]$$

$$u''(c) = -\left(\frac{g}{k^2T^2}\right)\left[\frac{-e^y}{(e^y + 1)^2} + \frac{2e^y}{(e^y + 1)^3}\right]$$

$$u''(c) = \left(\frac{ge^y}{k^2T^2}\right)\left[\frac{(e^y-1)}{(e^y+1)^3}\right] \quad (6.9)$$

We know from equation (6.5) that

$$\dot{c} = \frac{u'(c)}{u''(c)}(\beta - r)$$

Using the equations (6.8) and (6.9), we get

$$\dot{c} = -kT(\beta - r)\frac{e^y+1}{e^y-1} \quad (6.10)$$

Using the equation (6.7), the equation becomes

$$\frac{dc}{dt} = kT\frac{dy}{dt} \quad (6.11)$$

From the equations (6.10) and (6.11), we get

$$\frac{e^y-1}{e^y+1}dy = -dt(\beta - r) \quad (6.12)$$

We integrate

$$\int \frac{e^y-1}{e^y+1}dy = \int -(\beta - r)\, dt$$

$$\int \frac{e^y-1}{e^y+1}dy = \int (-1)dy + \int \frac{2e^y}{e^y+1}dy \quad (6.13)$$

We substitute
$$e^y = a$$

$$e^y dy = da$$

$$\int \frac{2e^y}{e^y+1}dy = 2\int \frac{1}{a+1}da = 2\ln(a+1) + const = 2\ln(e^y+1) + const \quad (6.14)$$

From the equations (6.13) and (6.14) we get

$$\int \frac{e^y-1}{e^y+1}dy = -y + 2\ln(e^y+1) + const$$

$$\int -(\beta - r)\, dt = -(\beta - r)t + const$$

$$-y + 2\ln(e^y+1) = -(\beta - r)t + const$$

$$Let\ \alpha = -(\beta - r)t + const \quad (6.15)$$

$$e^{-y+2\ln(e^y+1)} = e^\alpha$$

$$e^{-y}e^{2\ln(e^y+1)} = e^\alpha$$

$$e^{-y}e^{\ln(e^y+1)^2} = e^\alpha$$

$$e^{-y}(e^y + 1)^2 = e^\alpha$$

$$e^{-y}(e^{2y} + 2e^y + 1) = e^\alpha$$

$$e^y + 2 + e^{-y} = e^\alpha$$

$$y = \cosh^{-1} \frac{(e^\alpha - 2)}{2}$$

Using the equation (6.15), the equation becomes

$$\frac{c - \mu}{T} = \cosh^{-1} \frac{(e^{-(\beta-r)t+const} - 2)}{2}$$

$$c = \left[\cosh^{-1} \frac{(e^{-(\beta-r)t+const} - 2)}{2}\right] T + \mu$$

### 6.4 The model using Bose-Einstein consumption function

For function
$$u(c) = \frac{g}{e^{(c-\mu)/kT} - 1}$$

We apply the same substitution as in the equation (6.7)

$$u(c) = \frac{g}{e^y - 1}$$

$$u'(c) = -\frac{ge^y}{kT(e^y-1)^2} \quad (6.16)$$

$$u'(c) = -\left(\frac{g}{kT}\right)\frac{e^y - 1 + 1}{(e^y - 1)^2}$$

$$u'(c) = -\left(\frac{g}{kT}\right)\left[\frac{1}{(e^y - 1)} + \frac{1}{(e^y - 1)^2}\right]$$

$$u''(c) = -\left(\frac{g}{k^2T^2}\right)\left[\frac{-e^y}{(e^y - 1)^2} - \frac{2e^y}{(e^y - 1)^3}\right]$$

$$u''(c) = -\left(\frac{ge^y}{k^2T^2}\right)\left[\frac{-1}{(e^y - 1)^2} - \frac{2}{(e^y - 1)^3}\right]$$

$$u''^{(c)} = \left(\frac{ge^y}{k^2T^2}\right)\left[\frac{(e^y-1)}{(e^y-1)^3}\right] \quad (6.17)$$

Using the equation (6.5), we get

$$\dot{c} = \frac{u'^{(c)}}{u''^{(c)}}(\beta - r)$$

$$\dot{c} = -kT(\beta - r)\frac{e^y-1}{e^y+1}$$

$$\frac{dc}{dt} = -kT(\beta - r)\frac{e^y-1}{e^y+1} \quad (6.18)$$

But from the equation (6.7), the equation becomes

$$\frac{dc}{dt} = kT\frac{dy}{dt} \quad (6.19)$$

From the equations (6.18) and (6.19), we get

$$\frac{e^y+1}{e^y-1}dy = -(\beta - r)dt$$

We integrate

$$\int \frac{e^y+1}{e^y-1}dy = \int -(\beta - r)\,dt$$

$$\int \frac{e^y+1}{e^y-1}dy = \int(-1)dy + \int \frac{2e^y}{e^y-1}dy \quad (6.20)$$

$$e^y = a$$

$$e^y dy = da$$

$$\int \frac{2e^y}{e^y-1}dy = 2\int \frac{1}{a-1}da = 2\ln(a-1) + const = 2\ln(e^y - 1) + const \quad (6.21)$$

From the equations (6.20) and (6.21) we get

$$\int \frac{e^y-1}{e^y+1}dy = -y + 2\ln(e^y - 1) + const$$

$$\int -(\beta - r)\,dt = -(\beta - r)t + const$$

$$-y + 2\ln(e^y - 1) = -(\beta - r)t + const$$

$$\text{Let } \alpha = -(\beta - r)t + const$$

$$e^{-y+2\ln(e^y-1)} = e^\alpha$$

$$e^{-y} e^{2\ln(e^y-1)} = e^\alpha$$

$$e^{-y} e^{\ln(e^y-1)^2} = e^\alpha$$

$$e^{-y}(e^y - 1)^2 = e^\alpha$$

$$(e^{2y} - 2e^y + 1)e^{-y} = e^\alpha$$

$$e^y - 2 + e^{-y} = e^\alpha$$

$$y = \cosh^{-1} \frac{(e^\alpha + 2)}{2}$$

$$\frac{c - \mu}{T} = \cosh^{-1} \frac{(e^{-(\beta-r)t+const} + 2)}{2}$$

$$c = \left[\cosh^{-1} \frac{(e^{-(\beta-r)t+const} + 2)}{2}\right] T + \mu$$

For the Fermi-Dirac utility function, in addition to the utility function calculated using polynomial function, we find parameters T and μ. Both parameters explain partially the evolution of consumption. From the previous work of Yakovenko [21], we know that T =M/N and is the total amount of money that each individual or economic agent has. The consumption formula using Fermi-Dirac function is similar up to certain point to the one using polynomial distribution. The part of the expression inside the brackets (as coefficient of T) is a cosh function. Thus, it shows that consumption has fluctuations and is highly positively correlated with income, which is influenced by the cyclic behaviour of the economy. Also, μ is highly correlated with exports. We know also that exports increase the average income and hence, indirectly, the consumption. We notice that there is not a significant difference between the results using Fermi-Dirac distribution and Bose-Einstein distribution.

**6.5 Conclusions**

We show that Ramsey model, using utility function based on polynomial and Fermi-Dirac functions, is in agreement with the theoretical characteristics of consumption, which is described by cyclic behaviour. However, Fermi-Dirac comes up with additional explanations as parameters T and µ (income and exports) prove our functions to be right in describing utility and consumption. Thus, it is shown that consumption is linked with income and exports, in accordance with the correlations found in the third chapter. Also, this brings additional explanations to theoretical approach of time preference of the Austrian School.

# CHAPTER 7: FINAL CONCLUSIONS

We applied different statistical distributions from Physics to income considered more generally. Thus income considered more broadly gives a better insight, as wealth and expenditure are complementary expressions of the same phenomenon. One of the characteristics of this endeavour is that we approached the phenomenon of income distribution using new distributions and methodologies not used so far, new types of data from varied sources, and large amounts of data.

While most of the papers approaching this field use mean values for income and wealth, we extended our study to upper limit on income and lower (bound) limit on income. These methods of measurement were calculated on the basis of population divided in deciles for most of the countries except for the USA data. For the USA, we used different percentages of population calculated according to different mean values for income thresholds. It is noteworthy that regardless the methodology used for calculation, the distributions yielded similar results with no significant differences.

Regarding the economic variables, we emphasise that most papers use disposable nominal income. We extended our analysis to other types of income such as pension. Also, we used a more realistic notion in the approach of income distribution such as real income data, which give a better image of income distribution. Equally important, it is to the best of our knowledge the first approach of this phenomenon using expenditure expressed according to different methods of calculation and different angles of approach.

It is the first attempt to explain distributions utilised in Physics using large amount of data from many countries. Regarding the data, most papers so far generally applied statistical mechanics distributions to developed countries mostly because these countries provide data that are of reliable quality. Most papers of Yakovenko are about data from US, the UK, and Australia, which are developed countries with high income. We extended our applications to a larger pool of countries such as Brazil, Finland, France, Italy, Philippine, Romania, Singapore, the UK, Uganda, and USA. Even though most countries we analysed are developed countries with high income, we took into account also developing countries varying up to underdeveloped countries such as Uganda.

We used logistic distribution, Fermi-Dirac distribution, polynomial distribution (applied statically i.e. to annual values), and a dynamical polynomial distribution which dealt with dynamical evolution of income, wealth and expenditure distribution. The most important finding is that these distributions are robust in describing the distribution of income, wealth, and expenditure.

Logistic distribution is applicable to cumulated values for income and probability. It describes very well the distribution for income, expenditure and wealth. Its applicability to chaotic systems points as to why it is applicable as well to income distribution, as most of trade activities (which distribute the income) do not obey any rule and, therefore, are purely chaotic. Using logistic distribution to fit the annual data, it yielded a coefficient of determination above 99% which shows that the robustness of the distribution is very good.

Fermi-Dirac distribution is applicable since it is the solution of logistic function treated as a differential equation which describes the numerical evolution of animal population in a habitat and is applied on large scale in business and Economics. Also, Fermi-Dirac distribution, similarly with physical systems, is applicable considering that it can describe the occupancy of a certain level of income, expenditure, or wealth. This explanation regarding the application of Fermi-Dirac function we consider to be the second most important finding of our endeavour. The robustness of these distributions is measured using coefficient of determination ($R^2$), which when applied to annual values yielded values for most of the cases above 98%.

Polynomial distribution describes static or dynamical complex systems. Therefore, treating economic systems as a complex system, we applied polynomial distribution to income, expenditure, and wealth distribution. This distribution is applicable both to annual data or multiannual time intervals.

In the case of static polynomial distribution, the robustness of the third degree polynomial distribution is showed by coefficient of determination. Thus, $R^2$ for income and expenditure data was above 99% and for wealth was above 92%. Generally, all the values for parameters are characterised by stability. However, there are small variations from one year to another regarding the same variable. On larger time intervals, the variations regarding the same variable between the initial value and the final value are low, however they are bigger than year-to-year variations.

As for dynamical approach, in the consecutive years analysed which are characterised by economic stability the evolution for income, wealth, or expenditure the deciles have similar evolutions. Consequently, in the analysis of consecutive years the distribution is very similar with static polynomial distribution. The periods of crisis or economic recession distort the evolution of one or more deciles. Therefore, the shape of distribution changes very dramatically sometimes, making it close to impossible for polynomials to fit the data. This shortcoming of the dynamical polynomial distribution can be overcome by extending the time interval between the years considered for analysis or by increasing the degree of the polynomial. In the case of polynomial distribution applied dynamically, the values of the parameters change significantly given the big differences that occur over time intervals especially during crisis times.

In our opinion, this shows the power of logistic distribution, Fermi-Dirac distribution, and polynomial distribution when applied to macroeconomic variables. Fermi-Dirac distribution and polynomial distribution are also applicable as probability density functions to data describing income distribution. We were able to show that on US annual household income data, these being the only data available in such format so that households income to be expressed in different percentages according to different income thresholds arbitrarily chosen. We could see that they both fit the data very well.

In countries such as Finland, France, and Italy we have two sets of data regarding disposable income i.e. upper limit on income data set and mean income data set. Also, in the case of the UK we have data provided using lower limit on decile data. The results obtained from applying the abovementioned distributions to these data yielded different results for Fermi-Dirac distribution which resulted for higher values for coefficient of determination in the case of upper limit values than for the mean values. For the polynomial distribution and the logistic distribution the results were similar, so no significant discrepancies were noticed. We can draw the conclusion that Fermi-Dirac distribution is more sensible to the variations from the tenth decile of income, which is dependent on asset prices, unlike the rest of the population which depends mainly on wages.

Change of currency has also effects on the parameters fitting the data for all distributions considered. In the case of Fermi-Dirac, chemical potential is changed significantly. In the case of polynomial distribution, the values for parameters change significantly, except for the intercept. In the case of logistic distribution the *b* parameter has changed significantly regarding the values for coefficient of determination, there are no significant differences for time intervals of one year. For longer time intervals (decades in the case of Brazil), the values for coefficient of determination were higher compared to the values obtained for time intervals consisting of consecutive years.

Given the applicability of these statistical distributions to distribution of income, wealth, and expenditure calculated according to different methodologies, we conclude that this may point to a certain similarity between these economic variables regarding their underlying mechanism. Also, this confirms that the utilisation of these distributions may point to certain theoretical correlations as was pointed out by Clementi and Galegatti [111].

The distributions are not significantly affected by inflation regarding the goodness of the fit. Thus, the countries analysed with high inflation (Romania and Italy) had high coefficient of determination for the distributions fitting annual data with high level of inflation. It is worth mentioning that data were expressed in nominal terms for Italy and in real terms for Romania (for which we took in the account the inflation rate). No relevant change was observed in the values of the coefficients, when we compared the results from one year to the next provided that the rate of inflation was normal (under 5%). This is interesting especially since the inflation rate was very fluctuating especially in Romania compared to normal standards for an un-inflationary economy. Also, in the dynamical analysis the dynamical polynomial distribution fits well the data when analysing the contiguous years marked by inflation. The results are similar to the years unaffected by inflation.

The third most interesting finding of this article is the correlation between the values for parameters of Fermi-Dirac distribution and macroeconomic variables using the Pearson correlation coefficient (r). Thus, significant correlations were found between exports and chemical potential, degeneracy and Gini coefficient, and coefficient of activity and inflation when we applied it to cumulative income/probabilities set of data. Moreover, the upper limit on income and mean income data sets behave differently when it comes to analogies between variables of Fermi-Dirac distribution and macroeconomic variables. The most illustrative case for this is for France when we analysed the correlation between Gini index and degeneracy obtained from upper limit on income data set and mean income data set. Thus, in the first case it is significantly positively correlated and the latter is weakly negatively correlated (Chapter 3-table 2). For US data, which consisted of non-cumulated data, we applied Fermi-Dirac probability density function. Thus, besides the correlations found in line with Fermi-Dirac findings when applied to cumulative probabilities/income, we could find a significant negative correlation between temperature and income per capita.

The fourth important finding is that the data regarding pensions distribution (using cumulated values for pensions and probabilities) can be modelled using these distributions. This is surprising as the amount of pension for each individual is based not only market principles but on social principles as well. Moreover, in each country analysed the principles behind the amount of pension and share of private/public provision of pensions funding is different, therefore making it more surprising.

We tried to apply our empirical findings to a theoretical model such as Ramsey growth model. Thus, we maximised the utility function using Hamiltonian utilising Fermi-Dirac and polynomial functions as utility functions. The results contradict the consumption evolution as it is stated by the Austrian School regarding the relationship between time preference and interest rate when we used the polynomial function. Moreover, when we used a Fermi-Dirac utility function, the solution is in with line economic theory and with our previous finding showing that consumption is related to income and exports. Also, it shows that consumption exhibits a cyclic behaviour.

Polynomial distribution, logistic distribution, and Fermi-Dirac distribution are other distributions which join the list of other distributions that describe income distribution in broad sense, besides Boltzmann-Gibbs distribution of Yakovenko and Dragulescu [16], the ideal gas model of a closed economic system, which is characterised as fixed with regard to money and number of agents by Chakraborti et al. [44], kinetic exchange models that physicists have developed to understand the reasons and to formulate remedies for income inequalities by Chatterjee et al. [46], the k-generalized distribution as a descriptive model for the size distribution of income of Clementi et al. [47-49]. Also, they join the list of distributions describing income such as the ones introduced by Moura Jr. and Ribeiro [50] showed that the Gompertz curve combined with the Pareto power law provide a good description for whole income distribution. Moreover, the three distributions can describe the evolution for upper segment of income of population similarly to Tsallis [51] which claims to fit the entire income range.

Our endeavour can open the way for additional explorations regarding the applications of statistical physics distribution and complex systems theory to income distribution, macroeconomic models, and to the larger field of Econophysics.

The first area of exploration is about analysing more data regarding expenditure, wealth, and other types of income apart from disposable income. Also, more analysis is necessary using different methodologies for measuring income, wealth, and expenditure. Apart from the three types we have already used, this could be extended for example to median income.

Another area where these distributions could be applied is that of macroeconomic variables. The most promising area is that of taxes and these distributions are applicable to both the direct and the indirect taxes. We have tested this analysis on the data provided by Office of National Statistics from the UK and both distributions work.

A distinct area of analysis is about the correlations between the parameters of Fermi-Dirac distribution and macroeconomic variables as other macroeconomic characteristics may be considered in order to find further analogies. A possibility would be to use indexes of more complex nature for analogy with thermodynamic parameters. Another possible way is about finding other ways of looking at possible analogies. For example, the variables considered could be lagged considering that some effects occur slowly over the time.

Maybe the most important area of further explorations is about the theoretical connections between Perl - Verlhust model (logistic map) and Fermi-Dirac distribution. This research is truly of most interdisciplinary nature. Thus, this formula that was originally applied to numerical evolution of animal populations and then later on applied successfully to economic phenomena, and has as its differential solution the Fermi-Dirac function transcends natural, social, and exact sciences. This approach can lead to very interesting epistemological insights.